\documentclass[a4paper,12pt,tools]{report}
\usepackage{graphicx}
\usepackage{pdfpages}

\usepackage{enumitem,xparse}
\usepackage{longtable}
\usepackage{epsfig}
\usepackage[T1]{fontenc}
\usepackage[french]{babel}
\usepackage{fancyhdr}
\usepackage[ruled,vlined,french,titlenumbered]{algorithm2e}
\usepackage{amsmath}
\usepackage{amssymb}
\usepackage[utf8]{inputenc}
\newtheorem{mydef}{Definition}

\newtheorem{exemple}{Exemple}

\begin{document}

\includepdf[pages=-]{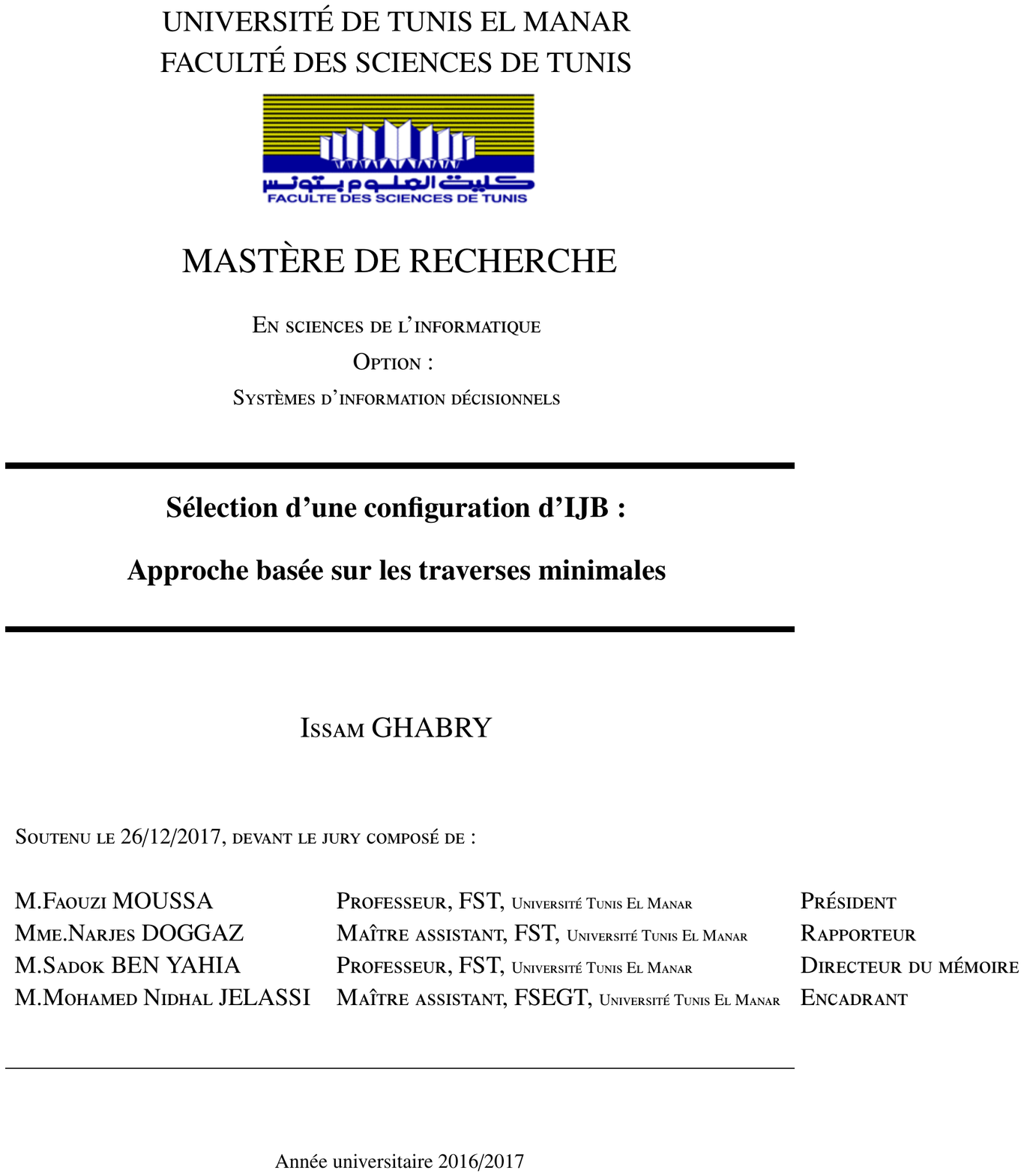}

\leavevmode\thispagestyle{empty}\newpage

\includepdf[pages=-]{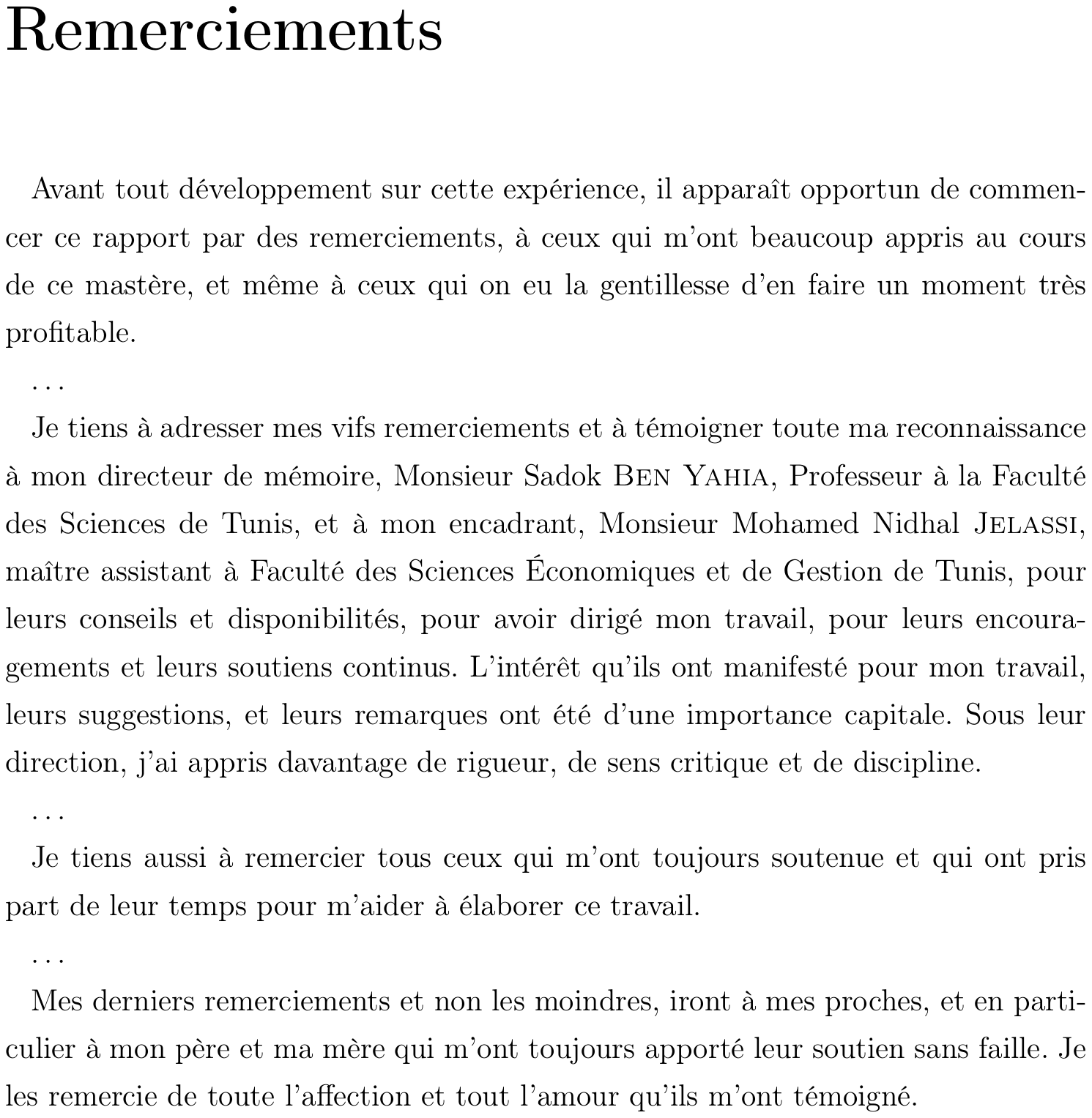}

\newpage
 \pagenumbering{roman}
\thispagestyle{empty} \tableofcontents
\newpage
\thispagestyle{empty} \listoffigures
\newpage
\thispagestyle{empty} \listoftables
\newpage

\pagenumbering{arabic}

\clearpage \addcontentsline{toc}{chapter}{Introduction générale}
\chapter*{Introduction générale}
\markboth{Introduction générale}{Introduction générale}
Dans un monde ultra-concurrentiel où l'innovation fait la différence, l'informatique décisionnelle (ou bien le Business Intelligence [BI] dans la terminologie anglo-saxonne), joue un rôle très important pour les entreprises. Il s’agit d’une panoplie de moyens et de méthodes informatique permettant l’exploitation, l’analyse et l’interrogation des données matérielles ou immatérielles de l’entreprise, afin d’en dégager de nouvelles informations qualitatives qui vont fonder des décisions, qu’elles soient tactiques ou stratégiques.

\indent L’objectif essentiel de l’informatique décisionnelle est d’avoir une vision globale de l’activité en vigueur, anticiper les actions, être en phase avec les attentes de ses clients et pouvoir ainsi adapter les bonnes stratégies, et faciliter la prise de décision et l'analyse décisionnelle des dirigeants pour un meilleur pilotage de l’entreprise. En d’autres termes, la BI unie les données, les technologies, les analyses et les connaissances humaines pour optimiser les perceptions de l’organisation et la conduire ainsi vers le succès.

\indent Une entreprise moderne brasse dans son système d’information une large volumétrie de données répartis dans plusieurs domaines applicatifs. Le fait que ces volumes sont trop importants et de l’hétérogénéité des sources de données, rend très difficile de donner un sens à ces données, de comprendre ce qu’elles expriment : des tendances sous-jacentes, des faiblesses ou des forces cachées, toute chose que l’on doit connaître afin de prendre de bonnes décisions. Toutes ces données sont stockées dans des nouvelles bases de données comme les entrepôts de données ($ED$), qui se caractérisent par : (\textit{i}) leurs sources de données distribuées et hétérogènes; (\textit{ii}) leur modélisation à travers un schéma en étoile; (\textit{iii}) la complexité des requêtes due à des opérations de jointure et d’agrégation; et (\textit{iv}) les exigences des utilisateurs en termes de temps de réponse des requêtes.

\indent Les décideurs sont ainsi des analystes, qui établissent généralement des études du comportement des clients, des produits, des sociétés concurrentes afin de mieux cibler leurs clients, de les fidéliser etc. Selon Bill Inmon, les $ED$ consistent en une collection de données dites données de production orientées sujet, intégrées, non volatiles, historisées et organisées pour supporter un processus d’aide à la décision \cite{Inmon02}. L'objectif principal de ces entrepôts est de permettre une analyse multidimensionnelle.

\indent Les entrepôts de données sont généralement modélisés par un schéma en étoile, qui présente classiquement une table de faits centrale et un ensemble de tables de dimensions. La table de faits contient les clés étrangères des tables de dimensions ainsi qu’un ensemble de mesures collectées durant l’activité de l’organisation. Les tables de dimensions contiennent des données qualitatives, qui représentent des axes sur lesquels les mesures ont été collectées. Dans ce type de modèle, il n'existe aucun lien direct entre deux tables de dimensions quelconques. Ainsi, si nous avons une requête qui fait appel à des attributs faisant partie de deux ou plusieurs tables de dimensions différentes, nous serons obligés de passer par la table de faits qui est très volumineuse (allant de quelques giga-octets à plusieurs téraoctets). Les requêtes $OLAP$ correspondantes à ce type de modèle sont par conséquent très complexes.

\indent Les requêtes définies sur un schéma en étoile (connues également par \textit{requêtes de jointure en étoile}) sont caractérisées par des opérations de sélection sur les tables de dimensions, suivies de jointures avec la table de faits. Aucune jointure n’existe entre les tables de dimensions. Toute jointure doit passer par la table de faits, ce qui rend le coût d’exécution de ces requêtes très important. Sans technique d’optimisation, leurs exécutions peuvent prendre des heures, voire des jours. Pour optimiser ces requêtes, l’administrateur est amené à effectuer une tâche importante: la conception physique. Durant cette phase, l’administrateur choisit un ensemble de techniques d’optimisation à sélectionner. Ces techniques appartiennent à deux catégories : redondantes comme les indexs et les vues matérialisées et non redondantes comme la fragmentation horizontale et le traitement parallèle.

\indent Ce travail vise alors à proposer des solutions destinées à optimiser les performances des bases de données en général et des entrepôts en particulier. Ce rapport est organisé en deux parties : dans la première partie, nous traitons les travaux déjà existants et dans la deuxième partie, nous introduisons notre propre contribution ainsi que son évaluation.

\indent Dans ce but, nous voulons proposer une nouvelle approche de sélection d'$IJB$ basée sur la notion de traverse minimale. Une étude comparative a été faite avec d’autres approches déjà proposées, qui permettent de déterminer les attributs indexables. Ces approches nous permettent une phase d'élagage qui réduit l’espace de recherche initial en ne gardant que les attributs qui répondent à certains critères de sélection tels que la fréquence d’utilisation, la cardinalité des tables, etc. Ces attributs sont susceptibles de réduire le temps de réponse des requêtes. Les tests réalisés sur les bancs d’essais $TPC$-$H$ et $SSB1$ validés sur le SGBD \textsc{Oracle} démontrent les bonnes performances de l’approche proposée.
\section*{Organisation du mémoire}

\indent L’organisation de notre rapport se présente comme suit : 

\begin{itemize}
\item Dans \textbf{le premier chapitre}, nous présentons l'entrepôt de données son architecture, sa modélisation et ses techniques d’optimisation.
\item Dans \textbf{le deuxième chapitre}, nous nous concentrons sur les $IJB$ en les définissant puis nous allons formaliser le problème de sélection des $IJB$ et nous finissons par la présentation de l’état de l’art.
\item Dans \textbf{le troisième chapitre}, nous allons présenter la théorie des hypergraphes avec quelques définitions sur ses concepts de bases, et par la suite nous introduisons notre approche basée sur les traverses minimales.
\item Dans \textbf{le quatrième chapitre}, nous validons notre approche par une étude expérimentale.
\end{itemize}

\chapter{Les entrepôts de données}
\section*{Introduction}
Un entrepôt de données, ou data Warehouse, est une vision centralisée et universelle de toutes les informations de l'entreprise \cite{Grim}. C'est une structure (similaire à une base de données) qui a pour objectif regrouper les données de l'entreprise pour des fins analytiques et pour offrir de l'aide à la prise de décisions stratégiques. La décision stratégique est une action vitale pour les décideurs de l'entreprise, vise à améliorer, quantitativement et/ou qualitativement, la performance de l'entreprise. En gros, c'est un gigantesque tas d'informations historisées, organisées, épurées, intégrées et provenant de plusieurs sources de données, servant aux analyses et à l'aide à la décision \cite{AB11}. L'entrepôt de données est l'élément central de l'informatique décisionnelle. L'informatique décisionnelle a connu et connaît aujourd'hui encore un essor important. Elle permet l’exploitation des données d’une organisation dans le but de faciliter la prise de décision. En effet, l'entrepôt de données est le meilleur moyen que les chercheurs et les professionnels ont trouvé pour modéliser de l'information pour des fins d'analyse.
\section{Concepts fondamentaux}
\subsection{Architecture des entrepôts de données}
Les bases de données traditionnelles, appelées bases de données opérationnelles, sont conçues pour créer des applications transactionnelles pour l'usage quotidien des entreprises. Elles visent à assurer des accès rapides et simultanés aux données tout en garantissant la cohérence. Elles sont basées sur le paradigme de traitement de transaction en ligne ($OLTP$). Ces bases de données sont conçues avec un degré de normalisation élevé en utilisant des dépendances fonctionnelles et des formes normales \cite{EN14}. Généralement, les transactions $OLTP$  renvoient quelques enregistrements \cite{EN14}. D'autre part, les exigences analytiques, imposées par les grandes entreprises, regroupent un grand nombre de données historisées de plusieurs tables (en utilisant des jointures). Cependant, les bases de données opérationnelles n'acceptent pas les requêtes analytiques. C'est parce qu'elles n'ont pas été conçues pour stocker les données historisées et les traiter efficacement. Ces limitations ont largement contribué à la naissance d'une nouvelle technologie qui est l'entrepôt de données qui supporte le traitement analytique en ligne ($OLAP$).\\
\indent L’entrepôt de données constitue le socle d’un système décisionnel pour une entreprise. Il stocke des données judicieuses aux besoins du processus d’aide à la décision. Contrairement à une base de données classique supportant des requêtes transactionnelles de type $OLTP$ (On-Line Transaction Processing), un entrepôt de données est conçu pour supporter des requêtes multidimensionnelles de type $OLAP$. La table \ref{tablll} présente une comparaison entre le système d’information décisionnel et le système d’information opérationnel.

\begin{small}

\begin{table}[]
\centering
\caption{Comparaison entre le SIO et le SID}
\label{tablll}
\begin{tabular}{l|l|l|}
\cline{2-3}
                                                                                               & Système opérationnel                                                                                (SIO) & Système décisionnel (SID)                                                                                    \\ \hline
\multicolumn{1}{|l|}{Conception}                                                               & \begin{tabular}[c]{@{}l@{}}Orientée Application,\\ Structure statique\end{tabular}                       & \begin{tabular}[c]{@{}l@{}}Orientée Sujet, Structure \\évolutive (en étoile, flocon)\end{tabular}      \\ \hline
\multicolumn{1}{|l|}{\begin{tabular}[c]{@{}l@{}}Principe\\ de conception\end{tabular}}         & \begin{tabular}[c]{@{}l@{}}Conception relationnelle \\ (Troisième forme normale)\end{tabular}            & Conception Multidimensionnelle                                                                         \\ \hline
\multicolumn{1}{|l|}{Objectif}                                                                 & Gestion \& Production                                                                                    & Consultation \& Analyse                                                                                \\ \hline
\multicolumn{1}{|l|}{\begin{tabular}[c]{@{}l@{}}Type\\ de données\end{tabular}}                & \begin{tabular}[c]{@{}l@{}}Détaillées, non agrégées, Récentes, \\ mises à jour, Normalisées\end{tabular} & \begin{tabular}[c]{@{}l@{}}Résumées, recalculées, agrégées,\\ Historiques, Dé-normalisées\end{tabular} \\ \hline
\multicolumn{1}{|l|}{\begin{tabular}[c]{@{}l@{}}Organisation\\ des données\end{tabular}}              & Par traitement                                                                                           & Par métier                                                                                             \\ \hline
\multicolumn{1}{|l|}{Utilisateur}                                                              & Employé                                                                                                  & Décideurs, analystes                                                                                   \\ \hline
\multicolumn{1}{|l|}{\begin{tabular}[c]{@{}l@{}}Interaction\\ avec l’utilisateur\end{tabular}} & \begin{tabular}[c]{@{}l@{}}Insertion, Modification, \\ Interrogation, Suppression\end{tabular}           & Interrogation                                                                                          \\ \hline
\multicolumn{1}{|l|}{Requêtes}                                                                 & \begin{tabular}[c]{@{}l@{}}Simples, prédéterminées, \\ données détaillées\end{tabular}                   & \begin{tabular}[c]{@{}l@{}}Complexes, ad-hoc, spécifiques, \\ agrégations et group by\end{tabular}     \\ \hline
\multicolumn{1}{|l|}{Transactions}                                                             & Courtes et nombreuses, temps réel                                                                        & Longues, peu nombreuses                                                                                \\ \hline
\multicolumn{1}{|l|}{\begin{tabular}[c]{@{}l@{}}Taille\\ de la base\end{tabular}}              & Plusieurs Giga-octets                                                                                    & Plusieurs Téraoctets                                                                                   \\ \hline
\end{tabular}
\end{table}

\end{small}
Bill Inmon \cite{Inmon02} a donné la définition suivante : "Un entrepôt de données est une collection de données orientées sujet, intégrées, non volatiles, historisées et utilisées pour supporter un processus d’aide à la décision". 
\begin{description}

\item[Orientées sujet : ]les données des entrepôts collectées doivent être orientées métier. Ainsi, elles sont organisées par sujet plutôt que par application. Par exemple, une chaîne de magasins d'alimentation organise les données de son entrepôt par rapport aux ventes qui ont été réalisées par produit et par magasin, par région, au cours d’un certain temps.
\item[Intégrées : ]Pour garantir une cohérence de l’information, les données originaires de diverses sources doivent être intégrées (Par exemple, faire correspondre différents formats (.xml, .txt, etc.) avant leur stockage dans l’entrepôt de données. 
\item[Non volatiles : ]Les données d’un entrepôt de données, contrairement à celles des bases de données classiques, doivent être permanentes. Ainsi, un rafraîchissement de l’entrepôt doit pouvoir ajouter de nouvelles données sans pour autant modifier ou perdre celles existantes.
\item[Historisées : ]Il est primordial que les données soient datées pour considérer leur évolution dans la prise de décision et les analyses décisionnelles étant donné qu’ils s’appuient sur tout l’historique des évolutions passées pour prédire celles prévues au futur.
\end{description}

L’objectif principal de l'entrepôt de données est l’aide à la prise de décision. Il collecte des données provenant de diverses sources de données hétérogènes, autonomes, évolutives et distribuées, les transforme et les nettoie et finalement les charge dans de nouvelles structures de données conçues pour prendre en charge les requêtes $OLAP$.
Les sources de données peuvent être regroupées en quatre catégories principales \cite{Ponniah10} : \textit{(i)} données de production; \textit{(ii)} données internes; \textit{(iii)} données archivées; et \textit{(iv)} données externes. Les données de production proviennent des différents systèmes opérationnels de l'entreprise. Les données internes comprennent les données stockées dans les feuilles de calcul, les documents, les profils de clients, les bases de données qui ne sont pas connectées aux systèmes opérationnels. Les données archivées ou légataires sont des données dites off-line car elles ne sont plus nécessaires aux applications transactionnelles. Elles permettent de disposer d'historique pour les analyses de tendance et la fouille de données par exemple. Les données externes telles que les données produites par des agences externes, des services de prévisions météorologiques, des réseaux sociaux, sont des données gérées et manipulées par d'autres entreprises. Elles peuvent être sous forme informatique (CD-ROM, Internet) ou papier (presse économique, rapports d'organismes et institutions spécialisés).  \\

\indent Il existe plusieurs stratégies pour la conception d’un entrepôt de données. Il s'agit de choisir une démarche parmi les trois suivantes : Approche descendante, approche ascendante, ou approche mixte.
\begin{itemize}
\item Approche descendante "Top down" (guidée par les besoins) introduite par Inmon \cite{Inmon02}: les besoins des utilisateurs sont identifiés d'abord et ce sont eux qui conduisent le processus de développement du l'$ED$.
\item Approche ascendante "Bottom up" (guidée par les données) introduite par Kimball \cite{Kimball96}: elle commence par un prototypage basé sur une technologie et appliquée à un sous-ensemble représentatif. On peut commencer par la construction d'un magasin de données, dite \textit{datamart}, et évaluer son utilisation avant de généraliser le développement.
\item Approche mixte : elle permet de combiner les aspects planification et stratégie de l'approche descendante et la rapidité d'implémentation de l'approche ascendante. 
\end{itemize}

Le processus de construction d’un entrepôt de données est composé de trois principales phases (cf. figure \ref{im1}):

\begin{description}
\item[Extraction : ] les données sont extraites à partir des sources hétérogènes.
Le processus d'$ETL$ est responsable de l'extraction et de la transformation des données provenant de différentes sources d'informations hétérogènes pour finalement les charger dans l'entrepôt cible. Cette phase joue un rôle crucial dans le processus de conception de l’$ED$. La qualité de l'entrepôt dépend fortement de l'$ETL$. Les différentes étapes de l'$ETL$ doivent comprendre les différents schémas (conceptuels, logiques et physiques) des sources de données. \\
L’$ETL$ comporte 3 étapes :
\begin{enumerate}
\item \textbf{E}xtraction des données à partir des sources (BD ou Fichiers) dans le but d’alimenter les tables de faits et les tables de dimensions.  Les données sont stockées dans des tables temporaires (staging area)
\item \textbf{T}ransformation : elle consiste à transformer les données sources selon les unités de mesure et les formats de l’$ED$,  homogénéiser les données sources, nettoyer les données et les dater.
\item \textbf{L}oad : chargement des données de la zone tampon vers l’$ED$.
\end{enumerate}

Il existe trois stratégies d’extraction des données:
\begin{itemize}

\item Push : la logique applicative de l’extraction se trouve dans le système source. Elle pousse les données vers la zone tampon.
\item Pull : la logique applicative de l’extraction se trouve dans le système cible ($ED$). Elle «tire» les données vers la zone d’attente.
\item Push-Pull : c’est une combinaison des deux stratégies précédentes.
\end{itemize}
La construction du processus $ETL$ est potentiellement l'une des tâches les plus importantes de la construction d'un entrepôt; Il est complexe, prend beaucoup de temps et prend la part du lion dans la conception et les efforts de mise en œuvre de tout projet d'entrepôt. En effet, la construction de l'$ETL$ nécessite plusieurs tâches et compétences en termes de modélisation et de mise en œuvre. Les exigences non fonctionnelles telles que la qualité et les performances sont prises en compte lors de la conception d'$ETL$

\begin{figure}[h!]
\begin{center}
\includegraphics[scale=0.5]{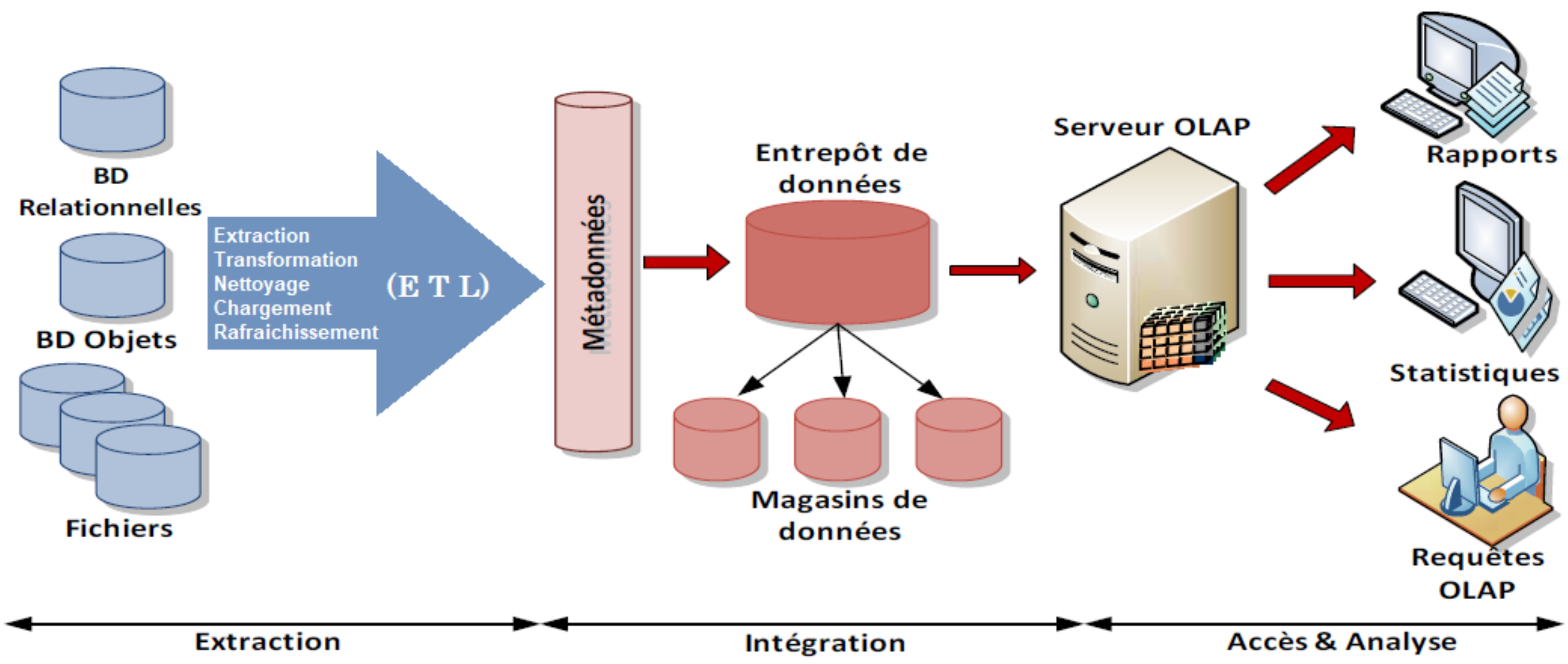}
\caption{Architecture d'entrepôt de données \label{im1}}
\end{center}
\end{figure}

\item[Intégration : ] une fois la tâche de l'organisation et de l'homogénéisation des données est terminée, les données seront  intégrées dans l'entrepôt a l'aide des métadonnées qui  décrivent les caractéristiques des données stockées : origine, date de dernière m-à-j, mode de calcul, procédure de transformation.  Les métadonnées contiennent :
\begin{itemize}
\item Les correspondances entre données sources et celles de l’$ED$.
\item Les règles d'extraction, création, stockage et mise à jour des données.
\item Un glossaire des objets manipulés par les utilisateurs.
\item Les requêtes, rapports, indexs et profils prédéfinis.
\item Les règles de rafraîchissement et de réplication entre l’$ED$ et les datamarts. éventuels
\end{itemize}
Elles sont utiles aussi bien aux utilisateurs (comprendre les données) qu’aux administrateurs (fournir des moyens d’exploitation et de maintenance du $ED$). \\
Les données peuvent être intégrées directement dans l’entrepôt de données ou bien elles peuvent être intégrées dans les datamarts. Les datamarts sont destinés à pré-agréger des données disponibles de façon plus détaillée que dans les $ED$, afin à traiter plus facilement certaines questions spécifiques, critiques, etc. C'est l'implémentation d'un $ED$ pour un domaine d'activité restreint e.g un sous-ensemble d'un $ED$. Une entreprise dispose souvent de plusieurs datamarts, un par département ou service. La présence de datamarts ou non dépend de l’architecture du SID (centralisée, répartie, dupliquée). La création d’un datamart peut être un moyen de débuter un projet de $ED$ (projet pilote).
\item[Accès et Analyse : ] ceci s’effectue sous une forme efficace et flexible. Durant cette phase, un serveur $OLAP$ se charge de présenter les informations demandées par les utilisateurs sous plusieurs formes : tableaux, rapports, statistiques, fouille de données, etc.  La fouille de données  permet d'exploiter le gisement informationnel contenu dans un $ED$ pour découvrir d'autres informations cachées comme par exemple les habitudes et le comportement des clients. Les informations déduites permettent de bien cibler les clients, les produits, les régions, etc. La fouille de données permet de comprendre ce qui s'est passé (Quoi), trouver une explication de ce qui est passé (Pourquoi) et déduire les actions à entreprendre.

\end{description}

\subsection{Modélisation des entrepôts de données }
Les données dans l’entrepôt doivent être organisées d’une façon simple et aisée pour faciliter leur exploitation ainsi que leur analyse par les décideurs. Il est indiscutable que ces analyses nécessitent l’exécution de requêtes complexes et particulières. Ces requêtes doivent produire comme résultat une représentation des données selon plusieurs axes d’analyses (analyse des habitudes, tendances, préférences d’achat, etc.). Les données manipulées dans le contexte d’entrepôts de données sont représentées sous la forme multidimensionnelle, qui est mieux adaptée pour le support des processus d’analyse \cite{Kimball96}. Ce modèle permet d’observer le sujet (le fait) et les différentes perspectives de l’analyse ou mesures d’intérêt (les dimensions).

\begin{figure}[h!]
\begin{center}
\includegraphics[scale=0.5]{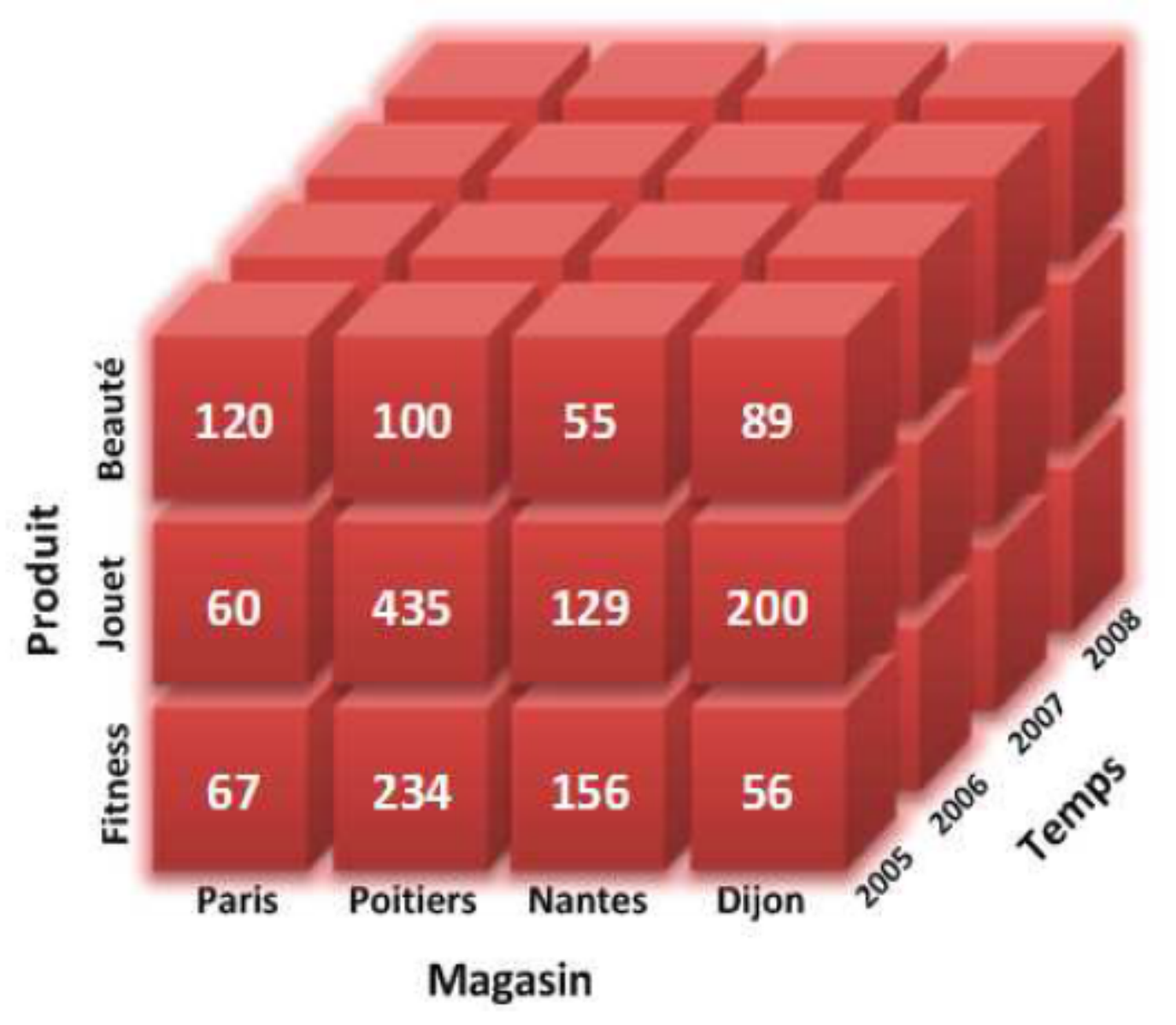}
\caption{Un exemple d'un cube de données \cite{KB09}\label{im2}}
\end{center}
\end{figure}

Le modèle multidimensionnel repose sur le concept de CUBE (ou hypercube) pour représenter les données (cf. figure \ref{im2}). Un cube organise les données en une ou plusieurs dimensions qui déterminent une mesure d’intérêt. Deux concepts fondamentaux caractérisent le modèle multidimensionnel : faits et dimensions
\begin{description}
\item[Un fait : ] il représente le sujet d’analyse et il est formé de mesures correspondantes aux informations de l’activité étudiée. En général et dans la plupart des cas il s’agit d’une valeur numérique. Ces mesures sont calculables à partir d’un grand nombre d’enregistrements en appliquant les opérations d’addition, de calcul du minimum/maximum ou de la moyenne, etc \cite{Teste00}.
\item[ La dimension : ]elle représente l’axe d’analyse de l’activité. Elle regroupe des paramètres pouvant influencer les mesures d’activités d'un fait. Si nous analysons  la quantité de produits vendus, il semble évident que cette quantité change selon la ville, le produit et le temps, en général elle dépend de la nature des données présentes dans la dimension. En outre, une dimension est munie d’une ou plusieurs hiérarchies permettant de faire l’analyse selon le niveau de granularité, i.e. le niveau de détail. La dimension $Temps$ par exemple peut être hiérarchisée, du plus fin au moins fin, comme suit : \textit{jour, semaine, mois, et année}. Les mesures dans la table de faits peuvent être par conséquent agrégées par \textit{mois, trimestre, semestre ou année}, selon le niveau de détail souhaité.
\end{description}
\indent Afin d’implémenter le modèle multidimensionnel approprié pour un entrepôt de données, deux modèles sont possibles : Un schéma relationnel $ROLAP$ (Relational On-Line Analytical Processing), à savoir le schéma en étoile, en flocon de neige ou en constellation ou bien un schéma multidimensionnel $MOLAP$ (Multidimensional On-Line Analytical Processing).
\subsubsection{Systèmes MOLAP}
C'est une forme d'hypercube multidimensionnel, qui permet de représenter les données sous la forme d'un tableau multidimensionnel avec le croisement de $n$ dimensions, où chaque dimension de ce tableau est associée à une dimension de l’hypercube de données ou bien à un axe. Les systèmes $MOLAP$ nécessitent un pré-calcul de toutes les agrégations possibles afin de les matérialiser dans les cellules du tableau multidimensionnel. 
\begin{figure}[h!]
\begin{center}
\includegraphics[scale=0.5]{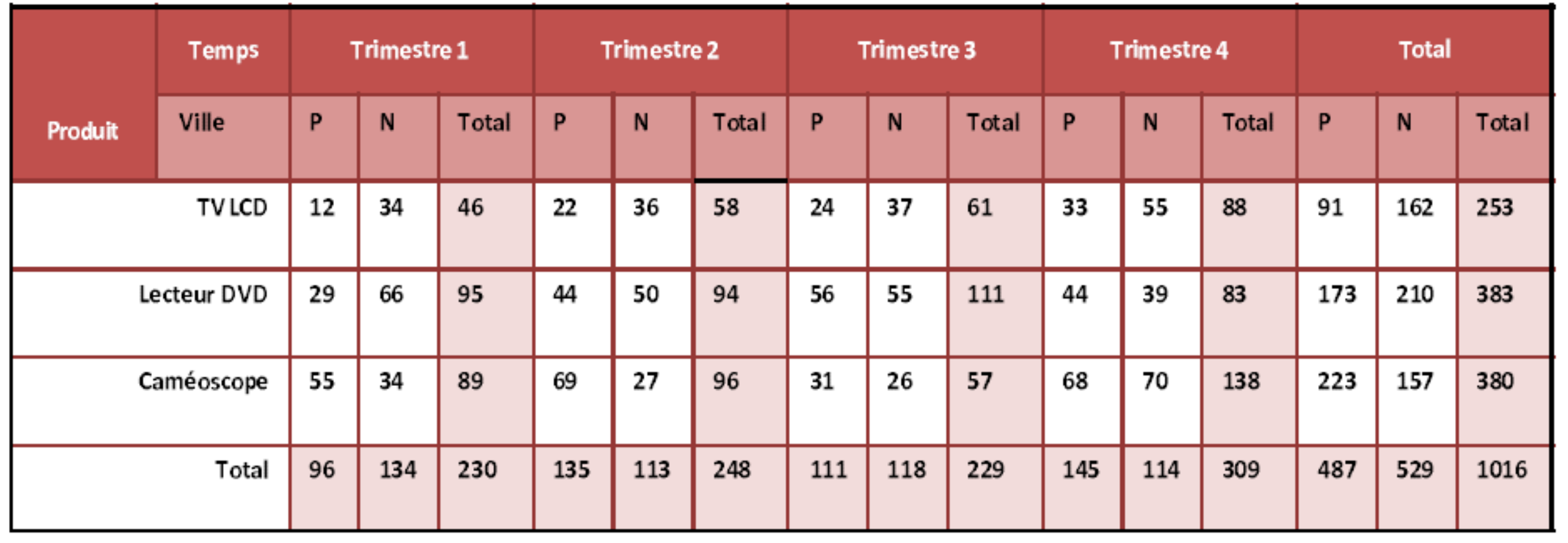}
\caption{Le système Molap \cite{KB09}\label{im3}}
\end{center}
\end{figure}

L’avantage principal de cette implémentation est le gain considérable en temps d’exécution des requêtes, vu que l’accès aux données est direct. Par contre, les opérations de mise à jour sont difficiles à effectuer compte tenu que les valeurs des cellules du tableau multidimensionnel doivent être recalculées après chaque opération de mise à jour.
\subsubsection{Systèmes ROLAP}
Les systèmes de type $ROLAP$ utilisent une représentation relationnelle du cube de données.
Chaque fait est exprimé à travers une table intitulée table de faits et chaque dimension est exprimée à travers une table intitulée table de dimension. Les attributs de la table de faits constituent les mesures à chercher et les clés étrangères vers les tables de dimension. Pour modéliser un système $ROLAP$ trois modèles ont été proposés, le schéma en étoile, le schéma en flocon de neige et le schéma en constellation.
\begin{description}
\item[Schéma en étoile :] un schéma en étoile consiste en une grande table de faits centrale attaché à plusieurs tables de dimensions à travers des clés étrangères. Les tables de dimensions sont nombreuses mais relativement petites par rapport à la table de faits et sont rarement mis à jour. Les tables de dimension contiennent généralement un nombre important d’attributs représentant des données qualitatives. La table de faits contient généralement un nombre très important d’instances. Chaque enregistrement dans la table de faits contient deux sortes d’attributs : \textit{(i)} un ensemble de clés étrangères référençant les tables de dimension; et \textit{(ii)} un ensemble de mesures qui peuvent être agrégées pour effectuer certains traitements.
\begin{figure}[h!]
\begin{center}
\includegraphics[scale=0.5]{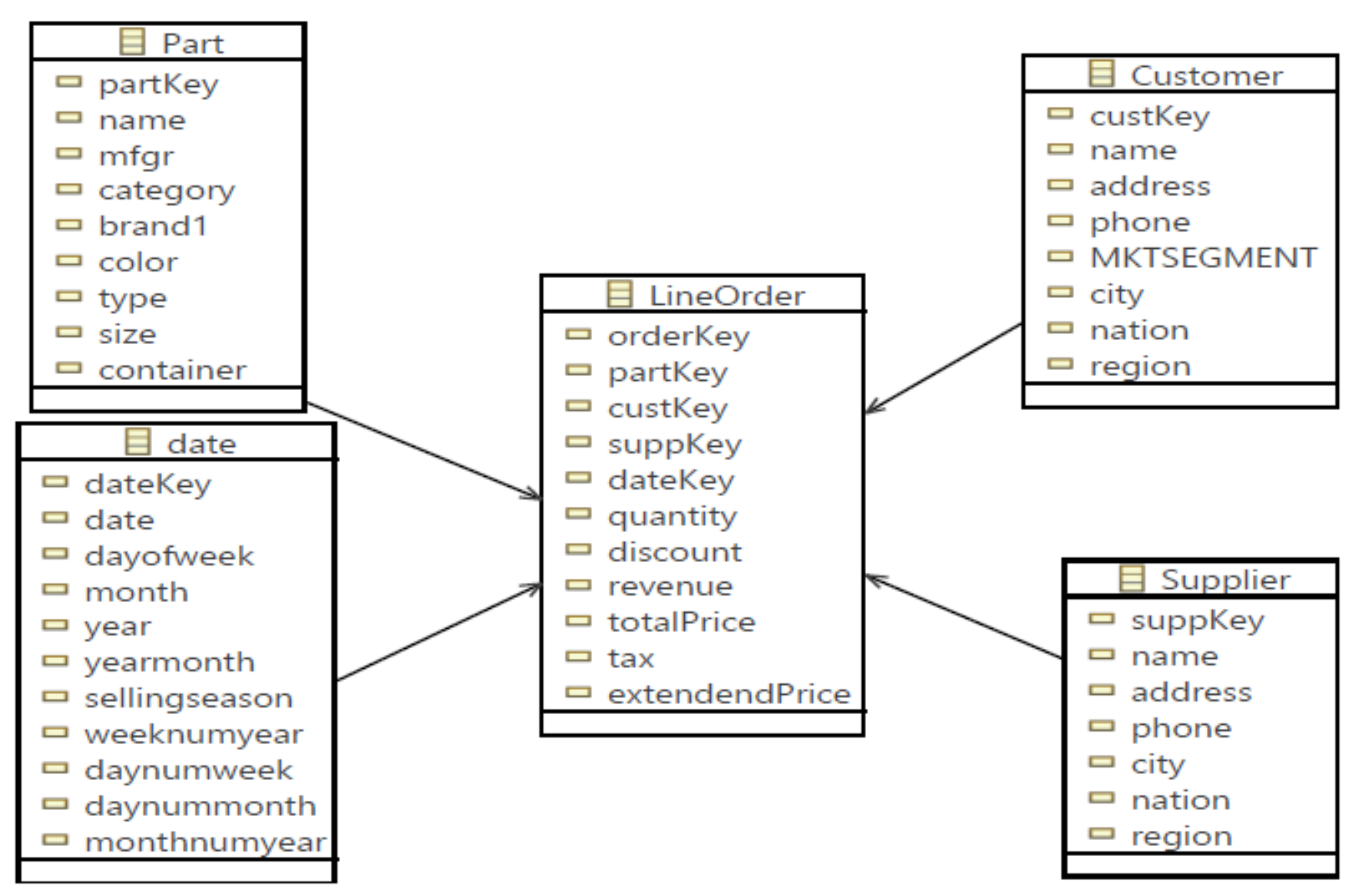}
\caption{Le schèma en étoile \cite{B16} \label{im4}}
\end{center}
\end{figure}

La table de faits est généralement normalisée, alors que les tables de dimension ne sont généralement pas normalisées de sorte que le nombre d'opérations de jointure nécessaires soit réduit. Pour éviter la redondance, les tables de dimensions d'un schéma en étoile peuvent être normalisées. Il y a un débat sur les avantages d'avoir de telles tables de dimensions normalisées, car, en général, ralentira le traitement des requêtes. Les requêtes qui interrogent ce schéma sont dites \textit{Requêtes de jointure en étoile}.
\item[Schéma en flocon de neige :] pour mettre en évidence la hiérarchie, le modèle en flocon de neige a été proposé.  La dé-normalisation des tables de dimension dans un schéma en étoile ne reflète pas les hiérarchies associées à chaque dimension. Ce schéma dérive du schéma en étoile précédent en gardant la table de faits au centre et les tables de dimension autour. Chaque table de dimension est éclatée en un ensemble de hiérarchies. Ce schéma normalise les dimensions, en réduisant la taille de chacune des relations et permettant ainsi d’intégrer la notion de hiérarchie au sein d’une dimension \cite{AGS97}.

\begin{figure}[h!]
\begin{center}
\includegraphics[scale=0.5]{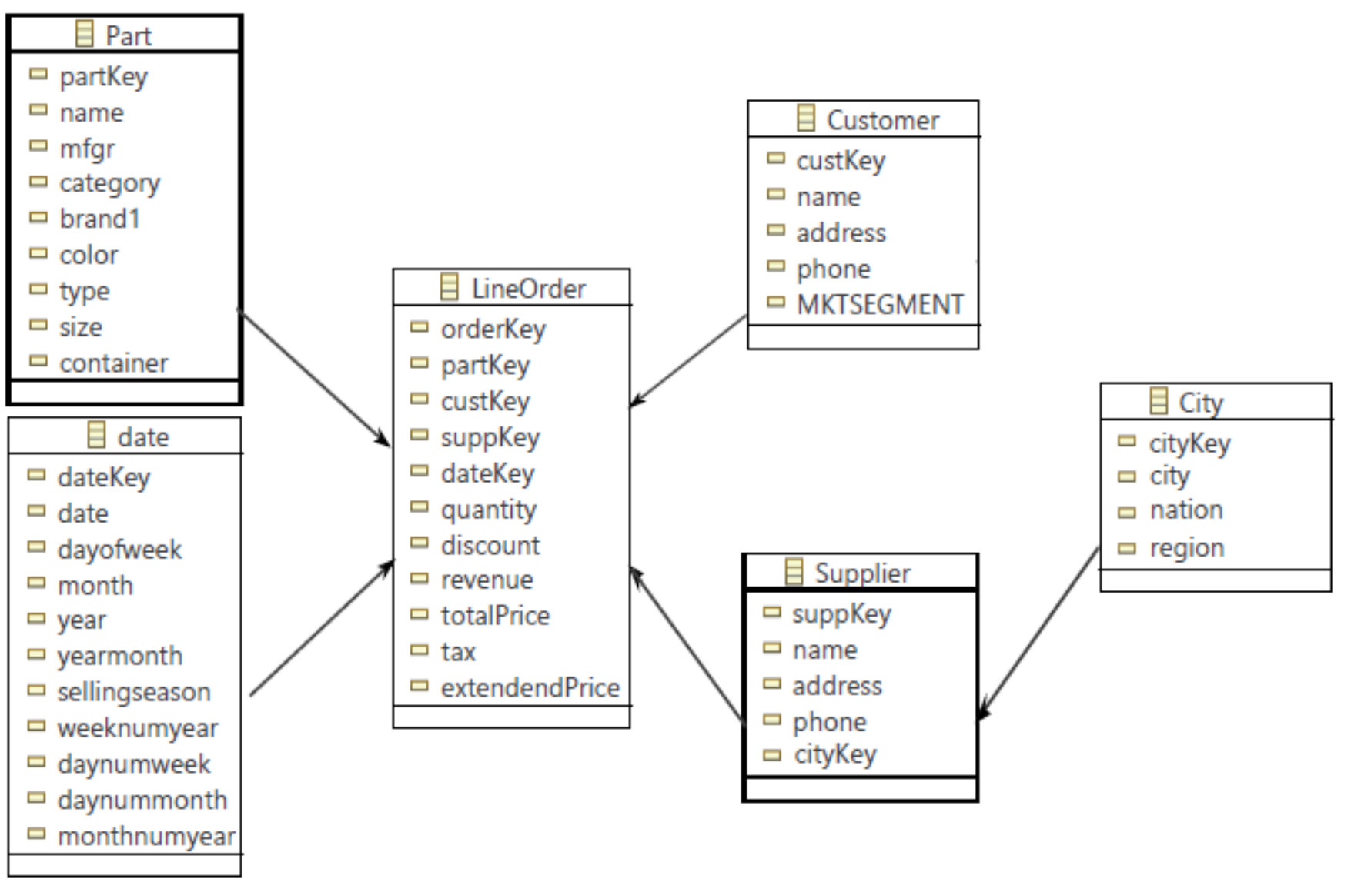}
\caption{Le schèma en flocon de neige \label{im5}}
\end{center}
\end{figure}
 Les tables dotées de la hiérarchie la plus fine sont directement liées à la table de faits. Les tables représentant les autres hiérarchies sont liées entre elles selon leur niveau de granularité du plus fin au moins fin.
\item[Le schéma en constellation :] la modélisation en constellation consiste à fusionner plusieurs modèles en étoile qui utilisent des dimensions communes. Un modèle en constellation comprend : plusieurs tables de faits, des tables de dimensions communes à ces tables de faits où chaque relation de faits a ses propres dimensions. Ce modèle offre une meilleure gestion des données creuses.

\begin{figure}[h!]
\begin{center}
\includegraphics[scale=0.5]{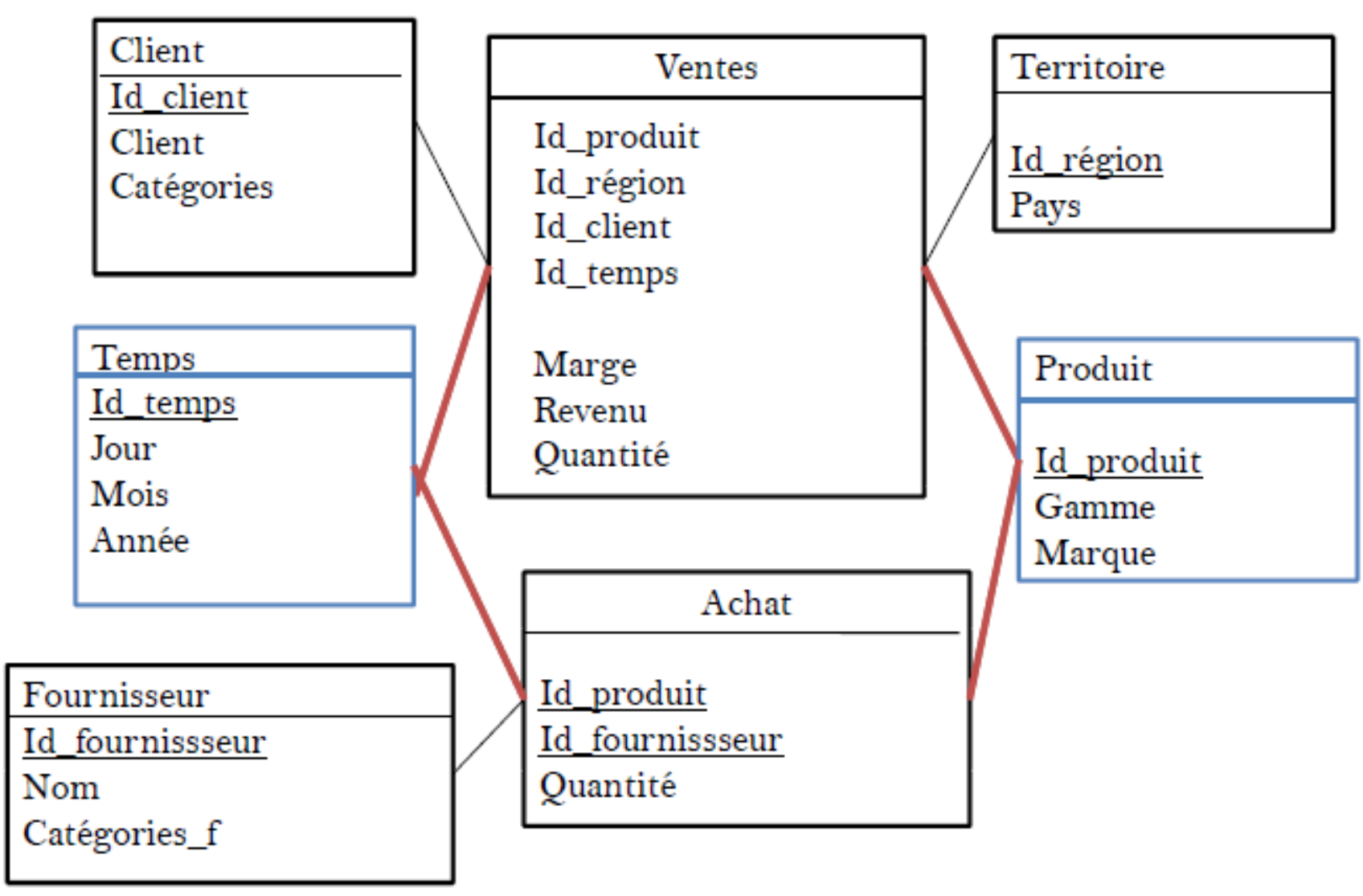}
\caption{Le schèma en flocon de neige \cite{B16}\label{im15}}
\end{center}
\end{figure}
\end{description}
Nous pouvons également citer une troisième approche, intitulée $HOLAP$, il s’agit d’un $OLAP$ hybride qui combine les deux approches. Il bénéficie de la capacité de stockage de $ROLAP$ et de la puissance de traitement de $MOLAP$.
\section{Les besoins d'entrepôt de données.}

Comme tout produit, le développement d'une application décisionnelle repose sur des besoins fonctionnels et non fonctionnels. Notez que les besoins fonctionnels décrivent les fonctionnalités, le fonctionnement et l'utilisation des applications au sein de l'$ED$ pour satisfaire les objectifs et les attentes des décideurs. Ils  identifient les principaux avantages que la technologie d'$ED$ apportera à l’entreprise et à ses utilisateurs. Ils décrivent les besoins des utilisateurs et de l’entreprise et les valeurs ajoutées. D'autres besoins fonctionnels sont associés aux utilisateurs de l'$ED$ (appelés \textit{besoins des utilisateurs}) décrivent les tâches que les utilisateurs doivent pouvoir accomplir grâce à l'application d'$ED$. Les besoins des utilisateurs doivent être collectés auprès de personnes qui utiliseront et travailleront avec cette technologie. Par conséquent, ces utilisateurs peuvent décrire à la fois les tâches qu'ils doivent effectuer avec l'$ED$. Ces exigences sont modélisées à travers plusieurs formalismes comme par exemple le diagramme de cas d'utilisation \emph{UML} \cite{B16}.

\indent Les besoins non fonctionnels, appelées \textit{exigences de qualité}, sont soit des exigences facultatives, soit des contraintes, qui sont détaillées dans l'architecture du système. Ils décrivent comment le système fera les objectifs suivants: la sécurité, la performance (par exemple, le temps de réponse, le temps de rafraîchissement, le temps de traitement, l'importation/exportation de données, le temps de chargement), la capacité (transactions par heure, stockage de mémoire), la disponibilité, l'intégrité des données, l'évolutivité, l'énergie, etc. Ce type d'exigences doit être validé pendant la majorité des phases du cycle de vie de l'$ED$.
\section{Techniques d’optimisation}
Vu la volumétrie gigantesque des données et la complexité des requêtes utilisées, un des éléments clés qui garantisse une bonne conception physique d'un entrepôt de données consiste à bien choisir les structures d'optimisation\cite{CN07}. Chacune d'elles est liée à un problème d'optimisation NP-Complet et supportée par certaines approches proposées par la communauté de recherche. 

\begin{figure}[h!]
\begin{center}
\includegraphics[scale=0.5]{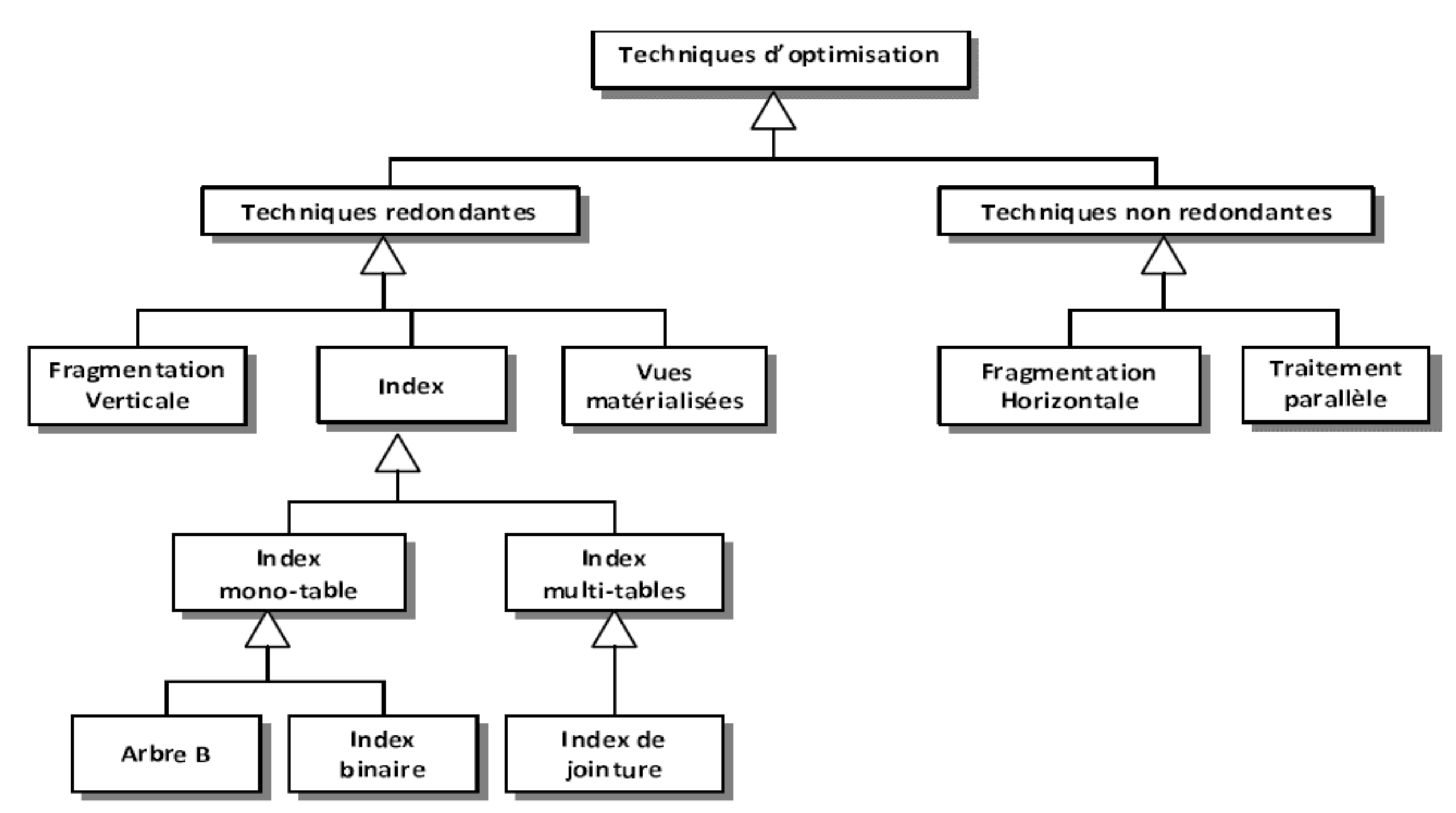}
\caption{Les techniques d’optimisation \cite{KB09}\label{im6}}
\end{center}
\end{figure}

Plusieurs techniques d'optimisation ont été proposées pour optimiser les requêtes de jointure en étoile. Ces techniques appartiennent à deux catégories :\textit{(i)} techniques d'optimisation redondantes comme les indexs et les vues matérialisées; et \textit{(ii)} techniques d'optimisation non redondantes comme la fragmentation horizontale et le traitement parallèle. 
\subsection{Techniques d’optimisation non redondantes}
Dans cette section, nous présentons deux techniques d’optimisation non redondantes, à savoir la fragmentation horizontale primaire définie et la fragmentation horizontale dérivée.

\subsubsection{La fragmentation horizontale primaire et dérivée}
La fragmentation horizontale consiste à partitionner les objets de la base de données (tables, vues et index) en plusieurs ensembles de lignes appelés \textit{fragments horizontaux}. Chaque ligne représente une instance de l’objet fragmenté. Les instances appartenant au même fragment horizontal vérifient généralement un prédicat de sélection. Chaque fragment horizontal $T_{i}$ d’une table $T$ est défini par une clause de sélection sur la table $T$ comme suit : $T_{i}$ = $\delta_{cli}(T)$. \\
La fragmentation horizontale constitue un aspect important dans la conception physique des bases de données \cite{SNY04}. Elle est considérée comme une technique d’optimisation non redondante du fait qu’elle ne réplique pas de données.

\begin{figure}[h]
\begin{center}
\includegraphics[scale=0.6]{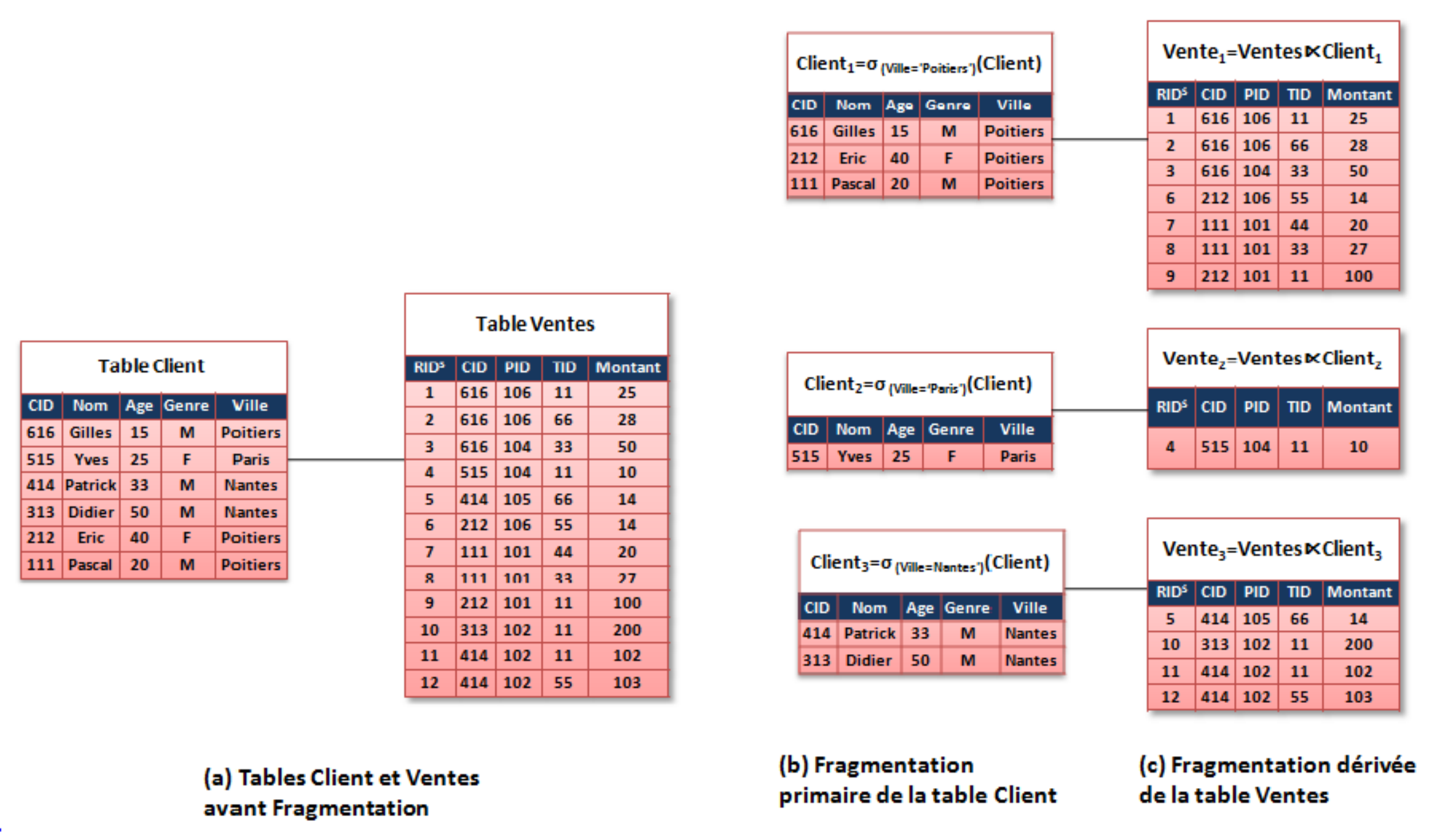}
\caption{La fragmentation horizontale primaire et dérivée \cite{KB09}\label{im7}}
\end{center}
\end{figure}
Elle a un impact important sur la performance des requêtes définies sur les données de volume important. Elle a aussi un impact significatif sur la facilité de gestion et la maintenance des données. Deux types de fragmentation horizontale sont disponibles : primaire et dérivée.\\
\indent La fragmentation horizontale primaire d’une table se base sur les prédicats de sélection définis sur cette table.\\
\indent La fragmentation horizontale dérivée utilise le lien existant entre deux tables pour fragmenter l’une d’entre elles en fonction des fragments de l’autre. De ce fait, la fragmentation horizontale dérivée d’une table se base sur les prédicats de sélection définis sur une autre table. Concrètement, la fragmentation dérivée d’une table $S$ n’est possible qu’à travers une jointure dans le sens qu’elle soit liée avec une table $T$ par sa clé étrangère. Une fois la table $T$ fragmentée par la fragmentation primaire, les fragments de $S$ sont générés par une opération de semi-jointure entre $S$ et chaque fragment de la table $T$. Les deux tables seront équi-partitionnées grâce au lien père-fils.
\subsection{Techniques d’optimisation redondantes}
Cette catégorie comprend les vues matérialisées, les indexs, la fragmentation verticale, etc. Ces techniques optimisent les requêtes, mais il faut prendre en considération le coût de stockage supplémentaire de la structure en question et le coût de sa maintenance dû aux différentes opérations de mise à jour (Insertion, Modification, Suppression) effectuées sur les données. Il est évident que plus la base de données est volumineuse, plus le temps de réponse sera lent surtout si nous partons du principe que cet accès s'effectue de façon séquentielle et aussi répétitive dû aux opérations de jointure des  requêtes. Dans ce contexte, nous reconnaissons l'importance des indexs de façon générale étant donné qu'ils rendent l'accès séquentiel en un accès direct. Il existe différents types d'indexs : Index B-Tree, Index de projection, Index de hachage, Index binaire, Index de jointure, Index de jointure en étoile, Index de jointure binaire, Index de jointure de dimension. \\
\indent Comme nous allons travailler sur les index de jointure binaire en particulier, alors nous allons étudier les différents indexs de plus près.
\subsubsection{Index B-Tree}

\begin{figure}[h!]
\begin{center}
\includegraphics[scale=0.5]{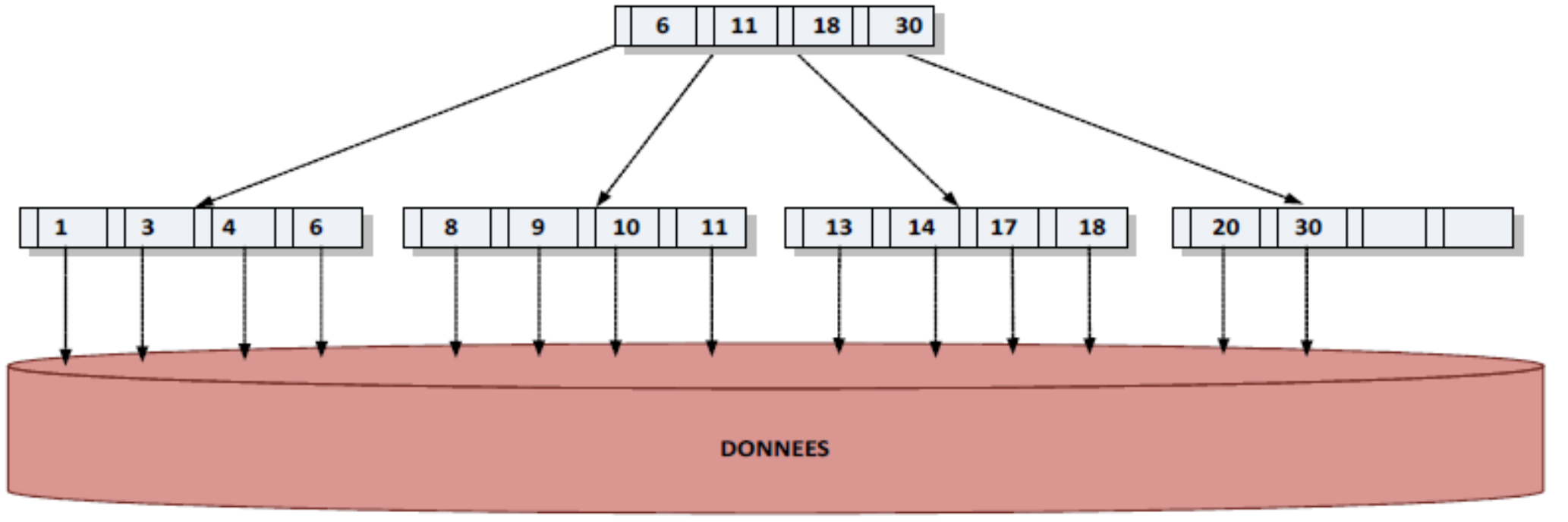}
\caption{Index B-Tree \cite{KB09}\label{im8}}
\end{center}
\end{figure}

La majorité des SGBD commerciaux utilisent l’index \textsc{B-Tree} comme étant l’index par défaut. Cet index est organisé sous forme d’un arbre à plusieurs niveaux. Chaque nœud d’un niveau pointe vers le niveau inférieur. Le niveau le plus bas qui comporte les nœuds feuilles  contient les entrées d’index ainsi qu’un pointeur vers l’emplacement physique de l’enregistrement correspondant (\textit{ROWID}). La figure \ref{im8} représente un exemple d’index B-arbre construit sur l’attribut $PID$ de la table Produit.

\subsubsection{Index de projection}

\begin{figure}[h!]
\begin{center}
\includegraphics[scale=0.5]{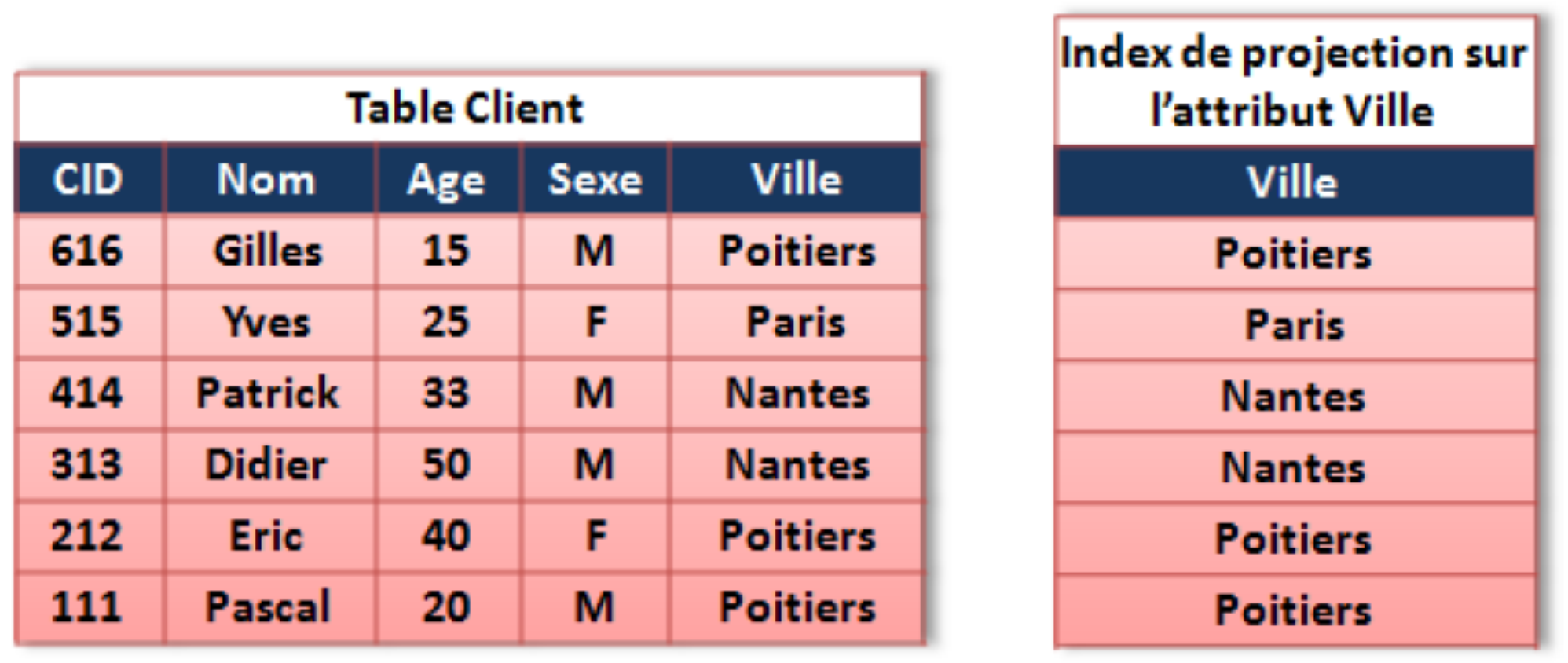}
\caption{Index de projection \cite{KB09}\label{im10}}
\end{center}
\end{figure}

Un index de projection est défini sur un ou plusieurs attributs d’une table. Il consiste à stocker toutes les valeurs de ces attributs dans l’ordre de leur apparition dans la table. Généralement, les requêtes accèdent à un sous-ensemble d’attributs d’une table. Si ces attributs sont contenus dans un index de projection, l’optimiseur ne charge que cet index pour répondre à la requête. La figure \ref{im10} montre un index de projection défini sur l’attribut $Ville$ de la table $Client$.

\subsubsection{Index de hachage}

L’index de hachage utilise une fonction de hachage. Cette fonction permet, à partir d’une valeur de clé $c$, de donner l’adresse $f(c$) d’un espace de stockage où l’élément doit être placé. La figure montre un index de hachage construit sur l’attribut $PID$ d’une table Produit. Dans ce type d’index, le choix de la fonction de hachage est la tâche la plus importante pour garantir une bonne performance de l’index. Par exemple, si la fonction donne la même valeur à un nombre important d’éléments, alors l’accès ressemblera à un balayage séquentiel et ainsi l’index perd son intérêt.
\begin{figure}[h!]
\begin{center}
\includegraphics[scale=0.5]{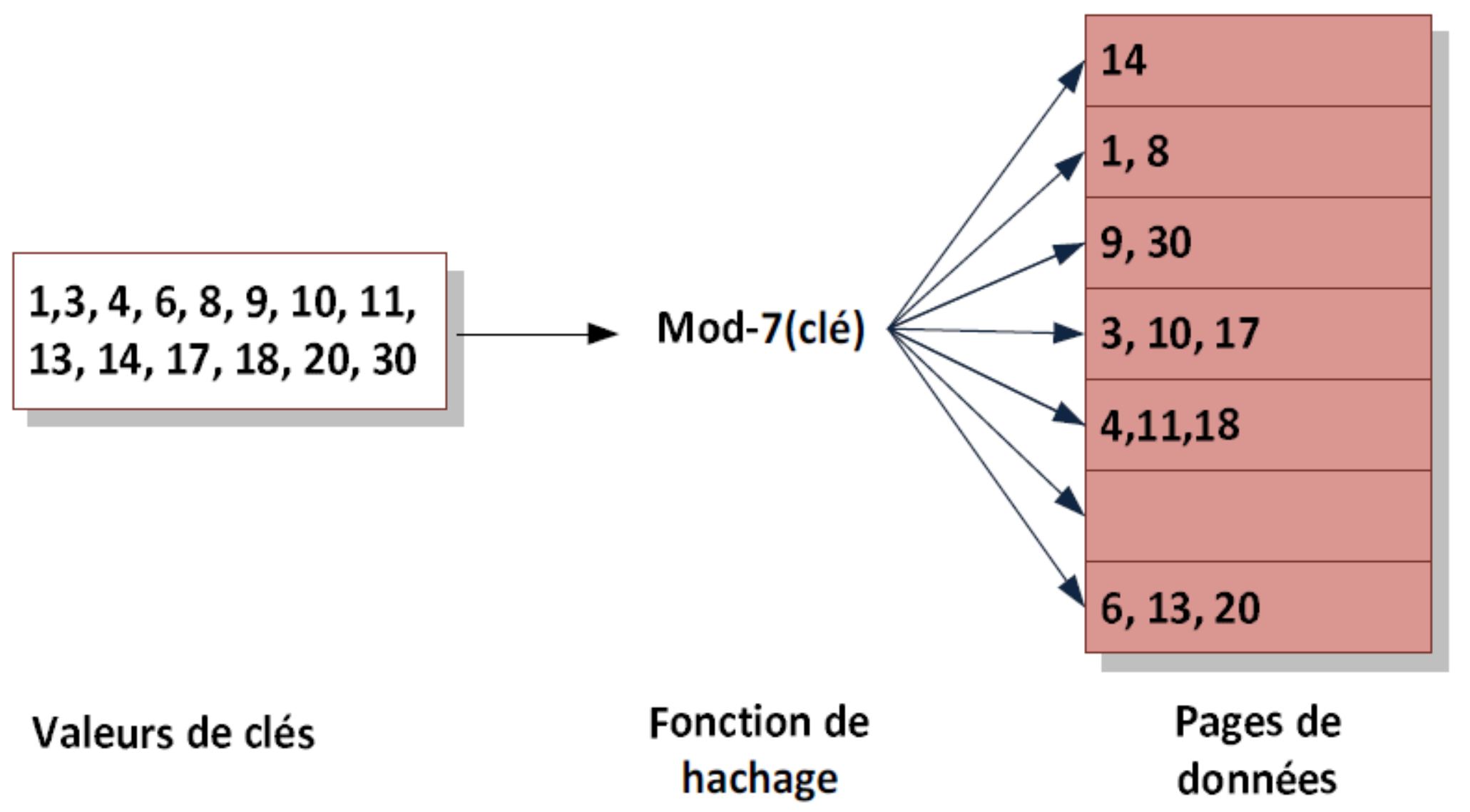}
\caption{Index de hachage \cite{KB09}\label{im100}}
\end{center}
\end{figure}
\subsubsection{Index binaire (bitmap index)}
L’index binaire utilise un ensemble de vecteurs binaires (contenant des valeurs 0 ou 1) pour référencer l’ensemble des n-uplets d’une table. Pour chaque valeur de l’attribut indexé, un vecteur de bits est stocké. Ce vecteur contient autant de bits qu’il y a de n-uplets dans la table indexée. L’index binaire a été considéré comme le résultat le plus important obtenu dans le cadre de l’optimisation de la couche physique des entrepôts de données \cite{GMR98}.  

\begin{figure}[h!]
\begin{center}
\includegraphics[scale=0.5]{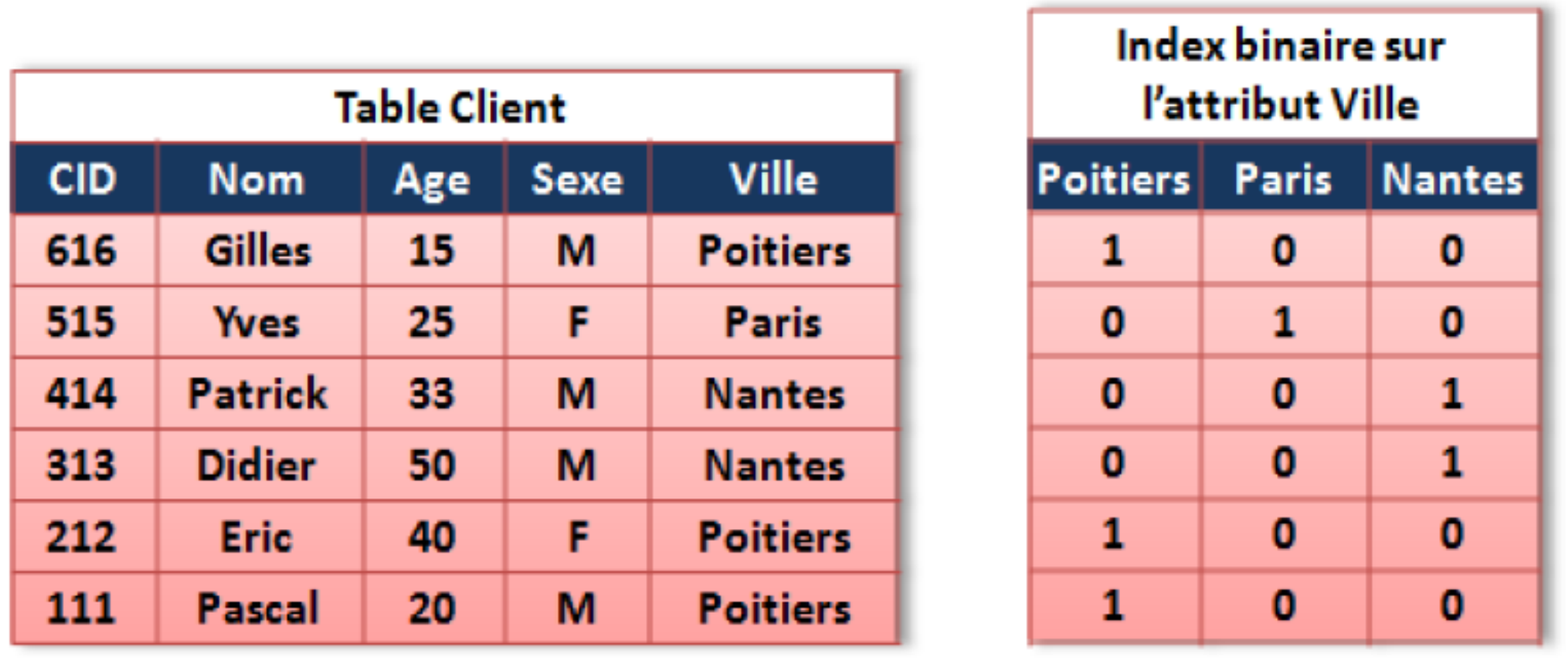}
\caption{Index binaire \cite{KB09}\label{im11}}
\end{center}
\end{figure}

Supposons un attribut $A$ ayant $n$ valeurs distinctes $v_{1}, v_{2}, . . ., v_{n}$ ($n$ est appelé \textit{cardinalité} de $A$) appartenant à une table $T$ composée de $m$ instances. La construction de l’index binaire $IB$ défini sur l’attribut $A$ se fait de la manière suivante :
\begin{enumerate}

\item  Créer $n$ vecteurs composés chacun de $m$ entrées ;
\item  Pour chaque n-uplet $i$, (1 $\leq i \leq$ $m$) dans $T$, si $i.A$ = $v_{k}$, alors mettre 1 dans le $i^{eme}$ bit du vecteur correspondant à $v_{k}$, mettre 0 dans les $i^{emes}$ bits des autres vecteurs.
Si une requête recherche les n-uplets vérifiant un prédicat d’égalité (par exemple $A$ = $v_{k}$,), alors il suffit de lire le vecteur associé à $v_{k}$, chercher les bits ayant la valeur 1, ensuite accéder aux n-uplets correspondant à ces bits. La figure \ref{im11} montre un exemple d’index binaire construit sur l’attribut $Ville$ de la table $Client$.

\end{enumerate}

\subsubsection{Index de jointure }
Quelque soit la base de données opérationnelle ou bien décisionnelle l’opération de jointure est toujours présente et surtout dans les requêtes $OLAP$. Elle est très coûteuse, puisqu’elle manipule de grands volumes de données. Plusieurs implémentations de la jointure ont été proposées dans les bases de données traditionnelles : les fonctions de hachage, les boucles imbriquées, etc. Ces implémentations perdent leur apport et restent limitées lorsque la taille des tables concernées par la jointure est importante. \emph{Valduriez} a proposé un index de jointure, qui pré-calcule la jointure entre deux tables \cite{Vald87}. L’index de jointure matérialise les liens existant entre deux tables en utilisant une table à deux colonnes chacune représentant l’identifiant d’une table. Soit $R$ et $S$ deux tables qui peuvent être jointes par les attributs $R.a$ et $S.b$. L’index de jointure est l’ensemble des n-uplets $<R.ID_{i}, S.ID_{j}>$ tel que les n-uplets de $R$ et $S$ ayant respectivement pour identifiant $IDi$ et $IDj$ vérifient la condition de jointure. \\
\begin{figure}[h!]
\begin{center}
\includegraphics[scale=0.5]{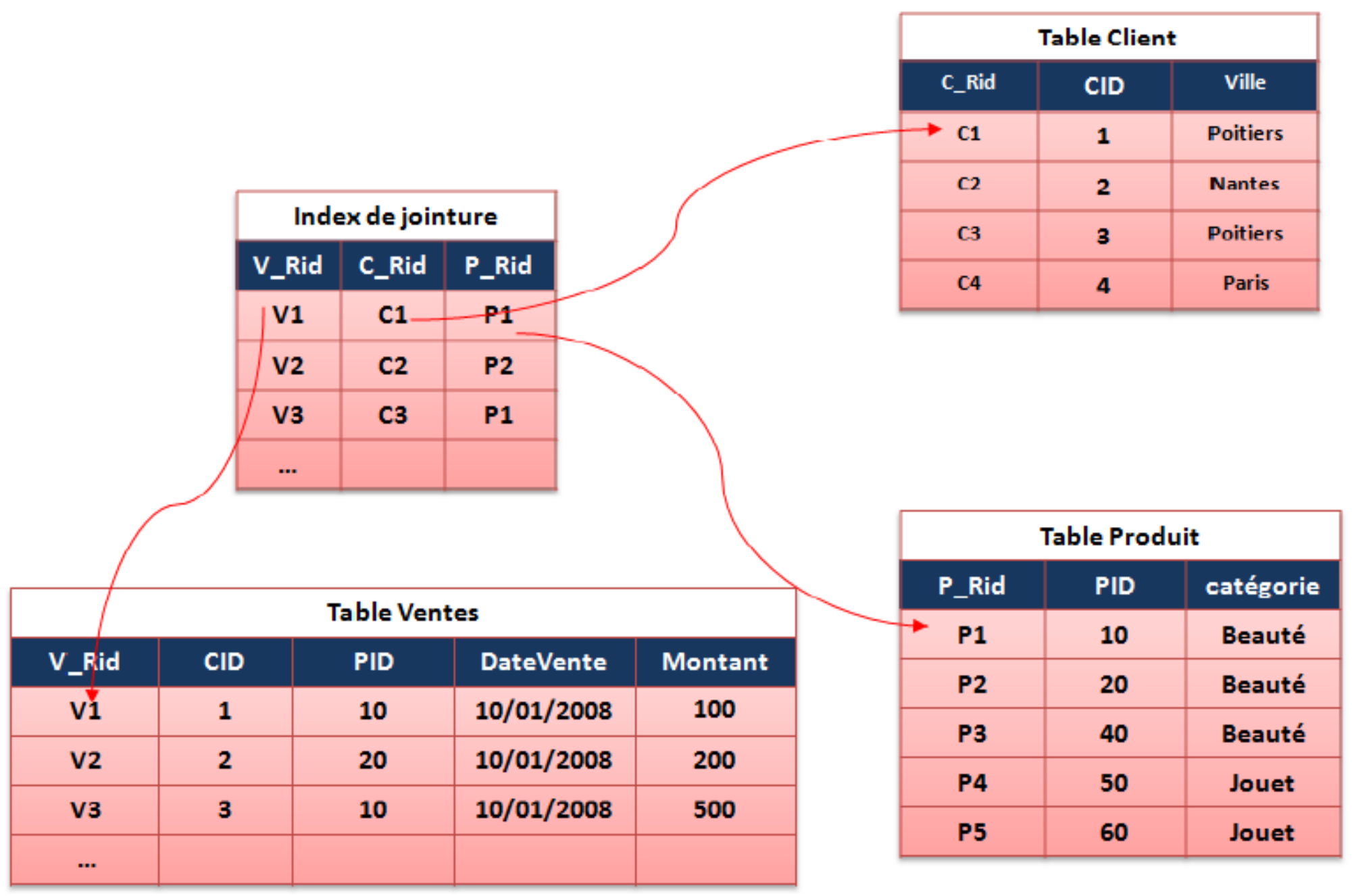}
\caption{Index de jointure \cite{KB09}\label{im12}}
\end{center}
\end{figure}
Notons que la taille de l’index de jointure dépend de la sélectivité de la jointure. Si la jointure est très sélective alors la taille de l’index est très petite. \\
L’exploitation de l’index de jointure $IJ$ entre $R$ et $S$ se fait de la manière suivante :

\begin{enumerate}
\item Charger l’index de jointure $IJ$ ;
\item Effectuer la semi-jointure $R \ltimes IJ$ ;
\item Effectuer la semi-jointure $S \ltimes IJ$ ;
\item Effectuer la jointure des deux résultats.
\end{enumerate}

\section*{Conclusion}
Nous avons exposé dans ce chapitre une présentation des entrepôts de données, leur différence avec les bases de données traditionnelles et leur architecture générale. Nous avons aussi vue la modélisation des entrepôts de données avec le modèle multidimensionnel qui peut être implémentée sur deux modèles : $MOLAP$ et $ROLAP$. Par la suite, nous avons abordé le sujet des techniques d’optimisation des entrepôts de données et nous avons vu les différentes approches. Dans le chapitre suivant, nous allons détailler une technique d’optimisation particulière qui est l’index de jointure binaire.
\chapter{Les index de jointure binaires IJB}
\section*{Introduction}
Pour réduire le temps de réponse très élevé des requêtes  et satisfaire les besoins des décideurs, l’administrateur de l’entrepôt est amené à faire une bonne conception physique. Dans cette dernière, il doit sélectionner un ensemble de techniques d’optimisation (voir Chapitre 1), qu’il considère pertinent pour l’ensemble des besoins des décideurs (exprimés sous forme de requêtes). Parmi les techniques d’optimisation les plus utilisées, on trouve les indexs en général et les indexs de jointure binaires $(IJB)$ en particulier, ces derniers servent à pré-calculer la jointure entre la table de faits et une ou plusieurs tables de dimensions. \\
\indent Pour bien présenter les $IJB$, leur utilité et les différentes techniques de sélection nous avons organisé ce chapitre en 4 sections. La section 1 présente les $IJB$ et la formalisation du problème de sélection des $IJB$. La section 2 donne la stratégie d’exécution d’une requête de jointure en étoile en présence d’un $IJB$ avec un exemple. La section 3 présente le modèle de coût que nous avons choisi pour évaluer les différentes techniques d’optimisation des $IJB$ présentes dans notre travail. La section 4 présente les approches existantes dans la littérature. 
\section{Concepts de bases}
\subsection{Définition de l'index de jointure binaire (IJB)}
Les $IJB$ sont une combinaison entre les indexs binaires et les indexs de jointure. Ils sont largement utilisés dans les environnements d'entrepôts de données. Ces environnements ont généralement une grande quantité de données et des requêtes ad hoc, mais un faible niveau de transactions $LMD$ simultanées. Pour de telles applications, l’$IJB$ permet de :
\begin{itemize}
\item Réduire le temps de réponse pour les grandes requêtes ad hoc,
\item Réduire les besoins de stockage par rapport à d'autres techniques d'indexation,
\item Améliorer les performances même sur le matériel avec un nombre relativement petit de CPU ou une mémoire limitée,
\item Maintenance efficace et facile.
\end{itemize}
\indent L'indexation complète d'une grande table avec un index B-tree traditionnel peut être prohibitive en termes d'espace car les indexs peuvent être plusieurs fois plus grands que les données dans la table. Les $IJB$ ne représentent généralement qu'une portion de la taille des données indexées dans la table. \\
\indent Les $IJB$ sont principalement destinés aux applications décisionnelles où les utilisateurs consultent les données plutôt que de les mettre à jour. Ils ne conviennent pas aux applications $OLTP$ avec un grand nombre de transactions simultanées essayant de modifier les données. La consultation parallèle et le $LMD$ traitent les $IJB$ comme ils le font avec des indexs traditionnel \cite{Orcl}. 

Les $IJB$ réduisent les exigences de stockage par rapport à d'autres techniques d'indexation sur la même quantité de données \cite{Wu99}, surtout lorsque l'attribut indexé comporte un petit nombre de valeurs distinctes (la cardinalité n'est pas élevée) \cite{AGS97}.
L’index de jointure binaire ($IJB$) est une variante des indexs de jointure. Il constitue une combinaison entre l’index de jointure et l’index binaire. Il a été proposé pour pré-calculer les jointures entre une ou plusieurs tables de dimension et la table de faits dans les entrepôts de données modélisés par un schéma en étoile \cite{PO95, OQ97}. L’$IJB$ est défini sur un ou plusieurs attributs appartenant à plusieurs tables, nous distinguons deux types d’$IJB$: les $IJB$ simples (mono-attribut) définis sur un seul attribut d’une table de dimension et les $IJB$ multiples (multi-attributs) définis sur plusieurs attributs issus d’une ou plusieurs tables dimensions, contrairement aux indexs binaires standards où les attributs indexés appartiennent à la même table. 
Supposons un attribut $A$ ayant $n$ valeurs distinctes $v_{1}$,$ v_{2}$, ... , $v_{n}$ appartenant à une table de dimension $D$. Supposons que la table de faits $F$ est composée de $m$ instances. La construction de l’$IJB$ défini sur l’attribut $A$ se fait de la manière suivante :
\begin{enumerate}
\item Créer $n$ vecteurs composés chacun de $m$ entrées ;
\item L'\textit{ième} bit du vecteur correspondant à une valeur $v_{k}$ est mis à 1 si le n-uplet de rang $i$ de la table de faits est joint avec un n-uplet de la table de dimension $D$ tel que la valeur de $A$ de cet n-uplet est égale à $v_{k}$. Il est mis à 0 dans le cas contraire. 
\end{enumerate}
Plus formellement :
$IJB_{k}^{j}$ = 1 si $\exists$ \emph{td} $\in$ $D$ tel que \emph{tf$_{j}$.D$_{id}$} =  \emph{$td.D_{id}$} $\wedge$ \emph{t$d.A$}=v$_{k}$ où $IJB_{k}^{j}$, \emph{$tf_{i}$},\emph{td}  représentent respectivement le $j^{ime}$ bit du vecteur correspondant à la valeur $v_{k}$, le n-uplet de $F $de rang $j$, un n-uplet de la table $D$, la clé étrangère de $D$.

\begin{figure}[h!]
\begin{center}
\includegraphics[scale=0.6]{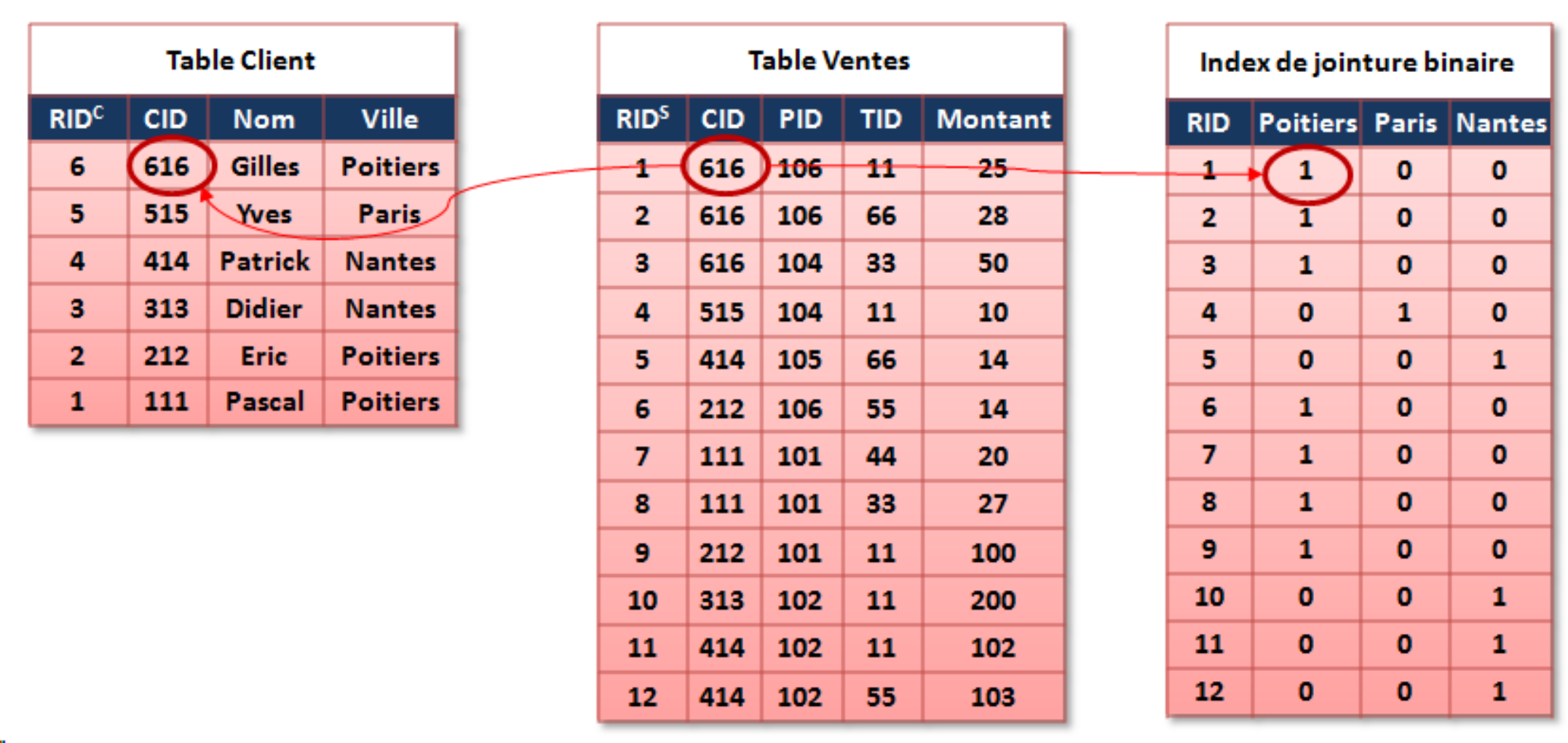}
\caption{Index de jointure binaire \cite{KB09}\label{h1}}
\end{center}
\end{figure}

Ce qui suit un exemple illustratif d’un $IJB$. 
\begin{exemple}
Soit la table de faits VENTES qui comprend 12 tuples, et soit l’attribut Ville de la table de dimension Client ayant 3 valeurs distinctes Poitiers, Paris et Nantes. La construction de l’index de jointure binaire $IJB$ défini sur l’attribut A s'effectue comme suit :
\end{exemple}

\begin{itemize}
\item Créer 12 vecteurs, chacun comprenant 3 entrées,
\item Le premier enregistrement de la table de faits $Ventes$ est joint avec le premier enregistrement de la table de dimension $Client$ tel-que ce client habite la ville $Poitiers$. Ainsi, la case correspondante à la ville de $Poitiers$ de la première ligne de cet $IJB$ est mise à 1, le reste sera mis à 0, et de même pour tous les autres enregistrements.
\end{itemize}

La syntaxe SQL relative à sa création est la suivante:
\begin{verbatim}
CREATE BITMAP INDEX Client_Ville_IJB
	ON Ventes(Client.Ville)
	FROM Ventes, Client
	WHERE Ventes.CID= Client.CID
\end{verbatim}

\subsection{Formalisation du problème de sélection d’index}
Le problème de sélection d'index est un problème difficile qui consiste à choisir la configuration d'index optimale susceptible d'optimiser les requêtes de jointure en étoile et réduire leur temps d'exécution. Cette configuration d'index doit respecter certaines contraintes imposées dans un entrepôt de données telles que l'espace de stockage,  le nombre d’index par table, le temps de maintenance et de mise à jour, etc. Par conséquent, ce type de problème est considéré dans la communauté de recherche comme un enjeu critique dans la conception physique étant donné qu'un $IJB$ est en général défini sur un ensemble d'attributs issus de plusieurs tables différentes. Pour formaliser ce problème, nous pouvons dire qu'en entrée, nous disposons de 3 paramètres essentiels :

\begin{enumerate}
\item Un ensemble de requêtes $Q $ = \{$Q_{1}$, $Q_{2}$, ..., $Q_{m}$\} qui forment la charge de travail,

\item Un schéma d’entrepôt de données renfermant une table de faits $F$ et un groupe de tables de dimensions $D_{i}$, 

\item Les contraintes de coût de stockage maximum à ne pas dépasser. 
\end{enumerate}

Le problème de sélection des $IJB$ consiste à déterminer une configuration d'index dont le coût d'exécution soit minimal tout en respectant l'ensemble des contraintes.
Le problème de sélection des $IJB$ est considéré comme un problème NP-Complet \cite{DCom78}. Par conséquent, élaborer un algorithme exhaustif fournissant une solution optimale et exacte en un temps raisonnable devient impossible. Dans ce contexte, la communauté de recherche s'intéresse à réaliser un algorithme s'appuyant sur des heuristiques pour s'approcher de la solution optimale représentée sous forme d'une configuration d'index basée sur un ensemble d'attributs candidats. \\

Pour construire une configuration d'index optimale, nous devons déterminer les attributs pertinents à indexer et cela peut s'effectuer de deux manières: 
\begin{description}
  \item[$\bullet$ Sélection manuelle : ]une technique qui s’appuie sur l’expertise humaine. En effet, cette tâche est effectuée par l'administrateur qui doit avoir une bonne pratique, l’expérience nécessaire, et un raisonnement fiable pour choisir les attributs candidats.
  
\item[$\bullet$ Sélection automatique : ]ce type de sélection passe par deux phases cruciales (cf. figure \ref{h2}) :

\begin{figure}[htbp]
\begin{center}
\includegraphics[scale=0.5]{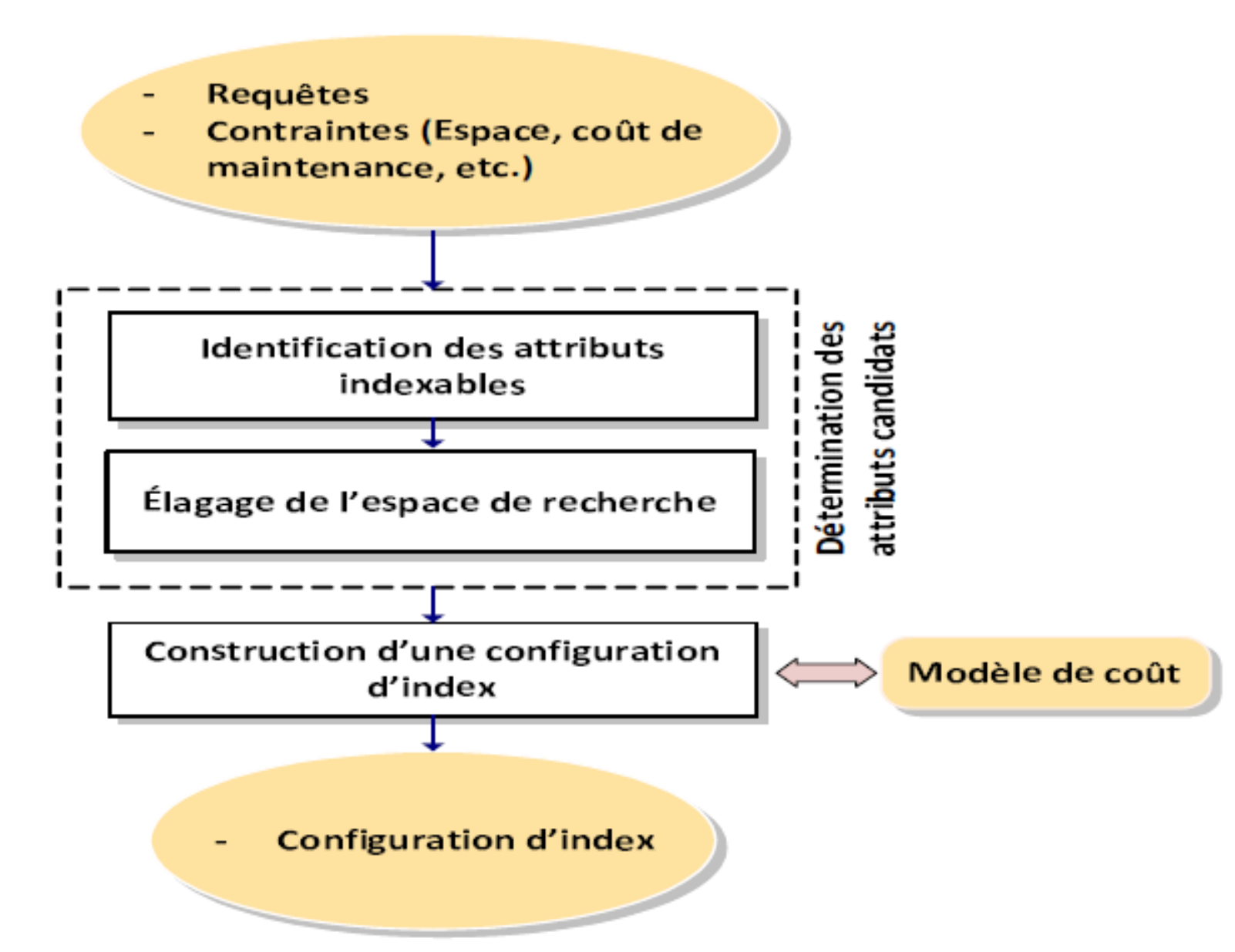}
\caption{Approche générique de sélection d'IJB \label{h2}}
\end{center}
\end{figure}

\begin{enumerate}
\item \textbf{Détermination des attributs candidats : }étant donné que l'espace de recherche des attributs candidats est exponentiel, une bonne étude des requêtes est essentielle pour le réduire en élaguant certains attributs qui sont moins intéressants que d'autres.

\item  \textbf{Construction de la configuration finale des indexs : }dans cette deuxième phase, nous nous basons sur des modèles de coût qui estiment le temps d'exécution des requêtes et nous permettent ainsi d'aboutir à une configuration finale qui renferme les meilleurs attributs à indexer.
\end{enumerate}

\end{description}

\section{Stratégie d’exécution en présence des IJB}
Une requête de jointure en étoile est caractérisée par un ensemble de sélections sur les tables de dimension, suivies de jointures avec la table de faits. Si tous les attributs des prédicats de sélection sont indexés, alors l’exécution d’une requête de jointure passe par 4 étapes \cite{KB09}, commence par la réécriture de la requête et se termine par la récupération des n-uplets de la table de faits. Dans ce qui suit, nous présentons une description détaillée des étapes \cite{KB09}:

\begin{enumerate}
\item Effectuer une réécriture de la requête qui consiste à séparer chaque jointure dans une sous-requête. Chaque sous-requête représente l’ensemble des sélections effectuées sur chaque table de dimension.
\item Pour chaque sous-requête, utiliser les $IJB$ définis sur la table de dimension pour trouver un vecteur de bits représentant les n-uplets de la table de faits référencés par la sous-requête. 
\item Effectuer une opération AND entre les vecteurs obtenus à partir des sous-requêtes pour trouver un seul vecteur référençant tous les n-uplets référencés par la requête.
\item Utiliser le vecteur résultat pour accéder à la table de faits et récupérer les n-uplets référencés par la requête globale.
\end{enumerate}

\indent Pour montrer cette stratégie d’exécution, nous présentons l’exemple suivant. 
\begin{exemple}
Soit l’entrepôt de données de la figure \ref{im11} de la page \pageref{im11}. Supposons la requête $Q$ définie par :
\end{exemple}
\begin{verbatim}
SELECT avg(montant)
FROM Ventes V, Client C, Produit P, Temps T
WHERE V.CID=D.CID AND V.PID=P.PID AND V.TID=T.TID AND
C.VILLE=(’Poitiers’ OR ’Nantes’) AND Mois =’Mars’ 
AND (P.TYPE=’Jouet’ OR ’Beauté’)
\end{verbatim}

Supposons l’existence de trois index mono-attributs, $IJB$-$Ville$, $IJB$-$Mois$ et $IJB$-$Type$ définis comme suit :
\begin{enumerate}
\item \begin{verbatim}
CREATE BITMAP INDEX IJB-VILLE
ON Ventes(Client.Ville)
FROM Ventes, Client
WHERE Client.CID=Ventes.CID
\end{verbatim}

\item \begin{verbatim}
CREATE BITMAP INDEX IJB-MOIS
ON Ventes(Temps.Mois)
FROM Ventes, Temps
WHERE Client.CID=Ventes.CID
\end{verbatim}

\item \begin{verbatim}
CREATE BITMAP INDEX IJB-TYPE
ON Ventes(Produit.Type)
FROM Ventes, Produit
WHERE Client.CID=Ventes.CID
\end{verbatim}

\end{enumerate}

Pour exécuter $Q$, nous suivons les trois étapes ci-dessus mentionnées.
\begin{enumerate}
\item La réécriture de $Q$ donne : \begin{verbatim}
SELECT avg(montant)
FROM Ventes
WHERE CID in 
(SELECT CID FROM Client WHERE AND VILLE=(’Poitiers’ OR ’Nantes’))
AND V.PID in 
(SELECT P.PID FROM Produit WHERE (P.TYPE=’Jouet’ OR ’Beauté’))
AND V.TID in 
(SELECT T.TID FROM Time WHERE T.Mois =’Mars’)
\end{verbatim}
\item Le calcul du vecteur de bits relatif à chaque sous-requête se fait comme suit (cf. figure \ref{h3}) :
\begin{figure}[htbp]
\begin{center}
\includegraphics[scale=0.5]{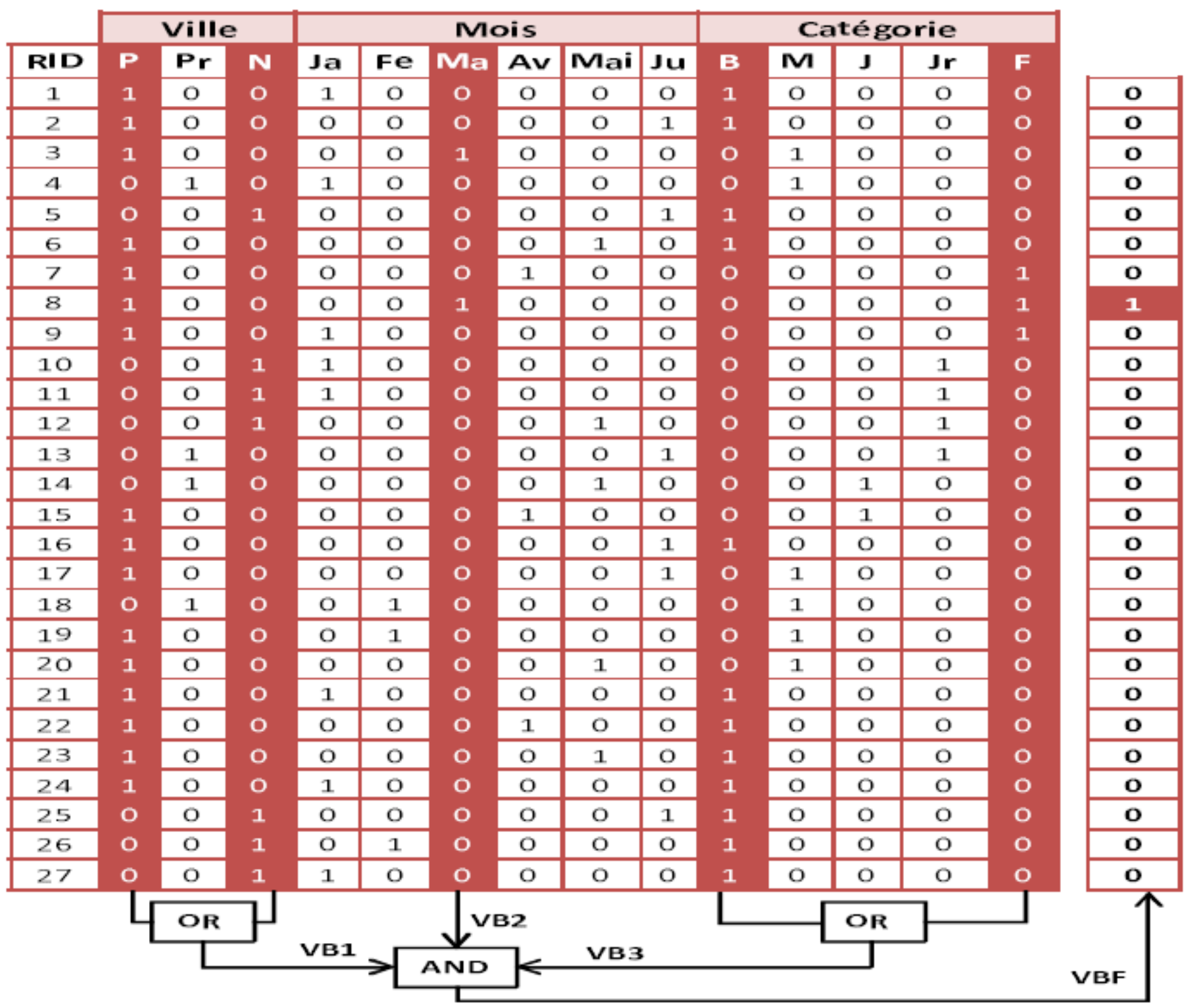}
\caption{Vecteur de bits d'un IJB \cite{KB09}\label{h3}}
\end{center}
\end{figure}

\begin{itemize}

\item  Sous-requête 1 : l'optimiseur calcule le vecteur vérifiant le prédicat \begin{verbatim}C.VILL E= ('Poitiers' OR 'Paris').
\end{verbatim}Pour cela, il effectue une opération $OR$ entre les deux vecteurs représentant les valeurs $Poitiers$ et $Paris$. Le résultat est sauvegardé dans un vecteur binaire $VB1$.
\item  Sous-requête 2 : chargement du vecteur relatif à la valeur \begin{verbatim}Mois ='Juin'.\end{verbatim}Le résultat est sauvegardé dans un vecteur binaire $VB2$.
\item Sous-requête 3 : Chargement des vecteurs relatifs aux valeurs $Jouet$ et $Beauté$. Il effectue une opération $OR$ pour avoir un vecteur résultat. Le résultat est sauvegardé dans un vecteur binaire $VB3$.

\end{itemize}

\item L’optimiseur calcule un vecteur binaire final $VBF$ qui est le résultat d’une opération $AND$ entre les trois vecteurs obtenus comme suit : $VBF=VB1$ $AND$ $VB2$ $AND$ V$B3$.

\item L’optimiseur accède à partir de $VBF$ aux n-uplets de la table de faits en utilisant les \textit{rowids} correspondant aux bits mis à 1 dans $VBF$. Il calcule la moyenne du montant de ces n-uplets.

\end{enumerate}

Cet exemple montre l’intérêt de ce type d’index pour l’optimisation de requêtes décisionnelles.

\section{Modèle de coût}
Généralement, le nombre d'index candidats est d'autant plus important que la charge en entrée est volumineuse. La création de tous ces indexs peut ne pas être réalisable en pratique à cause des contraintes systèmes telles que la taille de l'espace de stockage alloué aux indexs ou le nombre limité d'index par table. Pour contourner ces limitations, nous avons utilisé le modèle de coût proposé par \emph{Aouiche et al. }\cite{ADOB05}. À partir d’une requête $Q$ et d’une configuration d’index $CI$, ce modèle permet d’estimer la taille des indexs de la $CI$ ainsi que le coût d’exécution de $Q$ en termes de nombre d’entrées-sorties en présence de la $CI$. Il utilise un ensemble de paramètres dont les plus importants sont représentés sur le tableau \ref{tbl55}.

\begin{table}[]
\centering
\caption{Paramètres du modèle de coût}
\label{tbl55}
\begin{tabular}{|l|l|lll}
\cline{1-2}
Symbole   & Description                                                                                                  &  &  &  \\ \cline{1-2}
$|T|$     & \begin{tabular}[c]{@{}l@{}}Nombre de n-uplets de la table $T$ ou cardinalité\\ 			de l'attribut X\end{tabular} &  &  &  \\ \cline{1-2}
S$_{p} $        & Taille en octets d'une page disque                                                                           &  &  &  \\ \cline{1-2}
$PT$        & \begin{tabular}[c]{@{}l@{}}Nombre de pages nécessaires pour stocker la\\ 			table $T$\end{tabular}             &  &  &  \\ \cline{1-2}
S$_{pointeur} $ & Taille en octets du pointeur d'une page                                                                      &  &  &  \\ \cline{1-2}
$m$         & Ordre d'un B-tree &  &  &  \\ \cline{1-2}
$d$         & \begin{tabular}[c]{@{}l@{}}Nombre de bitmaps utilisés pour évaluer une\\ 			requête donnée\end{tabular}      &  &  &  \\ \cline{1-2}
$w(X)$      & \begin{tabular}[c]{@{}l@{}}Taille en octets d'un n-uplet de la table $T$ ou\\ 			de l'attribut $X$\end{tabular}  &  &  &  \\ \cline{1-2}
\end{tabular}
\end{table}

\subsection{Types de modèle de coût}
Nous différencions deux types de modèles de coût:

\begin{description}
\item [Modèle basé sur une fonction mathématique adhoc : ]cette catégorie de modèles définit une fonction de coût qui prend en entrée une requête et un plan d’exécution et donne en sortie le coût de ce plan. Ce coût est calculé en utilisant des formules qui prennent en compte un certain nombre de paramètres et de statistiques collectées sur la base de données, tout en considérant un certain nombre d’hypothèses.
\item [Modèle basé sur l’optimiseur des requêtes : ]ils font appel à l’optimiseur de requêtes du SGBD utilisé pour estimer le coût d’une requête. L’optimiseur reçoit la requête, il évalue les différents plans d’exécution et retourne le meilleur plan avec son coût. Le fait d’utiliser l’optimiseur de requêtes rend le calcul du coût plus fiable mais engendre deux inconvénients majeurs : \textit{(i)} le coût que l’optimiseur estime dépend du SGBD utilisé et \textit{(ii)} Le fait de recourir souvent à l’optimiseur engendre un coût d’exécution supplémentaire et dégrade la qualité de l’optimiseur qui passe plus de temps à estimer le coût d’exécution des requêtes qu’à les exécuter.
\end{description}
\subsection{Coût de stockage d’un IJB}
Un $IJB$ construit sur un ensemble d’attributs de dimension, stocke pour chaque n-uplet de la table de faits son identificateur de ligne (\textit{RowID}) ainsi qu’un ensemble de vecteurs binaires représentant chacune une valeur des attributs indexés. L’espace de stockage d’un $IJB$ dépend de deux paramètres : le nombre de n-uplets de la table de faits et la cardinalité des attributs indexables.
Soit un index $I$ défini sur $n$ attributs $A_{1},A_{2}, ..  ,A_{n}$. L’espace de stockage en octets de $I$ est calculé par la formule suivante : 
Taille($I$) = {\large $\frac{(|RowID| + \sum^{n}_{j-1} |A_{j}|) \times |F|}{8}$}
\subsection{Coût d’exécution}
Le coût d’exécution d’une requête est exprimé en nombre d'entrées-sorties nécessaires pour son exécution. Selon les attributs indexés dans $CI$, nous pouvons considérer trois scenarii possible pour l’exécution de $Q$ : \textit{(i)} aucun attribut indexable de $Q$ n'est couvert par $CI$; \textit{(ii)} tous les attributs indexables de $Q$ sont couverts par $CI$; et \textit{(iii)} quelques attributs indexables de $Q$ sont couverts par $CI$.
\begin{enumerate}
\item \textbf{Scénario 1 : Aucun attribut de $Q$ n'est indexé dans $CI$ :} dans le cas d'absence d'$IJB$ utilisés par $Q$, toutes les jointures de $Q$ peuvent être calculées en utilisant la jointure par hachage. Pour calculer le coût d’exécution de ces jointures, nous avons utilisé le modèle de coût développé et décrit dans \emph{Aouiche et al.} \cite{ADOB05} où il suppose que toutes les jointures sont obtenues par la jointure par hachage. Le nombre d'E/S nécessaires pour joindre le table R avec la table S est alors C-$_{hash} $ = 3 (pS + pR) (pX = Nombre de pages nécessaires pour stocker la table X).
\item \textbf{Scénario 2 : Tous les attributs de $Q$ sont indexés dans $CI$ : }ce cas représente la situation idéale où toutes les jointures dans $Q$ ont été pré-calculées dans $CI$. L’exécution de $Q$ dans ce cas passe par deux étapes importantes : le chargement des indexs, ensuite l’accès aux données. Par conséquent, deux coûts sont considérés : le coût de chargement des indexs et le coût d’accès aux données.
\begin{itemize}

\item \textit{Coût de chargement des indexs :} le coût de chargement d’un $IJB$ noté $CC(I)$ correspond au nombre de pages lues pour le charger. Il est calculé par la formule suivante :
$CC(I)$ = $\frac{Taille(I)}{PS}$

Le coût de chargement de l’ensemble d’index utilisés par une requête $Q$ est égal à la somme des coûts de chargement de ces index.
\item \textit{Coût d'accès aux n-uplets :} soit $N_{t}$ le nombre de n-uplets de la table de faits référencés par la requête $Q$. Le coût de lecture ($CL$) de ces n-uplets est donné par la formule suivante : $CL$ = $\Vert F \Vert (1- e^{-\frac{N_{t}}{\Vert F \Vert}}) $, \\
où  $\Vert F \Vert$ désigne le nombre de pages nécessaires pour stocker la table de faits $F$.
\end{itemize}
\item \textbf{ Scénario 3 : Quelques attributs de $Q$ sont indexés dans $CI$ :} dans ce scénario, l’exécution de $Q$ se fait en deux phases. Dans la première phase, les indexs utilisés par $Q$ sont chargés et utilisés pour trouver un ensemble de n-uplets de la table de faits. Le coût de cette phase est calculé comme celui du scénario 2. Il correspond au coût de chargement des indexs utilisés ainsi que celui de chargement des n-uplets de faits. Dans la deuxième phase, les jointures non encore effectuées (à cause de l'absence d’$IJB$ les pré-calculant dans $CI$) sont effectuées entre les n-uplets de faits, résultats de la première étape, et les tables de dimension non encore jointes. Le coût de cette étape est calculé en utilisant le modèle que nous avons défini dans le chapitre 3 (voir scénario 1). 
\end{enumerate}

\section{État de l’art }
Déterminer les attributs candidats à partir d’un entrepôt de données est un problème combinatoire et l’espace de recherche correspondant est gigantesque. Donc, on ne peut pas se permettre d’implémenter des solutions exhaustives ou énumératives. Pour réduire ce coût, nous passons comme première approche par une phase d’élagage (prunning) afin de réduire l’espace de recherche. Plusieurs travaux se sont intéressés à cette phase de d’élagage, qui s'appuie sur certains critères de sélection qui privilégie des attributs par rapport à d'autres tels que la fréquence d'utilisation, et la taille des tables, etc. \cite{GRS02}. Cependant, la sélection d'un ensemble optimal d’index est un problème très difficile \cite{CDN04} en raison du volume de données important et du nombre exponentiel d'attributs candidats pouvant être utilisés dans le processus de sélection. Pour réduire cette complexité, la plupart des techniques (\emph{Chaudhuri et Narasayya, 1997} \cite{CN97}; \emph{Valentin et al., 2000} \cite{VZZLS00}; \emph{Chaudhuri et al., 2004} \cite{CDN04}; \emph{Aouiche et al., 2005} \cite{ADOB05} ; \emph{Bellatreche et al., 2008} \cite{BMND08}) élaguent l’espace de recherche pour réduire le nombre d'attributs candidats, les indexs sont sélectionnés avec des algorithmes différents tels que des algorithmes glouton, des algorithmes linéaires, etc. même un algorithme génétique a été proposé pour la sélection d’$IJB$ simple \cite{BB12}. \\
Dans ce travail, nous nous focalisons sur la sélection des $IJB$ dans les entrepôts de données modélisés par un schéma en étoile.
\subsection{L'approche de Aouiche et al. 2005} 
L’algorithme \textsc{Close} (\emph{Pasquier et al.,1999} \cite{Pasc99}) de recherche des motifs fréquents fermés, notés $IFF$, est utilisé pour élaguer l’espace de recherche des $IJB$. Un motif fermé est défini comme suit :
\begin{mydef}
Soit  $ \Gamma = \{i_{1}, i_{2}, .., i_{m}\}$ un ensemble fini d'éléments, et $ D = \{T_{1}, T_{2}, .., T_{n}\}$ un ensemble de données contenant $N$ transactions, où chaque transaction $T_{k} \in D$ est un élément tel que $T_{k} \subseteq \Gamma$. 
Nous appelons k-itemset une séquence de $k$ éléments distincts \cite{Agr96}. 
Un ensemble $I$ est dit fréquent si son support est supérieur à un certain seuil appelé $minsup$. Le support d'un ensemble d'éléments est défini comme étant le nombre de transactions contenant ces éléments. Un motif est fermé s'il n'y a pas de super-ensemble avec la même fréquence \cite{GZ05}.
\end{mydef}

 Notons que les motifs générés servent à construire l’ensemble des $IJB$ candidats. Un algorithme glouton est ensuite exécuté pour sélectionner une configuration finale d’$IJB$. \emph{Aouiche et al.} \cite{ADOB05} considèrent seulement les fréquences d’accès des attributs comme critère de génération des motifs fréquents fermés.

\begin{figure}[htbp]
\begin{center}
\includegraphics[scale=0.7]{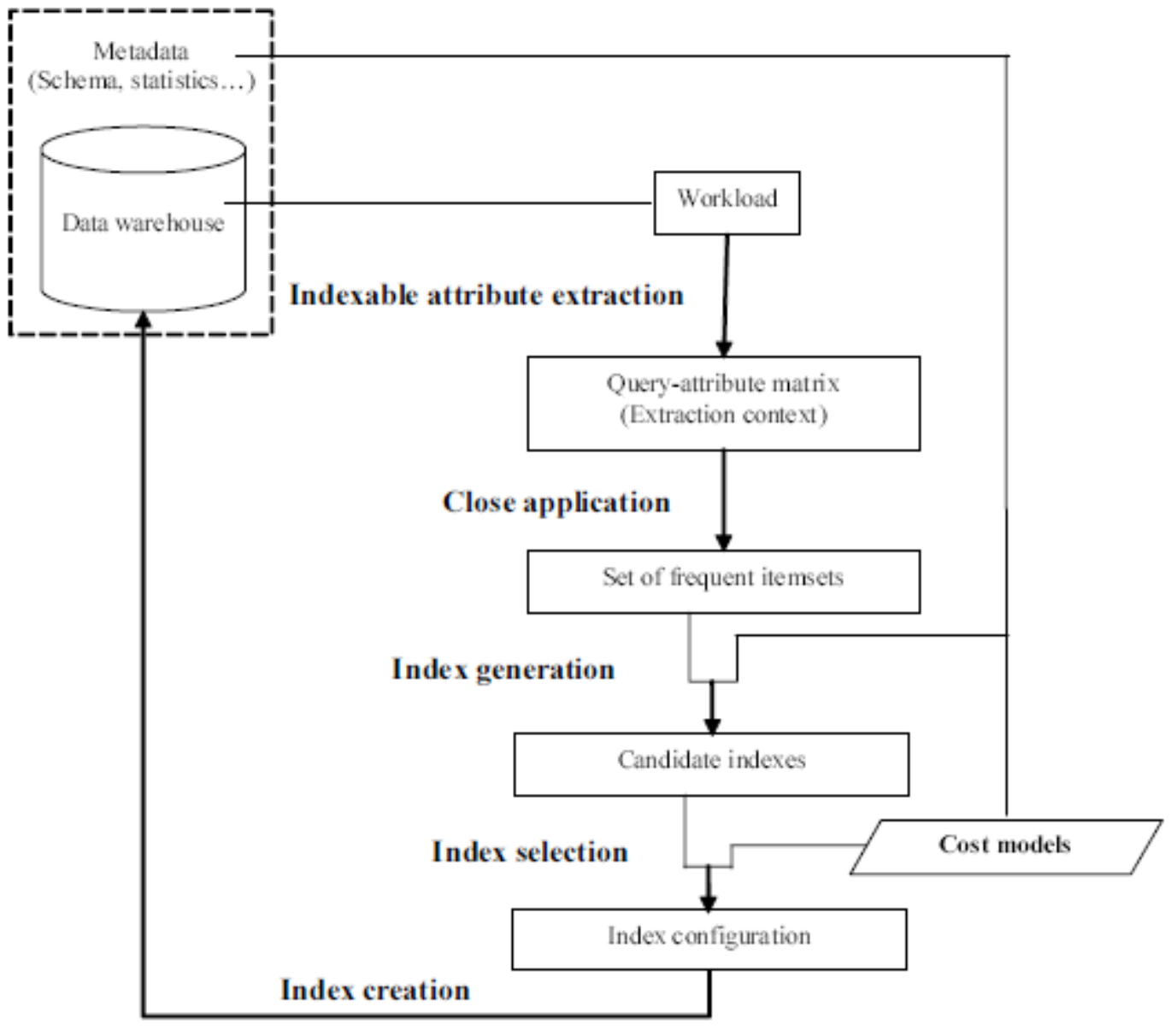}
\caption{L'approche Close de Aouiche \cite{ADOB05}\label{h4}}
\end{center}
\end{figure}
\begin{enumerate}[label=\roman*]
\item \textbf{Extraction de la charge de requêtes :} à partir du journal des transactions sauvegardé et maintenu automatiquement par le SGBD.
\item \textbf{Analyse de la charge :} la charge de requête obtenue est analysée afin d’extraire l’ensemble des attributs indexables. 
\item \textbf{Construction d’un contexte de recherche des motifs fermés :}  à partir des attributs retenus de l’analyse de la charge, construire une matrice Requêtes-Attributs qui a des lignes représentant les requêtes, et des colonnes représentant les attributs indexables.
\item \textbf{Application de l’algorithme \textsc{Close} \cite{Pasc99}.}
\item \textbf{Construction de l’ensemble des index candidats :} l’ensemble des index candidats est construit à partir des motifs fréquents fermés résultant de l’application de l’algorithme \textsc{Close}.
\item \textbf{Construction de la configuration d’index finale :} à partir de l’ensemble d’index généré dans l’étape précédente, un algorithme glouton est appliqué pour sélectionner une configuration d’index finale. Cet algorithme procède en plusieurs itérations. Dans la première itération, la fonction objective est calculée pour chaque index candidat. L’index $Imax$ maximisant la fonction objective est choisi pour former la configuration initiale. Durant chaque itération, un nouvel index vérifiant la même condition est ajouté à la configuration courante. L’algorithme s’arrête dans les cas suivants : 

\begin{itemize}
\item Aucune amélioration de la fonction objectif n’est possible.
\item Tous les indexs ont été sélectionnés.
\item L’espace de stockage disponible est saturé.
\end{itemize}
\end{enumerate}

\indent La fonction objective utilisée repose sur un modèle de coût, qui permet de calculer la taille des indexs sélectionnés ainsi que le coût d’exécution des requêtes en présence de ces indexs.

\indent La première fonction objective améliore les indexs qui offrent plus de profit lors de l'exécution des requêtes,  la seconde favorise  les indexs qui fournissent plus de bénéfices et occupent moins d'espace de stockage. La troisième combine les deux premières afin de sélectionner d'abord tous les indexs offrant plus de profit et ne gardent que ceux qui occupent moins d'espace de stockage lorsque cette ressource devient critique.

\subsection{Les travaux de Bellatreche 2008}

\emph{Bellatreche et al. }\cite{BMND08,NBM07} exposent une amélioration des travaux d'\emph{Aouiche et al }\cite{ADOB05}, ils proposent l'algorithme \textsc{DynaClose} qui est une
 adaptation de l'algorithme \textsc{Close} sur lequel se basent \emph{Aouiche et al}. en injectant  des paramètres d'élagage, ce qui permet d'arriver directement à l'index de   jointure binaire élu. L'algorithme proposé s'appuie sur une fonction permettant de    pénaliser les attributs fréquents définis sur des tables de petites tailles. \\
 \indent Cette approche commence d'abord par déterminer les attributs indexables, sachant qu'un attribut indexable est un attribut qui ne fait pas partie de la table de faits et qui n'est pas un attribut clé de la table dimension. Ensuite, elle détermine la matrice de contexte. Une fois la matrice créée, cette approche fournit le support relatif à chaque attribut indexable et les motifs pour chaque requête. Et enfin, elle calcule la valeur de la fonction  \textsc{Fitness} de ces motifs en suivant la formule suivante. Pour un motif fréquent $m_{i}$, la formule est définie par :\label{fitness} 

$Fitness(m_{i}) $= {\large  $\frac{1}{n}  \times ( \sum_{j = 1}^{n}  (\alpha_{j} \times sup_{j}))$}
\\
Où $n$ désigne le nombre d’attributs non clés, $j$ est l'attribut dans le motif à tester $m_{i}$.  $sup_{j}$ représente le support de $j$,  $\alpha_{j}$ est un paramètre de pénalisation déterminé par l’équation :
$\alpha_{j} = \frac{|D_{j}|}{|F|}$  ; sachant que les notations $|D_{j}|$ et $|F|$ expriment respectivement la taille  de la table de dimension $D$ à laquelle appartient l'attribut $j$ et la table de faits $F$ en nombre de pages. \\
\indent Suite à la détermination des motifs fréquents générés, l’approche \textsc{DynaClose} est utilisée dans la phase d'élagage de l’espace de recherche des indexs pour élaguer les motifs inaptes à produire un index de jointure. Par exemple, un motif fréquent et ne renfermant aucun attribut non clé de la table de dimension sera éliminé. La purification permet de générer un ensemble d’attributs indexables candidats. Cet ensemble est défini par l’union des attributs non clés appartenant aux motifs fréquents générés. 
\subsection{L'approche de Boukhalfa et al. 2010}
\emph{Boukhalfa et al.} \cite{BBB10} présentent une approche de sélection d’une configuration d’$IJB$ visant à réduire le coût d’exécution d’une charge de requêtes sous une contrainte de stockage. Cette approche permet à la fois d’extraire les $IJB$ mono-attribut et les $IJB$ multi-attributs. \\
L’extraction des $IJB$ Mono-attribut se fait en 3 étapes : 
\begin{enumerate}[label=\roman*]
\item \textbf{L’identification des attributs indexables :} sont choisis parmi les attributs indexables de faible et de moyenne cardinalité. 
\item \textbf{L’initialisation de la configuration :} configuration initiale composée d’un index mono-attribut défini sur l’attribut ayant une cardinalité minimum.
\item \textbf{L’enrichissement de la configuration actuelle par l’ajout de nouveaux index :} une amélioration itérative par l’ajout d’index définis sur d’autres attributs non encore indexés. L’algorithme s’arrête lorsque l’une des deux conditions suivantes sont satisfaites : aucune amélioration n’est possible et l’espace de stockage est consommé.
\end{enumerate}

\begin{figure}[htbp]
\begin{center}
\includegraphics[scale=0.6]{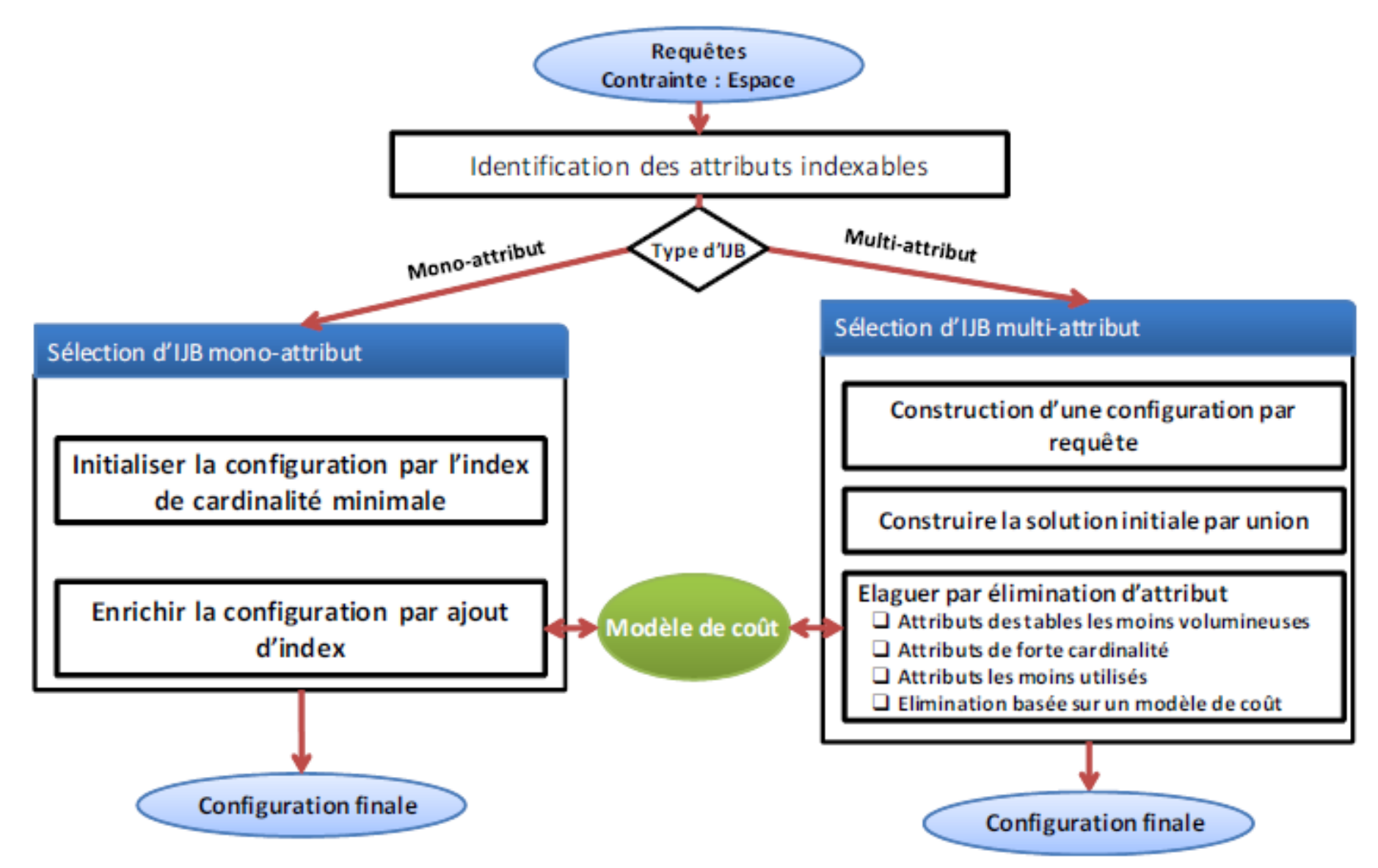}
\caption{L'approche de Boukhalfa et al. \cite{KB09}\label{h5}}
\end{center}
\end{figure}

L’extraction des $IJB$ Multi-attribut se fait en 4 étapes : 
\begin{enumerate}[label=\roman*]
\item \textbf{Identification des attributs indexables :} cette étape se fait de la même manière que dans l’approche de sélection des $IJB$ mono-attribut.
\item \textbf{Construction d’une configuration par requête :} dans cette étape, la charge de requête $Q$ est éclatée en $m$ sous-charges. Chaque sous-charge est composée d’une seule requête. Pour chaque requête $Q_{i}$ nous sélectionnons l’index qui couvre tous les attributs indexables utilisés par cette requête. 
\item \textbf{Construction d’une configuration initiale :} le but de cette étape est de construire une configuration initiale à partir de l’ensemble des index générés lors de l’étape précédente. Cette configuration initiale est générée en effectuant l’union des indexs obtenus afin que chaque requête puisse être optimisée via l'$IJB$ contenant tous les attributs indexables qu’elle utilise. Le nombre d’indexs dans la configuration initiale est inférieur ou égal au nombre de requêtes de la charge, car certaines requêtes partagent les mêmes index.
\item\textbf{ Construction d’une configuration finale :} vu le nombre important d’attributs utilisés par la configuration initiale, l’administrateur est confronté au problème du choix des attributs à éliminer. Plusieurs stratégies peuvent être suivies pour effectuer ce choix.
\begin{itemize}

\item Elimination des attributs de forte cardinalité (\textsc{AFC}),
\item Elimination des attributs appartenant aux tables moins volumineuses (\textsc{TMV}),
\item Elimination des attributs les moins utilisés (\textsc{AMU}),
\item Elimination des attributs apportant moins de réduction de coût (\textsc{MC}).
\end{itemize}
\end{enumerate}

\subsection{L'approche de Necir 2010}
\emph{Necir et al.} \cite{Necir10} exposent une amélioration de l’approche de \emph{Azefack} \cite{AAD06} qui utilise l'algorithme $GenMax$. L'idée fondamentale de cette approche est que seuls les attributs fréquents dans les requêtes pour lesquelles chaque super-ensemble est infréquent sont considérés dans la configuration de l'index. 
\emph{Necir et al.} proposent une approche opérant en quatre étapes. Dans la première, ils effectuent un prétraitement de la charge de travail des requêtes. Dans la seconde, ils appliquent un nouvel algorithme d'élagage, appelé \textsc{BJIMax} qui identifie l'ensemble des $IMF$. Un $IMF$ est défini selon la définition \ref{defimf}.
\begin{mydef}\label{defimf}
Soit  $ \Gamma = \{i_{1}, i_{2}, .., i_{m}\}$ un ensemble fini d'éléments, et $ D = \{T_{1}, T_{2}, .., T_{n}\}$ un ensemble de données contenant $N$ transactions, où chaque transaction $T_{k} \in D$ est un élément tel que $T_{k} \subseteq \Gamma$. 
Nous appelons k-itemset une séquence de $k$ éléments distincts \cite{Agr96}. 
Un ensemble $I$ est dit fréquent si son support est supérieur à un certain seuil appelé $minsup$. Le support d'un ensemble d'éléments (en valeur absolue) est défini comme étant le nombre de transactions contenant ces éléments. Un ensemble est maximal s'il n'a pas de super ensemble qui est fréquent \cite{GZ05}.
\end{mydef}
Ensuite, les $IMF$ définies générées par \textsc{BJIMax} sont nettoyées pour éviter les $IJB$ erronées. Enfin, ils exécutent un algorithme glouton pour sélectionner l'ensemble des index gagnants parmi les candidats sélectionnés. 
\begin{enumerate}[label=\roman*]
\item \textbf{Extraction de la charge de requêtes :} à partir du journal des transactions sauvegardé et maintenu automatiquement par le SGBD. 
\textbf{Analyse de la charge : }la charge de requête obtenue est analysée afin d’extraire l’ensemble des attributs indexables. 
\textbf{Construction d’un contexte de recherche des motifs fermés : }est une matrice qui a des lignes représentant les requêtes, et des colonnes représentant les attributs indexables.
\item \textbf{Génération des $IMF$ :} L’algorithme est inspiré de l'algorithme \textsc{GenMax} de \emph{Gouda et Zaki, 2005} \cite{GZ05} qui adopte une stratégie de recherche en profondeur et est basé sur une nouvelle métrique d'élagage appelée \textit{poids}. Ce dernier pénalise l'$IMF$ définie sur un petit profit comme suit: 

$Weight(X) $={\large  $\frac{sup(X)}{n} \times \lceil \sum^{n}_{i=1} profit(A_{i}) \rceil$}
\\
Où $n$, $sup(X)$ représentent respectivement le nombre d'attributs et le support de l'attribut $A_{i}$ dans $IMF (X)$. Le profit est calculé comme suit: \\
{\large $profit(A_{i}) = \frac{\Vert T_{i} \Vert \times LT}{PS}$}

où $T_{i}$, $LT$ et $PS$ représentent respectivement le nombre d'instances de $T$, la longueur d'une instance de $T$ et la taille de la page (en octets).
\emph{Necir et al.} définissent une nouvelle contrainte de valeur de seuil minimale, appelée $MinWeight$ calculée comme suit: $MinWeight$ = $MinSup$ $\times$ $Minprofit$ où $Minprofit$ représente la valeur minimale du profit.

\item \textbf{ Sélection des IJB candidats : }
l'algorithme élimine par la suite de l'ensemble des $IMF$ générées par la dernière étape, celles qui ne respectent pas les exigences $IJB$. Rappelons que chaque $IJB$ valide composé de $N$ attributs distincts contient 2 * $N$ attributs clés qui représentent les prédicats de jointure.
\end{enumerate}
\subsection{Travaux de Whang 1985}

\emph{Whang et al.} \cite{Whang85} ont proposé une approche ascendante et une approche descendante pour la sélection d’indexs. Ces deux approches sont implémentées par deux algorithme \textsc{Drop} et \textsc{Add}. L’approche descendante considère un état initial contenant tous les indexs possibles et durant chaque itération, l’index engendrant la plus grande décroissance du coût d’exécution de la charge de requête est éliminé. Quand l’élimination d’un seul index à la fois ne permet pas de réduire le coût d’exécution des requêtes, l’algorithme \textsc{Drop} élimine deux index à la fois, ensuite trois index, et ainsi de suite. L’approche ascendante quant à elle considère un état initial vide dans lequel aucun index n'est et durant chaque itération de l’algorithme \textsc{Add}, un index réduisant le coût d’exécution des requêtes sera ajouté à l’état courant. Le processus s’arrête lorsque tous les indexs sont créés ou aucune réduction de coût n’est possible. 
\subsection{Bilan et discussion}
Nous avons présenté dans cette section les principaux travaux sur le problème de sélection des indexs. La majorité des approches proposées commencent par l’identification des attributs indexables, qui peut être manuelle ou automatique, ensuite elles utilisent des algorithmes de sélection (algorithme glouton ou dirigé par des techniques de data mining) afin de générer la configuration d’index finale. La qualité de cette configuration est mesurée soit par un modèle de coût mathématique, soit par le modèle de l’optimiseur du SGBD, que nous avons expliqué dans la section précédente. Le tableau \ref{tbll54} présente une comparaison entre les principaux travaux en se basant sur plusieurs critères tel que la nature des $IJB$ , qu'ils soient mono-attribut ou multi-attribut, le type de l'algorithme de sélection et le modèle du coût utilisé pour mesurer la qualité de la configuration finale.

\begin{table}[]
\centering
\caption{Récapitulatif des travaux de l'état de l'art}
\label{tbll54}
\begin{tabular}{|l|l|l|l|}
\hline
\textbf{Approche}                           & \textbf{\begin{tabular}[c]{@{}l@{}}Sélection\\ 			des attributs candidats\end{tabular}} & \textbf{\begin{tabular}[c]{@{}l@{}}Algorithme\\ 			de sélection\end{tabular}} & \textbf{\begin{tabular}[c]{@{}l@{}}Modèle\\ 			de coût\end{tabular}} \\ \hline
Wang et al. \cite{Whang85}       & Manuelle                                                                                & Glouton                                                                       & Mathématique                                                         \\ \hline
Aouiche et al. \cite{ADOB05}     & Automatique                                                                             & \begin{tabular}[c]{@{}l@{}}Data Mining \\  + Glouton\end{tabular}            & Mathématique                                                         \\ \hline
Bellatreche et al. \cite{BMND08} & Manuelle                                                                                & \begin{tabular}[c]{@{}l@{}}Data Mining \\ + Glouton\end{tabular}            & Mathématique                                                         \\ \hline
Boukhalfa et al. \cite{BBB10}   & Manuelle                                                                                & Glouton                                                                       & Mathématique                                                         \\ \hline
Necir et al. \cite{Necir10}      & Manuelle                                                                                & \begin{tabular}[c]{@{}l@{}}Data Mining \\ + Glouton\end{tabular}            & Mathématique                                                         \\ \hline
\end{tabular}
\end{table}

\section*{Conclusion}
Les $IJB$ sont des structures redondantes permettant de pré-calculer les jointures dans un entrepôt de données modélisé par un schéma en étoile. Ces indexs sont caractérisés par une représentation binaire permettant l’utilisation des opérations logiques pour évaluer des conjonctions ou disjonctions de prédicats contenus dans les requêtes de jointure en étoile. Ces indexs sont généralement recommandés pour les attributs de faible cardinalité.
Nous avons étudié dans ce chapitre les $IJB$ et leur problème de sélection. Nous avons ensuite expliqué la stratégie d’exécution d’une requête en présence d’un $IJB$. Nous avons aussi détaillé le modèle de coût utilisé en présentant les différentes formules et les scénarios possibles. Et nous avons terminé par la présentation de quelques approches existantes dans la littérature. Dans le chapitre suivant nous allons proposer notre approche de sélection des $IJB$.
\chapter{Nouvelle approche de sélection d'IJB : utilisation des traverses minimales}
\section*{Introduction}
L'étude de l'état de l'art concernant la conception physique a conduit à une constatation essentielle : les approches existantes n'améliorent pas suffisamment les requêtes parce que la majorité de ces dernières ne sont pas concernées par la configuration finale choisie. Nous focalisons alors sur cette limite pour introduire une nouvelle approche qui se base sur une notion clé dans la théorie des hypergraphes appelée \textit{traverse minimale} $TM$. Cette dernière est un ensemble de sommets qui intersecte toutes les hyperarêtes d'un hypergraphe et elle est dite minimale si elle l'est au sens de l'inclusion \cite{JLB13}. À partir de cette structure, nous déterminons les attributs candidats, nous en créons les $IJB$ relatifs et nous améliorons ainsi la performance du système. Nous entamons ce chapitre par une présentation de la théorie des hypergraphes et ses notions de base avec une mise en évidence de l’utilité des traverses minimales. Ensuite, nous expliquons ce que représente la $TM$ grâce auquel nous pouvons exploiter les solutions potentielles. En outre, nous détaillons les différentes étapes que suit notre politique adoptée pour contourner le problème d’optimisation des requêtes. Enfin, nous concluons ce chapitre par un exemple illustratif de notre approche. 
\section{La théorie des Hypergraphes }
Les hypergraphes généralisent la théorie de graphe en définissant la notion des hyperarêtes, qui peuvent contenir un, deux ou plusieurs sommets, contrairement aux arêtes classiques qui ne peuvent joindre que deux sommets \cite{JLB13}. D'un point de vue théorique, les hypergraphes permettent de généraliser certains théorèmes de graphes, voire d'en factoriser plusieurs en un seul. Les hypergraphes sont parfois préférés aux graphes puisqu'ils modélisent mieux certains types de contraintes, d'un point de vue pratique. Dans cette section, nous allons présenter quelques notions de bases et des définitions essentielles sur les hypergraphes et les traverses minimales nécessaires à l'introduction de la problématique de l'extraction des traverses minimales, en se basant essentiellement sur les définitions proposées par Berge \cite{Ber89}.

\subsection{Les hypergraphes }

Un hypergraphe $H = (S, E)$ est constitué de deux ensembles $S$ et $E$, et est défini comme suit.

\begin{figure}[htbp]
\begin{center}
\includegraphics[scale=0.6]{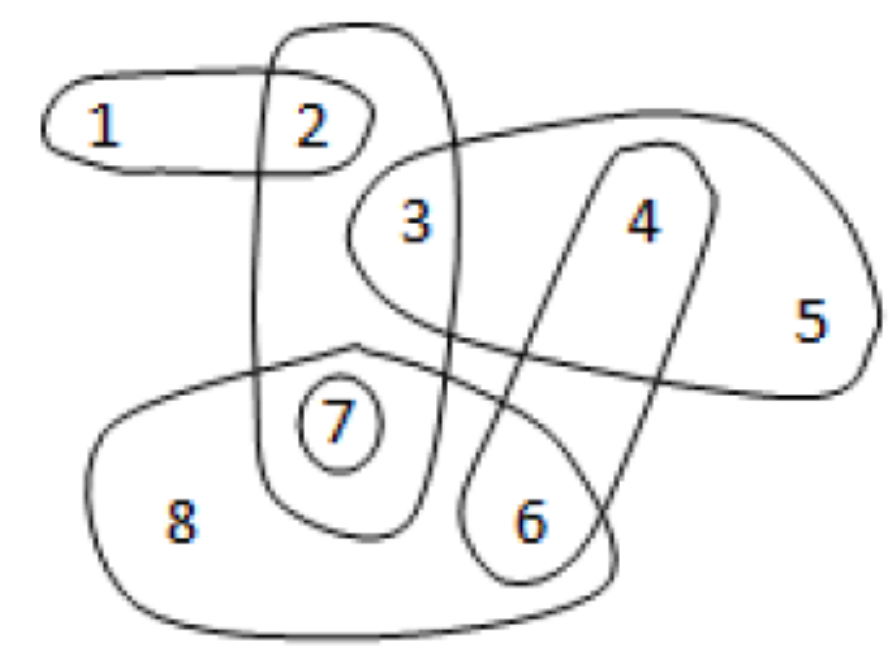}
\caption{Hypergraphe \label{g1}}
\end{center}
\end{figure}

\begin{mydef}\label{defhyp} hypergraphe \cite{Ber89} : Soit le couple $H = (S, E)$ avec $S$ = \{$\textrm{s}_\textrm{1}$, $\textrm{s}_\textrm{2}$, . . . , $\textrm{s}_\textrm{n}$\} un ensemble fini et $E$ = \{$\textrm{e}_\textrm{1}$, $\textrm{e}_\textrm{2}$, . . . , $\textrm{e}_\textrm{m}$\} une famille de parties de $E$. H constitue un hypergraphe sur $S$ si : 
\begin{enumerate}
\item e$_{i}$ $\neq$ $\emptyset$, i$\in \{1,....,m\};$
\item $\bigcup_{i=1,...,m} e_{i} = S $

\end{enumerate}
\end{mydef}
Les éléments $s_{i}$ de $S$ sont appelés \textit{sommets} de l'hypergraphe et les éléments $e_{i}$ de $E$ sont appelés \textit{hyperarêtes} de l'hypergraphe. Un hypergraphe est dit d'ordre $n$ si $|S|$ = $n$ où $n$ est le nombre de sommet,  et la taille d'un hypergraphe est égale au nombre d'occurrences des sommets dans ses hyperarêtes. L’exemple de la figure \ref{g1} illustre un hypergraphe $H = (S, E)$ d'ordre 8 et de taille 14 tel que $S$ = \{1, 2, 3, 4, 5, 6, 7, 8\} et $E$= \{\{1, 2\}, \{2, 3, 7\}, \{3, 4, 5\}, \{4, 6\}, \{6,7, 8\}, \{7\}\}.

\subsection{Traverse minimale}
\subsubsection{Définitions }

Une traverse d'un hypergraphe $H$ est un ensemble de sommets $s$ $\subseteq$  $S$ qui intersecte chaque hyperarête de $H$ au moins une fois et est défini comme suit.

\begin{mydef}\label{deftm} Traverse minimale \cite{Ber89} : Soit un hypergraphe $H = (S, E)$. L'ensemble des traverses de $H$, noté $\gamma$ $\mathcal{(H)}$, est égal à : $\gamma$ $\mathcal{(H)}$  = \{T $\subset$  S | T $\bigcap$  $\textrm{s}_\textrm{i}$  $\neq$   $\varnothing$, $\forall$ i = 1, . . . , | E |\}.
Une traverse $T$ de $\gamma$ $\mathcal{(H)}$  est dite minimale s'il n'existe pas une autre traverse $S$ de $\gamma$ $H$ incluse dans $T$ :  $S$ $\in$  $\gamma$ $\mathcal{(H)}$  s.t. $S$ $\subset$  $T$.
\end{mydef}
Nous noterons $\mathcal{(MH)}$, l'ensemble des traverses minimales définies sur H. Dans l'exemple illustratif de la figure \ref{g1}, l'ensemble $\mathcal{(MH)}$  des traverses minimales de l'hypergraphe est égal à : \{ \{1, 4, 7\}, \{2, 4, 7\}, \{1, 3, 6, 7\}, \{1, 3, 6, 9\}, \{1, 5, 6, 7\}, \{2, 3, 6, 7\}, \{2, 3, 6, 9\}, \{2, 5, 6, 7\}, \{2, 4, 6, 9\}, \{2, 4, 8, 9\}, \{2, 5, 6, 9\}, \{1, 3, 4, 8, 9\}\}. À partir d'un hypergraphe $H = (S, E)$, l'ensemble des traverses minimales $\mathcal{(MH)}$  permet la construction de l'hypergraphe transversal, qui est un hypergraphe où toutes ses hyperarêtes sont des traverses minimales.

\begin{mydef} \label{defnt}
Soit un hypergraphe $H = (S, E)$, le nombre minimum de sommets d'un ensemble transversal est appelé \textit{le nombre de transversalité} de l'hypergraphe $H$ et est désigné par : $\tau$ $\mathcal{(H)}$  = $min \{|T|$, tel que $T$ $\in$ $\mathcal{(MH)}$ \} \cite{Ber89}.
\end{mydef}
Ainsi, dans l'exemple illustratif de la figure \ref{g1}, le nombre de transversalité de l'hypergraphe $H$ est égal à 3 car la plus petite traverse minimale de $\mathcal{MH}$ est composée de 3 sommets. 
La détermination d'un nombre de transversalité apparaît dans de nombreux problèmes combinatoires \cite{Ber89}.
\subsubsection{Problème de l'extraction des traverses minimales}
L'extraction des traverses minimales d'un hypergraphe est un des problèmes les plus importants en théorie des hypergraphes. C'est un problème algorithmique central et particulièrement difficile et la question de sa complexité exacte reste toujours ouverte. Plusieurs travaux se sont attachés à proposer diverses méthodes pour le traiter \cite{Ber89}. Trouver une traverse minimale d'un hypergraphe est une tâche aisée mais calculer l'ensemble de toutes les traverses minimales pose plusieurs problèmes dans la mesure où le nombre de sous-ensembles de sommets à tester est très grand. Les travaux de recherche, pour faire sauter les différents verrous scientifiques que posait l'extraction des traverses minimales d'un hypergraphe, se sont attachés à réduire l'espace de recherche. Néanmoins, le coût du calcul reste substantiellement élevé et les algorithmes existants se sont heurtés à des temps d'exécution conséquents et à l'incapacité de traitement lorsque le nombre de transversalité de l'hypergraphe d'entrée est grand.
\subsubsection{Domaines d’applications des traverses minimales}
\begin{figure}[htbp]
\begin{center}
\includegraphics[scale=0.4]{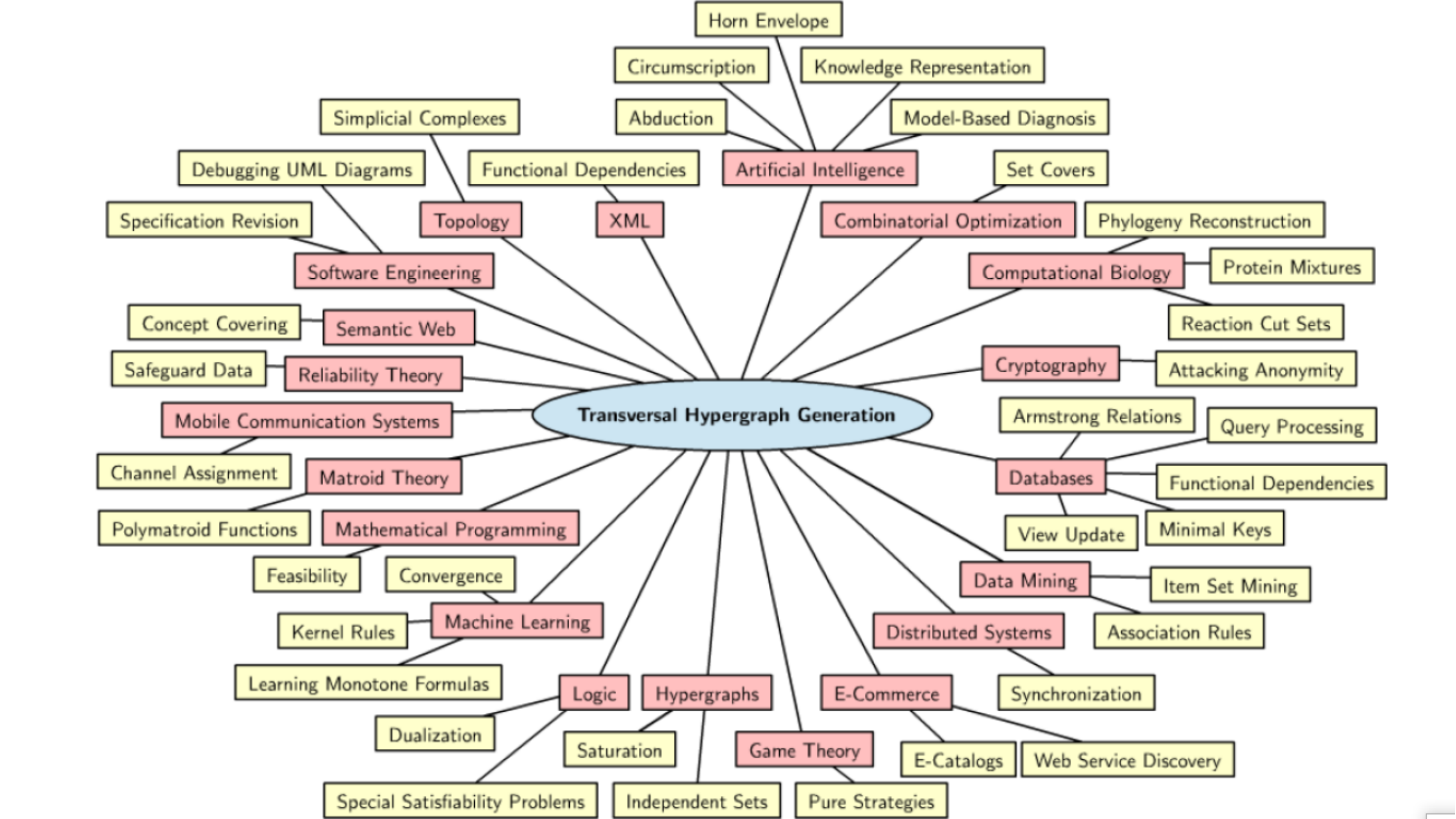}
\caption{Les domaines d’applications des traverses minimales \cite{Hag08} \label{g2}}
\end{center}
\end{figure}

L'intérêt pour l'extraction des traverses minimales s’augmente ces dernières années, en raison de la multitude et de la diversité des domaines d'application où le recours aux traverses minimales peut constituer une solution. Le large éventail des domaines d'application, comme le résume la figure \ref{g2}, donne ainsi une importance plus grande aux traverses minimales et motive l'intérêt qu'elles suscitent. Dans ce qui suit, nous en donnerons un aperçu et nous citerons les problèmes les plus connus, où les traverses minimales sont applicables.

\indent Parmi les domaines d’applications des traverses minimales, nous pouvons citer la logique, par le fait de rendre la détermination de la dualité $FNC$/$FND$ au problème de calcul des traverses minimales d'un hypergraphe. De son coté, les bases de données ont eu leur part de l’utilisation  des traverses minimales pour aboutir à des solutions de ses différents problèmes, comme par exemple, étant donné une relation et un ensemble de clés, décider de l'existence d'une autre clé est un problème, équivalent à celui de la recherche des traverses minimales. Les dépendances d'inclusion \cite{MP03}, qui sont une généralisation des clés étrangères dans un modèle relationnel, peuvent ainsi être déduites en adaptant les techniques de calcul des traverses minimales d'un hypergraphe. Celles-ci peuvent, par ailleurs, présenter des solutions aux problèmes de réécriture des requêtes, d'exécution des requêtes et d'actualisation des vues. Ces dernières, dont le rôle est très important dans la présentation des données à partir des bases, peuvent en effet être gérées en se basant sur les traverses minimales. L'inférence des dépendances fonctionnelles représente aussi un domaine d'application fort intéressant des traverses minimales. 
 Sans oublier le domaine de fouille de données et l’utilisation des traverses minimales pour la génération des règles associatives, des motifs fermés ou encore des motifs fréquents.\\
\indent Le survol des domaines d'application des traverses minimales démontre l'intérêt, de plus en plus, croissant pour les traverses minimales et, dans la littérature, plusieurs algorithmes dédiés à leur calcul ont été proposés.
\subsubsection{Techniques d’extraction de traverses minimales. }
Plusieurs auteurs se sont intéressés au problème de l'extraction des traverses minimales d'un hypergraphe. Dans cette partie, nous présentons un survol sur ces différentes approches, en mettant en exergue leurs points forts et leurs limites. 
Le nombre de traverses minimales d'un hypergraphe pouvant être exponentiel en la taille de l'hypergraphe, la question de la mise en place d'un algorithme résolvant le problème de l'extraction des traverses minimales d'un hypergraphe $H$ avec une complexité polynomiale en |$H$| reste néanmoins ouverte. \\
Dans ce qui suit nous allons présenter l’algorithme de \emph{Berge}, qui est le premier auteur à s’être intéressé à ce problème en proposant un algorithme pour le contourner, et par la suite nous détaillons l’algorithme \textsc{Mmcs} de \emph{Murakami et T.Uno} qui est le meilleur algorithme de la littérature en termes de performances et temps de traitement.  
\begin{description}
  \item[$\bullet$ L’algorithme de Berge : ] 
\emph{Berge} est le premier à s'être intéressé au problème du calcul des traverses minimales et à avoir proposé un algorithme pour le résoudre. Cet algorithme, dont le principe est simple, commence par calculer l'ensemble des traverses minimales de la première hyperarête, qui est équivalent à l'ensemble des sommets contenus dans cette dernière. Ensuite, il met à jour cet ensemble des traverses minimales en ajoutant les autres hyperarêtes, une à une, de manière incrémentale. Ainsi, l'algorithme de \emph{Berge} construit des hypergraphes partiels au fur et à mesure qu'il ajoute des hyperarêtes. Néanmoins, l'algorithme a toujours besoin de stocker les traverses minimales intermédiaires avant de passer à l'étape suivante consistant à ajouter une nouvelle hyperarête. \\
L'un des inconvénients majeurs de l'algorithme de \emph{Berge} est la consommation excessive en mémoire d’où plusieurs chercheurs ont cherché à l’améliorer. Parmi les améliorations les plus connues proposées récemment figurent celles introduites par\emph{ Dong et al.} \cite{DL05}, \emph{Bailey et al.} \cite{BMR03} et \emph{Kavvadias et Stavropoulos }\cite{KS05}.

  \item[$\bullet$ Algorithme MMCS de Murakami et T.Uno (2013): ] 

\emph{Murakami et Uno} proposent les algorithmes de type \textsc{Shd}, \textsc{Mmcs} et \textsc{rs}, qui visent à réduire l'espace de recherche \cite{MU13}. En ce sens, ces algorithmes sont destinés à traiter des hypergraphes de grande taille constitués par un très grand nombre d'hyperarêtes. Les algorithmes de type \textsc{Shd} adoptent une stratégie de parcours en profondeur de l'espace de recherche qui, dans le cas de \textsc{rs}, est équivalente à celle de l'algorithme de \emph{Kavvadias et Stavropoulos}. La principale différence entre ce dernier et \textsc{rs} repose sur l'élimination des itérations redondantes où aucun sommet n'est ajouté à un ensemble de sommets générés auparavant. De plus, \emph{Murakami et Uno} introduisent deux nouveaux concepts, i.e, le test de la transversalité (\textsc{uncov}) et les hyperarêtes critiques (\textsc{crit}), et ce afin d'optimiser les tests sur la minimalité effectués sur l'ensemble des traverses générées.\\
\indent Les auteurs proposent aussi divers lemmes, dans \cite{MU13}, pour optimiser le calcul de la fonction \textsc{crit}, qui est la clé de leur approche. Les algorithmes de type \textsc{Shd} se basent donc sur la même approche. L'algorithme \ref{unoalgo}  décrit le pseudo-code de l'algorithme \textsc{Mmcs}. Cet algorithme est récursif et fournit en sortie des traverses minimales en série. Pour tester un ensemble de sommets $X$, les  algorithmes cherchent, de façon itérative, les sous-ensembles de $X$ et effectuent un appel récursif pour chacun tout en mettant à jour les ensembles \textsc{crit} et \textsc{uncov}.  En opérant de cette manière, \emph{Murakami et Uno} permettent à leur algorithme de balayer l'espace de recherche en profondeur en cherchant seulement les sous-ensembles du candidat courant. La méthode et les étapes pour la recherche des sous-ensembles d'un candidat sont détaillées dans \cite{MU13}. 

\begin{algorithm}[!h]
{
\LinesNumbered
\textit{\textbf{Var. Globale}} : \emph{uncov} (initialisé à $\xi$), $Cand$ (initialisé à $\mathcal{X}$), \emph{crit}[$x$] initialisé à $\emptyset$ pour chaque $x$\\
\Entree{$H$ = $(\mathcal{X}$, $\xi)$: Hypergraphe, $X$ : ensemble de sommets}
\Sortie{$T$ tel que $T$ $\in$ $\mathcal{M}_H$}

\Deb
{
\Si{\emph{uncov} = $\emptyset$}{\Retour{$X$}}
Choisir une hyperarête $e$ à partir de \emph{uncov};\\
$C$ = $Cand$ $\cap$ $e$;\\
$Cand$ = $Cand$ $\backslash$ $C$;\\
\PourCh{$x$ $\in$ $C$}{\textsc{Update\_crit\_uncov}($x$, \emph{crit, uncov});\\
\Si{\emph{crit}(f, $X$ $\cup$ $x$) $\neq$ $\emptyset$ \textbf{pour chaque} f $\in$ $X$}{\textsc{Mmcs}($X$ $\cup$ $x$);\\ $Cand$ = $Cand$ $\cup$ $x$;}
Restaurer les valeurs de crit et \emph{uncov} d'avant la ligne $8$;
}

}
}

  \caption{L'algorithme \textsc{Mmcs} \cite{MU13}}
  \label{unoalgo}
\end{algorithm}
L'étude expérimentale effectuée par les auteurs a montré que les algorithmes de type \textsc{Shd} (et notamment \textsc{mmcs}) présentaient des performances très intéressantes et s'imposaient comme les algorithmes les plus performants dans la littérature.
\end{description}

\section{Approche de sélection d’IJB basée sur les traverses minimales}
L’intérêt pour la sélection des $IJB$ est en nette croissance grâce notamment aux solutions qu'elles offrent au niveau de la conception physique des entrepôts de données et au niveau de l’optimisation des requêtes et du système en général.  Dès le début nous avons décidé de choisir les $IJB$ comme structure d’optimisation qui est une technique très bien répandue dans les entrepôts de données.

\subsection{Motivations}
Les motivations sur lesquelles nous nous sommes basés pour considérer l’utilisation des traverses minimales sont les suivants: \textit{(i)} leur rapidité et efficacité d’extraire des attributs indexables sans la nécessité du paramétrage de la part de l’administrateur au préalable; \textit{(ii)} s’assurer que la majorité des requêtes sera améliorée en termes de temps d'exécution et \textit{(iii)} diminuer le risque de négliger des attributs candidats intéressants à l’optimisation des requêtes décisionnelles.

\subsection{Démarche de sélection automatique d'index}
Dans cette section, nous allons implémenter notre approche proposée dont la principale caractéristique est l’utilisation des traverses minimales sur un contexte d’extraction, qui consiste en une matrice Requêtes-Attributs appelé \textit{hypergraphe}, telle que chaque hyperarête  représente une requête et chaque sommet représente un attribut qui a fait l’objet d’un prédicat de sélection ou de projection. \\

\begin{figure}[htbp]
\begin{center}
\includegraphics[scale=0.6]{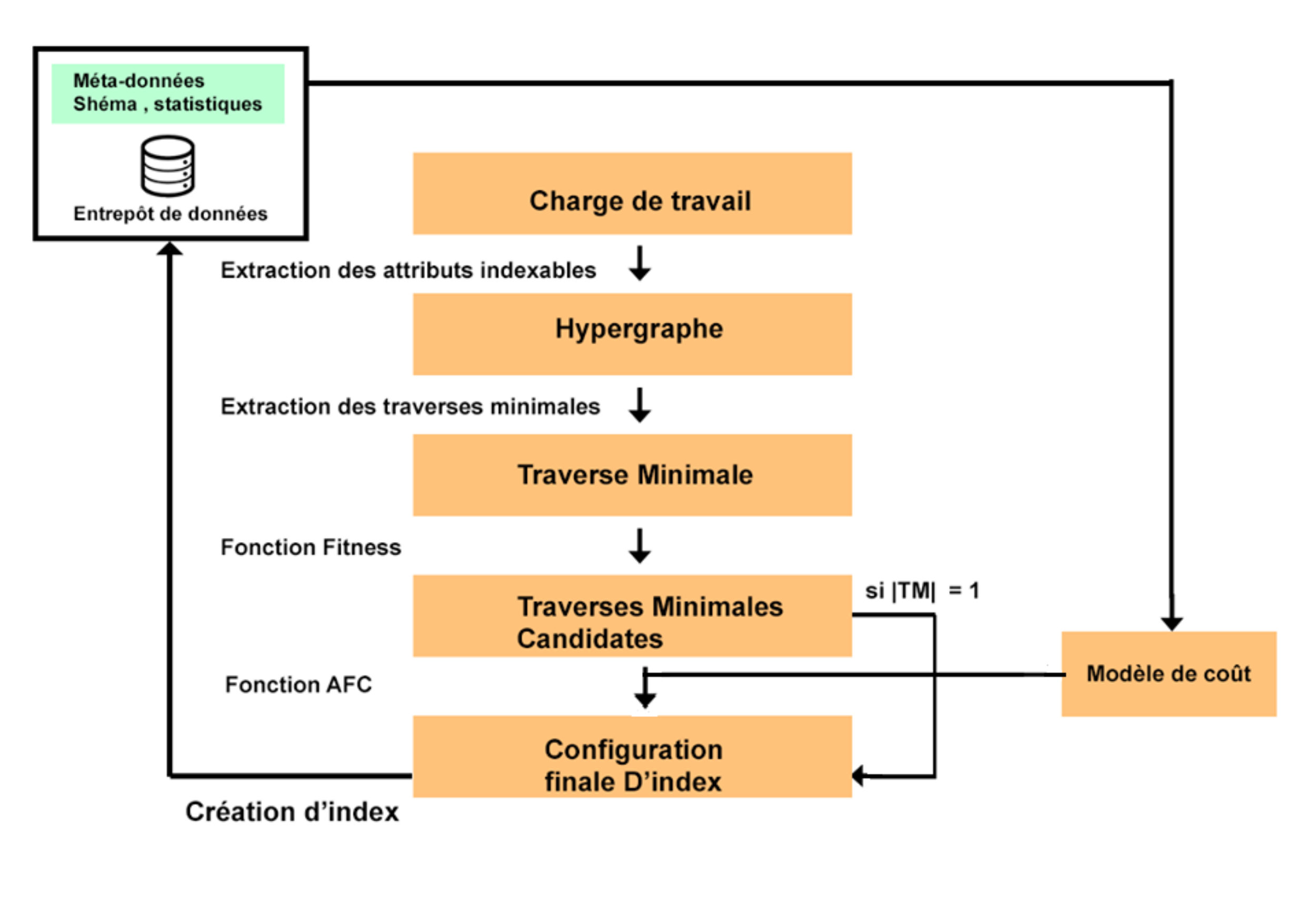}
\caption{Approche de séléction d'IJB \label{g3}}
\end{center}
\end{figure}

\indent Après l'analyse de la charge et la construction de l'hypergraphe à travers les attributs  susceptibles d’être des indexs de la charge de travail, le processus mis en œuvre par notre approche commence par le calcul du nombre de transversalité, qui est égal au cardinal de la plus petite traverse minimale (voir définition \ref{defnt} page \pageref{defnt}). Par la suite, il enchaîne par l’extraction de toutes les traverses minimales qui ont une taille égale au nombre de transversalité. Une fois il a toutes les traverses minimales, il va calculer pour chacune d’elles la fonction \textsc{Fitness}, détaillée dans le chapitre précédent (page \pageref{fitness}), et ne garder que les $TMs$ qui ont la valeur maximale, si à ce niveau il retourne plus qu’une seule traverse minimale dans ce cas il va calculer la somme des cardinalité de chaque attribut des $TMs$ et ne retenir que la traverse minimale qui minimise la somme de cardinalité car la principale cause de l’explosion de la taille des $IJB$ est la cardinalité des attributs indexés. La figure \ref{g3} montre l'enchaînement des étapes et leurs relations. \\
Dans ce qui suit, nous décrivons avec plus de détailles toutes les étapes.

\begin{description}

  \item[$\bullet$ Analyse de la charge : ]les requêtes SQL présentes dans la charge de travail sont traitées par un analyseur syntaxique automatique afin d'en extraire tous les attributs susceptibles d'être des indexs, appelés \textit{les attributs indexables}. Ces attributs sont ceux présents dans les clauses \textit{Where} des requêtes. Ces attributs servent à la recherche dans les requêtes d'interrogation.

    \item[$\bullet$ Construction de l’hypergraphe : ]à partir des attributs extraits dans l'étape précédente, nous construisons une matrice (requêtes-attributs) dite \textit{hypergraphe} qui a pour hyperaretes les requêtes de la charge et pour sommets les attributs à indexer.

  \item[$\bullet$ Extraction des traverses minimales : ]une fois l’hypergraphe est prêt, nous passons au processus de l’extraction des traverses minimales, pour cette étape, nous profitons des récents travaux de la littérature dédiés au calcul des traverses minimales. Pour ce faire, nous utilisons le meilleur algorithme existant en termes de performances, i.e., \textsc{Mmcs} de \emph{Murakami et Uno} \cite{MU13}. Notre choix est uniquement dicté par les temps de traitement intéressant de ce dernier. Nous avons automatiquement configuré l’algorithme de \textsc{Mmcs} pour qu’il ne génère que les traverses minimales de plus petite taille à travers la fonction \textsc{GetMinTransversality} introduite par \emph{Jelassi et al. }\cite{JLB13}, qui retourne  le nombre de transversalité d’un hypergraphe que nous allons la détailler par la suite.

  \item[$\bullet$ Construction de l'ensemble d'indexs candidats : ]après la génération de l’ensemble des traverses minimales, une phase de génération d’index candidats est nécessaire. Ainsi, nous procédons comme suit :

\begin{enumerate}
    \item \textbf{Application de la fonction \textsc{Fitness}} \\
    Au lieu de se baser sur la fréquence d’apparition présentée par le support comme un critère de détermination des attributs fréquents, nous nous inspirons de l’approche de \emph{Bellatreche  et al.} \cite{BMND08} en utilisant la fonction \textsc{Fitness}. Cette dernière permet de pénaliser chaque $TM$, qui contient des attributs appartenant à des petites tables de dimensions, en prenant compte des cardinalités des tables de dimensions et la taille de la page système. Notre fonction \textsc{Fitness} est calculée comme suit :  \[ \text{Fitness(tm) = }\	\sum_{i=1}^{n}sup_{i}\times\alpha_{i} \]
    
Où $n$ désigne le nombre d’attributs non clés de la traverse minimale $tm$, $i$ est l'attribut dans la traverse minimale $tm$.  $sup_{i}$ représente le support de $i$,  $\alpha_{i}$ est un paramètre de pénalisation déterminé par l’équation :
$\alpha_{j} = \frac{|D_{j}|}{|F|}$  ; sachant que les notations $|D_{j}|$ et $|F|$ expriment respectivement la taille  de la table de dimension $D$ à laquelle appartient l'attribut $j$ et la table de faits $F$ en nombre de pages. \\    
    
    \end{enumerate}
Si après avoir exécuté la fonction \textsc{Fitness}, plusieurs $TMs$ sont retenues dans ce cas nous allons faire appel à une autre métrique pour choisir la meilleure configuration possible.
\begin{enumerate}   
\setcounter{enumi}{1}
    
    \item \textbf{Élimination des attributs de forte cardinalité} \\
La principale cause de l’explosion de la taille des $IJB$ est la cardinalité des attributs indexés. Dans cette stratégie, les attributs sont triés par ordre décroissant par rapport à leur cardinalité. Ensuite, nous allons calculer, pour chaque $TM$ retenue la somme de cardinalité de ses attributs et ne garder que la $TM$ qui a le plus petit total en termes de somme de cardinalité. Cette stratégie est inspirée de la stratégie $AFC$ introduite par \emph{K.Boukhalfa} \cite{BBB10}.
\end{enumerate}

 \item[$\bullet$ Construction de la configuration finale : ]une fois toutes les étapes précédentes soient terminées, la seule traverse minimale retenue va servir à construire la configuration finale d’indexs après avoir éliminé les sommets qui représentent des attributs non-indexables, i.e., des  attributs clés ou bien des attributs appartenant à la table de faits.
\end{description}
\indent Les travaux antérieurs ont fait recours au modèle de coût pour calculer le coût en termes d’entrées/sorties de chaque index et ne garder que ceux qui maximise la fonction objective pour choisir la configuration finale d’index. Dans notre travail, nous gardons les attributs indexables de la traverse minimale résultante et nous les considérons comme étant la configuration finale sans tenir compte d'aucune contrainte. Le modèle de coût reste un outil important si nous voulons ajouter ou retirer des indexs de la configuration finale. 
\subsection{L’algorithme TM-IJB}
Par conséquent, pour choisir les meilleurs attributs pertinents en indexation, nous suivons un processus itératif, qui consiste à commencer par la détermination du nombre de transversalité (ligne 2). Ensuite la génération de toutes les traverses minimales de plus petite taille à travers l’algorithme \textsc{MMCS} de \emph{Murakami et Uno} que nous avons détaillé dans la section précédente (ligne 3). Ensuite, le processus enchaîne par le calcul de fonction \textsc{Fitness} de chaque $TM$ et ne garder que celles qui ont la plus grande valeur (voir ligne 4).  Le processus vérifie le nombre de $TM$ retenues qui ont la valeur maximale de \textsc{Fitness}. Si elle est supérieure à 1, donc il passe a calculer la somme de cardinalités pour chaque $TM$ et retourne celle qui la minimise. Autrement l'algorithme retourne la $TM$ actuelle (ligne 8). 
Le pseudo code de notre approche est donné dans l'algorithme \ref{algtm}.
\begin{algorithm}[!h]
{
\LinesNumbered
\Entree{Hypergraph (S,E) ;}
 \Sortie{Traverse Minimale F : configuration finale d'IJB}
\Deb
{
X = GetMinTransversality (H) ;\\
TM = MMCS(S,E,X) ; \\
F = max(Fitness(TM)) ;\\
\Si{(|F| $>$ 1)}{AFC(F);\\ \Retour{min(AFC(F));}}

 }
\Retour{(F)}
}

  \caption{\textsc{TM-IJB}}
  \label{algtm}
\end{algorithm}

\subsubsection{GetMinTransversality }
La fonction \textsc{GetMinTransversality} est introduite par \emph{Jelassi et al.}\cite{JLB13}. Cette fonction est une approche parmi de nombreuses contributions proposées par notre laboratoire \textsc{Lipah}. Parmi les approches qui ont utilisé des techniques de fouille de données pour résoudre des différents problèmes scientifique, il est important de citer (\cite{lpah1}, \cite{lpah2}, \cite{lpah3}, \cite{lpah4}, \cite{lpah5}, \cite{lpah6}, \cite{lpah8}, \cite{lpah9}, \cite{lpah10}, \cite{lpah11}). Ces travaux, qui se sont distingués, ont utilisé une panoplie de méthodes tels que la classification, les règles d'associations et les motifs fréquents fermés pour palier les différents problèmes des différents domaines de recherche y compris l'utilisation des traverses minimales pour la sélection des IJBs que nous traitons dans ce travail.  \textsc{GetMinTransversality} permet d'obtenir le nombre de transversalité de l'hypergraphe. Pour ce faire, la fonction parcourt les sommets, un par un (ligne 3). Pour chaque élément $s$ de $S$, \textsc{GetMinTransversality} supprime les hyperarêtes de $E$ qui contiennent $s$ (ligne 5). Les hyperarêtes restantes sont stockées dans $E'$. La fonction invoque ensuite la fonction \textsc{hyp\_empty()}. \textsc{hyp\_empty} est une fonction récursive qui stocke les sommets ayant le plus grand support dans $E'$ et les traitera, un par un, en supprimant à chaque fois les hyperarêtes auxquelles appartient le sommet traité.  Cette fonction récursive s'arrête lorsque $E'$ sera vide. La valeur retournée correspond au nombre d'appels à la fonction \textsc{hyp\_empty} nécessaires pour que $E'$ soit égal à l'ensemble vide. 
\begin{algorithm}[!h]
{
\LinesNumbered
\Entree{Matrice d'incidence $IM_{H}$ associée à $H$ = ($\mathcal{X}$, $\xi$)}
 \Sortie{$T$: Une plus petite traverse minimale de $H$ ; $k$ : Nombre de transversalité de $H$}
\Deb
{
$k$ = $\mid \xi \mid$;\\
$T$ = $\emptyset$;\\
\PourCh{$x$ $\in$ $\mathcal{X}$}{$i$ = 1;\\
$T_{tmp}$ = $\emptyset$;\\
$T_{tmp}$[$i$] = $x$;\\
$\xi'$ = $\xi$ $\backslash$ $\{e \in \xi \mid x \in e\}$;\\
($n$, $T_{tmp}$) = \textsc{hyp\_empty}($\xi'$, $\mid \xi \mid$, i, $T_{tmp}$);\\
\Si{$n$ $<$ $k$}{$k$ = $n$;\\ $T$ = $T_{tmp}$;}
}
\Retour{($k$, $T$)}
}
}
  \caption{\textsc{GetMinTransversality}}
  \label{proc}
\end{algorithm}
Pour chaque sommet $s$ traité,  l'ensemble $E'$ est réactualisé à toutes les hyperarêtes de l'hypergraphe auxquelles nous supprimons celles qui contiennent $s$. Au final, \textsc{GetMinTransversality} retourne le nombre minimum d'itérations permettant de "vider" la matrice d'incidence. La valeur de $k$, retournée par la fonction, correspond ainsi au nombre de transversalité de l'hypergraphe d'entrée $H$.
\section{Exemple illustratif }
Nous considérons l'exemple suivant avec une partie d'un schéma en étoile contenant deux tables de dimensions CHANNELS (désignés par $Ch$) et CUSTOMERS (désignés par $C$) et une table de faits SALES (désignées par $S$). Les cardinalités de ces tables (nombre d'instances) sont: ||SALES|| = 16260336; ||CHANNELS|| = 5 et ||CUSTOMERS|| = 50000. Supposons que les cinq requêtes sont exécutées le plus souvent sur le cube de données correspondant (voir tableau \ref{req}). Dans les requêtes du tableau, deux opérations de jointure principales sont utilisées: une entre SALES et CUSTOMERS ($J1$), et une autre entre SALES et CHANNELS ($J2$). A la base, le coût de $J1$ est supérieur à $J2$ puisque la taille des CUSTOMERS (50 000 instances) est supérieure à la taille des CHANNELS (5 instances).

\begin{table}[]
\centering
\caption{Requêtes de test}
\label{req}
\begin{tabular}{|l|}
\hline
\begin{tabular}[c]{@{}l@{}} (1) select S.channel\_id, sum(S.quantity\_sold)
 from S, C\\ where S.channel\_id=C.channel\_id and C.channel\_desc='Internet'  \\ group by
  S.channel\_id \end{tabular}                      \\ \hline
\begin{tabular}[c]{@{}l@{}}(2) select S.channel\_id, sum(S.quantity\_sold), sum(S.amount\_sold) from S, C\\ where S.channel\_id=C.channel\_id and C.channel\_desc ='Catalog'  \\ group by S.channel\_id\end{tabular} \\ \hline
\begin{tabular}[c]{@{}l@{}}(3) select S.channel\_id, sum(S.quantity\_sold),sum(S.amount\_sold) from S, C\\ where S.channel\_id=C.channel\_id and C.channel\_desc ='Partners' \\ group by S.channel\_id\end{tabular} \\ \hline
\begin{tabular}[c]{@{}l@{}}(4) select S.cust\_id, avg(quantity\_sold) from S, C \\ where S.cust\_id=C.cust\_id and C.cust\_gender='M' group by S.cust\_id \end{tabular}                \\ \hline
\begin{tabular}[c]{@{}l@{}}(5) select S.cust\_id, avg(quantity\_sold) from S, C \\ where S.cust\_id=C.cust\_id and C.cust\_gender='F' group by S.cust\_id  \end{tabular}\\ \hline
\end{tabular}
\end{table}

\begin{figure}[htbp]
\begin{center}
\includegraphics[scale=0.7]{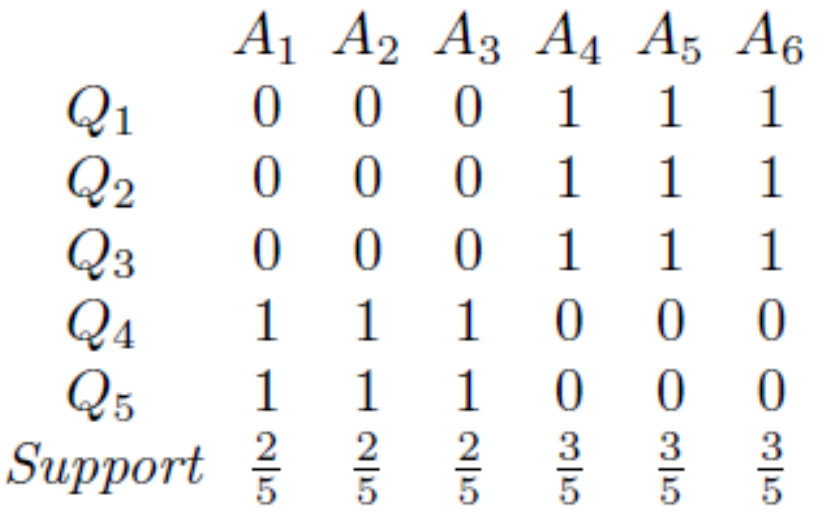}
\caption{Matrice de données \label{g4}}
\end{center}
\end{figure}

\indent La taille de l'instance de CUSTOMERS, CHANNELS et SALES est: 24, 24 et 36, respectivement, et la taille de la page $PS$ = 65536. Nous considérons les mêmes cinq requêtes que celles définies dans le tableau \ref{req}. Pour faciliter la construction de la matrice de contexte, nous renommons les attributs indexables comme suit: SALES: \textit{cust\_id} = $A1$, CUSTOMERS: \textit{cust\_id} = $A2$; CUSTOMERS: \textit{cust\_gender} = $A3$, CHANNELS: \textit{channel\_id} = $A4$, SALES: \textit{channel\_id} = $A5$; CHANNELS: \textit{channel\_desc} = $A6$. La matrice est donnée par la figure \ref{g4}.

\indent Une fois la construction de l’hypergraphe est terminée, le processus mis en place par notre  approche commence par le calcul du nombre de transversalité, qui est égal au cardinal de la plus petite traverse minimale. Dans cet exemple, le nombre de transversalité est égal à 2. \\
Par la suite il enchaîne par l’extraction de toutes les traverses minimales qui ont une taille égale au nombre de transversalité, qui sont dans cet exemple 9 traverses minimales {(1,4), (1,5), (1,6), (2,4), (2,5), (2,6), (3,4), (3,5), (3,6)}. Dès que l’algorithme termine la génération de toutes les traverses de taille 2, il calcule pour chacune d’elles la fonction \textsc{Fitness}. Sans oublier de mentionner que seuls les attributs $A3$ et $A6$ qui sont respectivement \textit{cust\_gender} et \textit{channel\_desc} sont des attributs indexables. Vu que les attributs $A1$, $A2$, $A4$ et $A5$ sont des attributs non-indexables du fait que $A$1 et $A5$ appartiennent à la tables de fait SALES et $A2$ et $A4$ sont des attributs clés des tables CUSTOMERS et CHANNELS. Nous rappelons la formule du calcul de notre fonction fonction \textsc{Fitness} : \label{rapfit} 

 \[ \text{Fitness(tm) = }\	 \sum_{i=1}^{n}sup_{i}\times\alpha_{i} \]

À partir de notre exemple trivial, seulement 4 parmi les 9 traverses minimales seront prises en considération car elles contiennent des attributs indexables et qui sont {$TM1$(1,6), $TM2$(3,4), $TM3$(3,5), $TM4$(3,6)}. Nous appliquons la fonction \textsc{Fitness} pour la traverse minimale $TM2$(3,4) comme un titre d’exemple. \\

\textsc{Fitness}(TM2) =       	\begin{large}sup(A$_{3}$) $\times \alpha_{3}$ \\             \end{large} 

=  \begin{Large}$\frac{2}{5}   \times \frac{\frac{50000 \times 24}{65536}}{\lceil\frac{\Vert1626033\Vert \times 36 }{65536}\rceil}    $ \\\end{Large}

= \begin{Large}$\frac{2}{5} \times \frac{19}{894}$ \\\end{Large}

= 0.0085

\indent Et de même pour toutes les autres traverses minimales, et à la fin nous obtenons les résultats suivants,  $TM1$(1,6) = 0.0001, $TM2$(3,4) = 0.0085, $TM3$(3,5) = 0.0085, $TM4$(3,6) = 0.0086 et selon ces derniers résultat nous allons retenir seulement $TM4$(3,6) par ce qu’elle a la plus grande valeur. Vu qu'après l’application de la fonction \textsc{Fitness} une seule traverse minimale a été retenue, alors  nous allons s'abstenir de faire appel à la fonction de calcule de cardinalité. \\
 \indent Si la traverse minimale $TM4$ n’existait pas, l’algorithme va retourner $TM2$ et $TM3$ et dans ce cas il passe automatiquement au calcul de cardinalité de chaque traverse. Dans ce scénario, $TM2$ aura un total de 16310336 par contre $TM3$ aura un total de 50005 et selon ces valeurs l’algorithme va retourner $TM3$ comme une configuration finale.\\
\indent Dans notre exemple, 2 $IFF$ peuvent être générés, $IFF_{1}$ = \{$A_{1}$, $A_{2}$, $A_{3}$\} avec un support égal à 0,4 et $IFF_{2}$ = \{$A_{3}$, $A_{5}$, $A_{6}$\} avec un support de 0,6. Avec $minsup$ = 3 (en support absolu), la solution retournée par l'approche de \emph{Aouiche et al.} est l'attribut $A_{6}$ de l'$IFF_{2}$. Ceci, est dû au fait qu'il existe trois occurrences du même prédicat de sélection défini sur cet attribut dans les cinq requêtes. D’où le résultat est un $IJB$ défini sur CHANNELS et SALES à l'aide de l'attribut \textit{channel\_desc}.
 Cependant, aucun index de jointure binaire n'est proposé entre les tables SALES et CUSTOMERS puisque l'attribut \textit{cust\_gender} n'est pas aussi fréquent que $minsup$. En conséquence, seule la jointure $J2$ sera optimisée, mais pas l'ensemble de requêtes globales.  Quant à la solution retournée par l'approche \textsc{DynaClose} de \emph{Bellatreche  et al.}, nous pouvant remarquer que $IFF_{1}$ contient seulement un attribut non-clé qui est $A_{3}$, et
donc \textsc{Fitness}($IFF_{1}$) ne prendra en compte que l'attribut $A_{3}$ qu'il s'agit d'un $IJB$ défini sur CHANNELS et CUSTOMERS à l'aide de l'attribut \textit{cust\_gender}. Cet $IJB$ permet d'optimiser la jointure $J1$. En conséquence, seule la jointure $J1$ sera optimisée, mais pas l'ensemble de requêtes globales. 
 
La traverse minimale $TM2$(3,6) nous permet d'optimiser la jointure $J1$ et la jointure $J2$, d’où elle permet d’améliorer la totalité des requêtes de la charge de travail. 

\section*{Conclusion}
Dans ce chapitre nous avons présenté la théorie des hypergraphes, les traverses minimales  et leur utilité dans les différents domaines. Ensuite, nous avons décrit l’approche que nous avons proposée, avec la présentation des différentes étapes que nous avons considérées pour la génération de la configuration finale de l’$IJB$ et nous avons terminé par un exemple illustratif. \\
Dans le chapitre suivant, nous allons valider notre approche par une étude expérimentale et la comparer avec d’autres approches.
\chapter{Étude expérimentale et évaluations}
\section*{Introduction}
Dans l'objectif de valider notre  approche, présentée dans le chapitre précédent, et prouver sa performance, nous expérimentons le temps d’exécution d’un jeu de données sur un entrepôt de données et nous analysons les résultats obtenus sur un échantillon de données. En ce sens, le présent chapitre présente l’environnement de travail utilisé et confronte les résultats obtenus par l’approche basée sur les traverses minimales avec celles figurantes dans l'état de l'art qui se basent sur des techniques de data mining (e.g : \textsc{Close} et \textsc{Dynaclose}), le but étant de mettre en évidence que notre  approche est meilleure. L'étude expérimentale se base ainsi sur le temps d’exécution des requêtes en présence de l’$IJB$ et le taux de réduction par rapport aux requêtes exécutées sans index ainsi le modèle de coût, expliqué également dans le deuxième chapitre.
\section{Environnement de travail }
Notre objectif principal consiste à étudier l’impact de l’utilisation des traverses minimales. Pour se faire,  l'implémentation des approches existantes ainsi que l’approche que nous proposons qui traite les requêtes en tant que un hypergraphe nécessitent l’usage de :
\begin{itemize}
\item 8 GB de RAM et un processeur Intel 1.8 Ghz base up to 3 Ghz avec 3 Mo de mémoire cache.
\item Un banc de test TPC-H disposant de 1Go de données.
\item Charge principale contenant 22 requêtes de jointure en étoile.
\item Un banc de test SSB1  disposant de 1Go de données.
\item Charge principale contenant 30  requêtes de jointure en étoile.
\item Le SGBD \textsc{Oracle} Version 12c, servant à stocker des données volumineuses.
\item Java version 7 sous l’IDE Eclipse Kepler nécessaire pour implémenter les algorithmes.

\end{itemize}
\section{Jeux de données}
\begin{enumerate}[label=\alph*.]
\item \emph{\textbf{Le banc de test TPC-H}}  \\
\indent L’entrepôt de données sur lequel nous effectuons nos tâches est constitué d’une table de faits \emph{LineItem} et de 7 tables de dimension \emph{Customers, Nation, Orders, Part , PartSupp (Part Supplier), Supplier, Region} comme le montre le schéma de la figure \ref{f9}. Les tailles de ces tables sont détaillées dans le tableau \ref{t1}. Nous avons considéré 46 attributs candidats à l’indexation vu que l’entrepôt contient 15 attributs clés. Notons que nous avons identifié les attributs par un numéro de 1 jusqu’à 61. La charge de travail comporte un jeu de requêtes fourni par le banc de test contenant 22 requêtes (voir Annexe A). Plusieurs types de requêtes ont été considérées : requêtes de type count(*) avec et sans agrégation, requêtes utilisant les fonctions d’agrégation comme \textit{Sum}, \textit{Min}, \textit{Max}, requêtes ayant des attributs de dimension dans la clause SELECT, etc.

\begin{figure}[htbp]
\begin{center}
\includegraphics[scale=0.4]{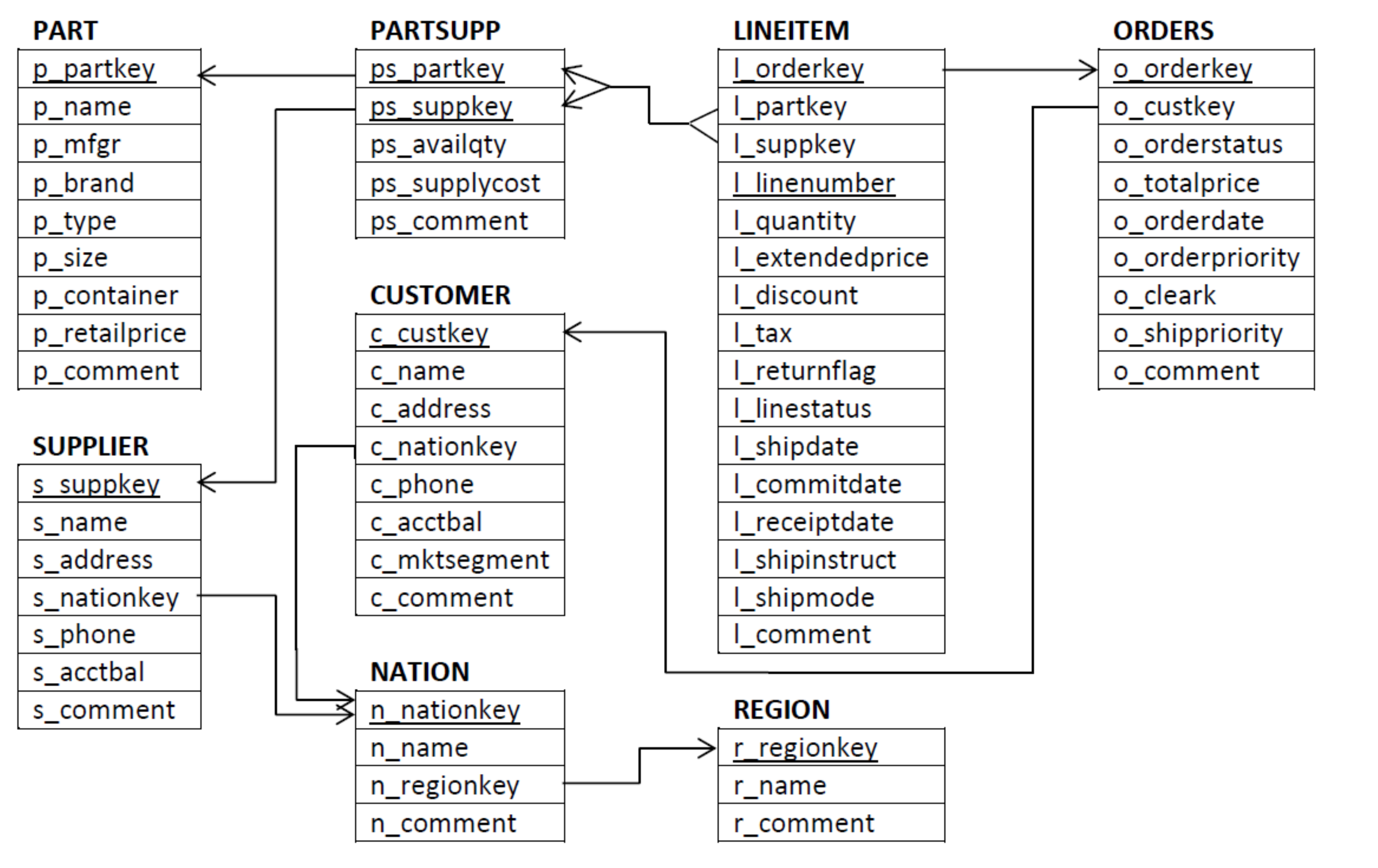}
\caption{Schéma du banc de test TPC-H\label{f9}}
\end{center}
\end{figure}

\begin{table}[]
\centering
\caption{Taille des tables de TPC-H \label{t1}}

\begin{tabular}{|l|c|c|c|c|}
\hline
Table                                                                 & \multicolumn{1}{l|}{\begin{tabular}[c]{@{}l@{}}Nombre \\ d'instances\end{tabular}} & \multicolumn{1}{l|}{\begin{tabular}[c]{@{}l@{}}Taille de la table \\ en nombre de pages\end{tabular}} & \multicolumn{1}{l|}{Espace} & \multicolumn{1}{l|}{\%} \\ \hline
Customers                                                             & 150,000                                                                            & 3397                                                                                                  & 31.3 MB                     & 3\%                     \\ \hline
LineItems                                                             & 6,001,215                                                                          & 105866                                                                                                & 726,7 MB                    & 63\%                    \\ \hline
Nation                                                                & 25                                                                                 & 1                                                                                                     & 5 KB                        & 0\%                     \\ \hline
Orders                                                                & 1,500,000                                                                          & 23766                                                                                                 & 190,3 MB                    & 17\%                    \\ \hline
Part                                                                  & 200,000                                                                            & 3746                                                                                                  & 30,5 MB                     & 3\%                     \\ \hline
\begin{tabular}[c]{@{}l@{}}PartSupp\end{tabular} & 800,000                                                                            & 16294                                                                                                 & 164,0 MB                    & 14\%                    \\ \hline
Supplier                                                              & 10,000                                                                             & 207                                                                                                   & 1,9 MB                      & 0\%                     \\ \hline
Region                                                                & 5                                                                                  & 1                                                                                                     & 1 KB                        & 0\%                     \\ \hline
\end{tabular}
\end{table}
Notons que la taille d’une page est 8096 octet, c'est-à-dire 8 Ko. 

\item \emph{\textbf{Le banc de test SSB1}}  \\
\indent Le deuxième entrepôt de données sur lequel nous effectuons nos tâches est constitué d’une table de faits \emph{Lineorder} et de 4 tables de dimension \emph{Date, Supplier, Customer, Part}. Les tailles de ces tables sont détaillées dans le tableau \ref{t2}. Nous avons considéré 48 attributs candidats à l’indexation vu que l’entrepôt contient 9 attributs clés. Notons que nous avons identifié les attributs par un numéro de 1 jusqu’à 57. La charge de travail comporte un jeu de requêtes contenant 30 requêtes (voir Annexe B). Plusieurs types de requêtes ont été considérées : requêtes de type count(*) avec et sans agrégation, requêtes utilisant les fonctions d’agrégation comme \textit{Sum, Min, Max}, requêtes ayant des attributs de dimension dans la clause SELECT, etc. Le schéma de l’entrepôt est illustré par la figure \ref{f10}.

\begin{figure}[htbp]
\begin{center}
\includegraphics[scale=0.55]{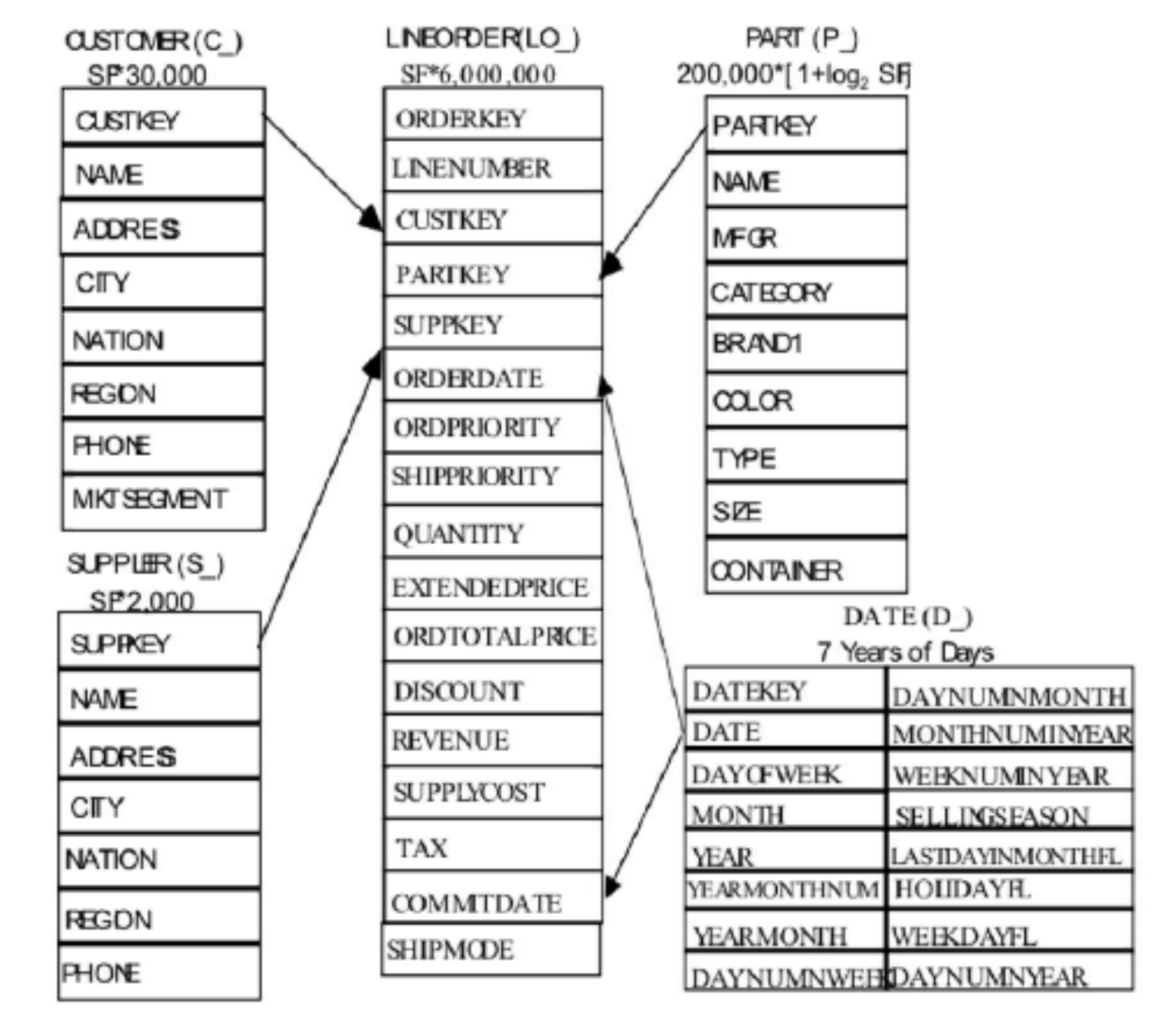}
\caption{Schéma du banc de test SSB1 \label{f10}}
\end{center}
\end{figure}

\begin{table}[]
\centering
\caption{Taille des tables de SSB1 \label{t2}}
\label{my-label}
\begin{tabular}{|l|c|c|}
\hline
Table     & \multicolumn{1}{l|}{Nombre d'instances} & \multicolumn{1}{l|}{\begin{tabular}[c]{@{}l@{}}Taille de la table \\ en nombre de pages\end{tabular}} \\ \hline
Lineorder & 6000000                                 & 123047                                                                                                \\ \hline
Dates     & 2556                                    & 58                                                                                                    \\ \hline
Part      & 200000                                  & 4102                                                                                                  \\ \hline
Supplier  & 2000                                    & 42                                                                                                    \\ \hline
Customer  & 30000                                   & 696                                                                                                   \\ \hline
\end{tabular}
\end{table}
Notons que la taille d’une page 8096 octet, c'est-à-dire 8 Ko.
\end{enumerate}

\section{Évaluation des approches existantes}
\indent Tout d'abord, nous avons effectué des expériences pour définir la valeur appropriée de $minsup$ qui permet la génération d'un grand ensemble d'$IFF$. Les résultats montrent que la valeur minimale appropriée doit être définie sur 0.1.
\subsection{Approche Close}
\indent Cette approche utilise un contexte d’extraction qui consiste en une matrice Requêtes-Attributs. Dans cette matrice, chaque ligne représente une requête et chaque colonne représente un attribut qui a fait l’objet d’un prédicat de sélection ou de projection. Notons qu’on a identifié les attributs par un numéro de 1 jusqu’à 57 pour simplifier la manipulation du fichier texte en entrée.\\

\begin{figure}[h!]
\begin{center}
\includegraphics[scale=0.6]{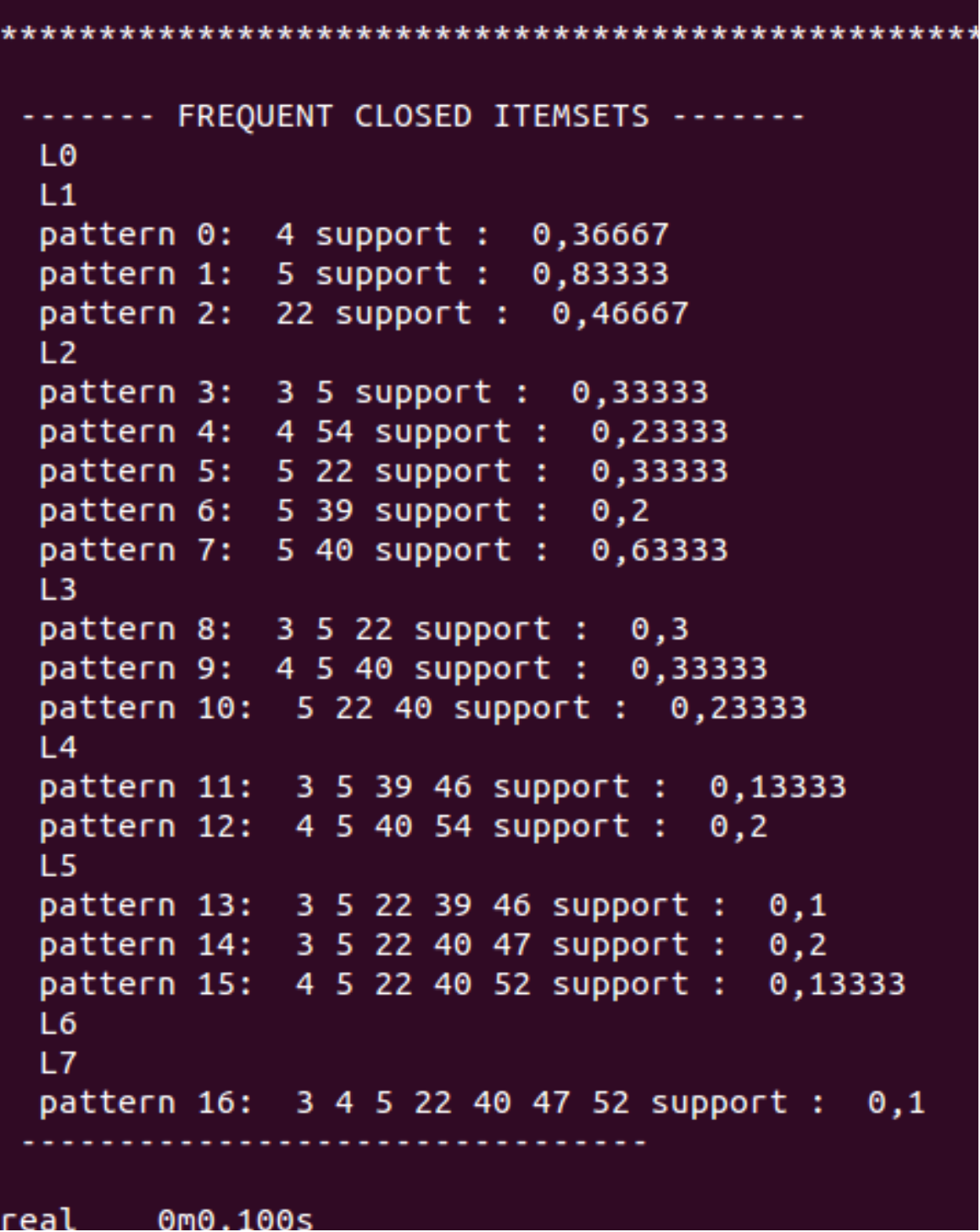}
\caption{Réalisation de l'approche Close \label{f1}}
\end{center}
\end{figure}

\indent Une fois la matrice d’usage est prête et le seuil minimum \textit{minsup} est fixé,  $minsup=0.1$, \textsc{Close} détermine les motifs fréquents fermés, les attributs proposés sont  affichés dans la figure. Or un attribut indexable est un attribut non clé qui appartient à la table de dimension. Ainsi, le seul attribut indexable que \textsc{Close} a pu générer porte le numéro 22 qui a un support égale à 0.46667, comme le montre la figure \ref{f1}. 
    
\subsection{Approche Dynaclose}
Cette approche prend également en input une matrice de contexte Uses($Qi,Aj$) où :
\begin{itemize}
  \item $Uses(Qi,Aj)$= 1 Si la requête $Q_{i}$ utilise un prédicat défini sur l’attribut $A_{j}$
  \item $Uses(Qi,Aj)$= 0 Sinon
  \end{itemize}

\begin{figure}[h!]
\begin{center}
\includegraphics[scale=0.75]{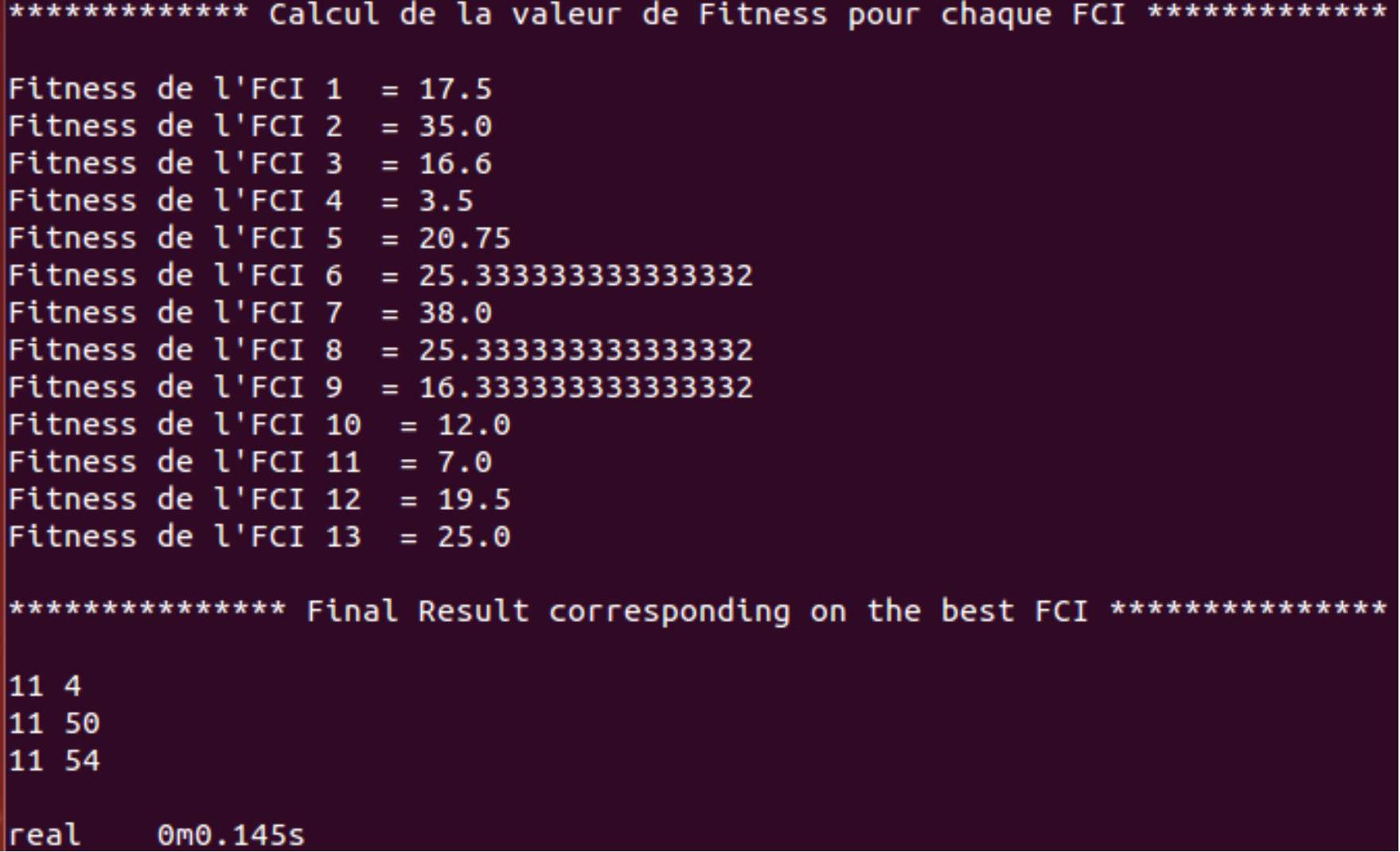}
\caption{Réalisation de l'approche DynaClose \label{f2}}
\end{center}
\end{figure}

\indent Sur cette base, nous déterminons le support relatif à chaque attribut indexable, et les motifs non dupliqués pour chaque requête $Q_{i}$ ; tel que le motif d’une requête correspond aux attributs dont la case Uses($Q_{i}$,$A_{j}$) est égale à 1. Nous calculons pour chacun de ces motifs, la valeur de \textsc{Fitness} dont la formule est précisée dans le chapitre précédent(cf. page \pageref{rapfit}). \\
Enfin, le meilleur motif à choisir serait celui qui a la valeur de \textsc{Fitness}(motif) la plus grande. Le motif proposé est constitué des attributs dont les identifiants sont 11, 54, le seul attribut indexable dans ce groupe est le 54, comme le montre la figure \ref{f2}.

\section{L’approche TM-IJB}
\begin{figure}[h!]
\begin{center}
\includegraphics[scale=0.65]{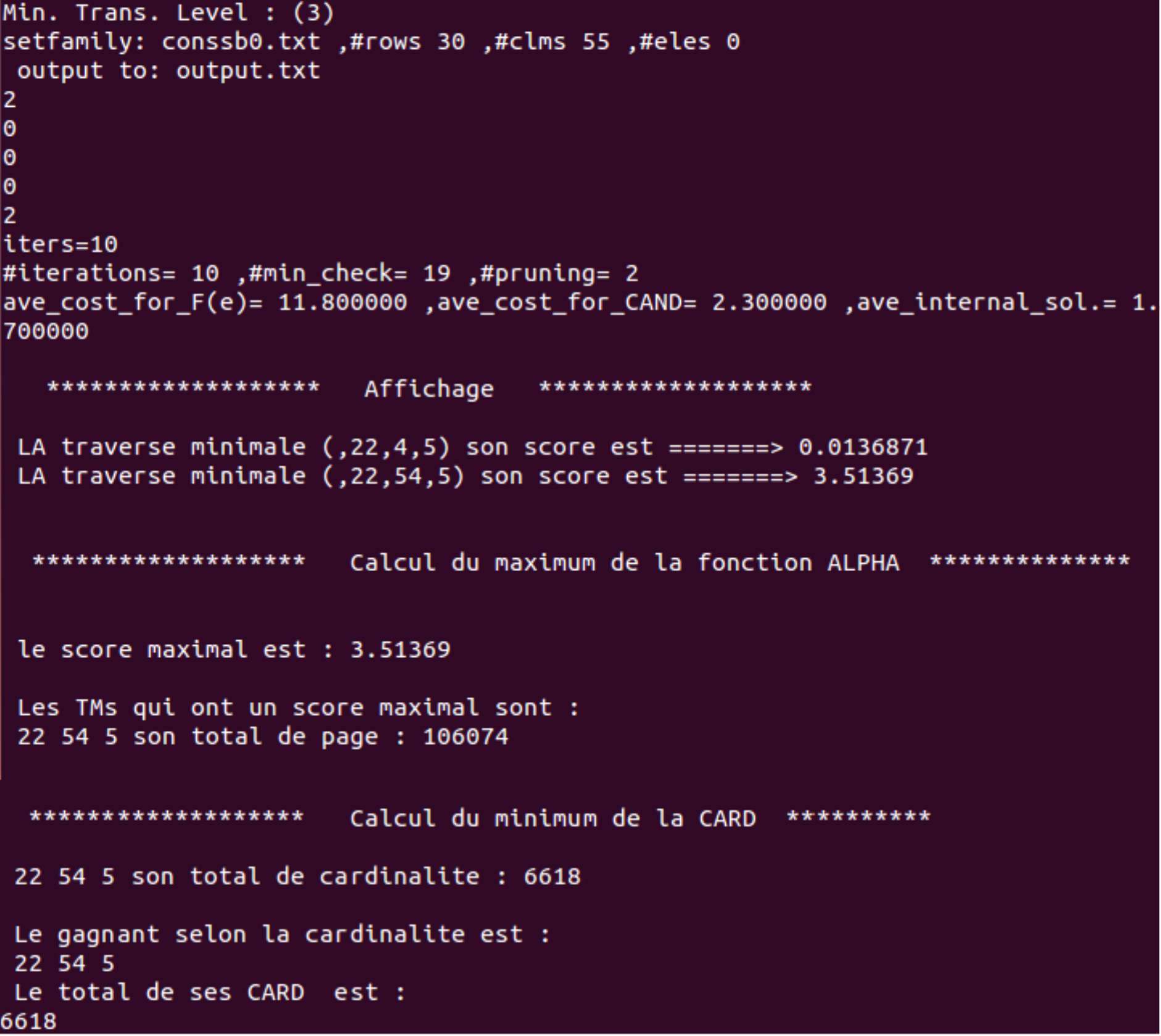}
\caption{Réalisation de l'approche TM-IJB \label{f3}}
\end{center}
\end{figure}

\indent La principale caractéristique de notre approche est l’utilisation des traverses minimales sur un contexte d’extraction qui consiste en une matrice Requêtes-Attributs appelé \textit{hypergraphe}, où chaque hyperarête représente une requête et chaque sommet représente un attribut qui a fait l’objet d’un prédicat de sélection ou de projection. \\
\indent Dès que  l’hypergraphe soit  prêt après l'analyse de la charge de travail et l'extraction des attributs indexables, le processus mis en œuvre par notre  approche commence par le calcule du nombre de transversalité, qui est égal au cardinal de la plus petite traverse minimale. Par la suite, il enchaîne avec l’extraction de toutes les traverses minimales qui ont une taille égale au nombre de transversalité. Une fois qu'il a toutes les traverses minimales il va calculer pour chacune d’elles la fonction \textsc{Fitness(tm)} et ne garder que les $TMs$ qui ont la valeur maximale. Si à ce niveau, il retourne plusieurs traverses minimales, alors il va calculer la somme des cardinalités de chaque attribut des $TMs$ et ne retenir que la traverse minimale qui minimise la somme de cardinalité. En effet, la principale cause de l’explosion de la taille des $IJB$ est la cardinalité des attributs indexés. \\
\indent La figure \ref{f3} montre que le nombre de transversalité est égal à 3. Ainsi, notre algorithme va extraire toutes les traverses minimales de taille 3, qui sont dans cet exemple seulement deux (4,5,22)  et (5,22,54). Ensuite, il va calculer la fonction \textsc{Fitness} pour chaque traverse et ne garder que celle qui a le meilleur résultat qui est dans ce cas la traverse minimale (5,22,54) comme le montre la figure \ref{f3}. Vu qu’après l’exécution de la fonction \textsc{Fitness}, une seule traverse minimale a été retenue donc l’algorithme ne va pas faire recours au calcul de cardinalité.

\section{Étude expérimentale}
\indent Dans cette partie, nous allons implémenter les approches existantes \textsc{Close} et \textsc{DynaClose}.\\
Les expériences ont été menées selon quatre étapes: (1) identification de la valeur de \textit{minsup} qui donne un nombre important d'$IFF$ pour les approches (\textsc{Close, DynaClose}); (2) évaluation du temps d’exécution en exécutant les requêtes en absence de tous les index; (3) évaluation des différentes approches (\textsc{Close, DynaClose, TM-IJB}) en termes de temps d’exécution en exécutant les requêtes en présence des configurations d’indexs adéquates de chaque approches; et (4) le calcul du coût théorique en entrées/sorties.\\
\indent Dans notre cas, nous allons nous concentrer sur l’optimisation des requêtes sans contraintes de stockage vu que de nos jours le coût des supports de stockage désormais moins cher et plus disponible. Nous pouvons même utiliser le stockage dans les nuages avec son avantage de l’élasticité et le ‘pay per use’.  \emph{Perriot et al.} ont proposé un modèle de coût que nous pouvons également l'utiliser pour les $IJB$  \cite{Per13}.   \\
Pendant l’étude expérimentale nous avons utilisé 2 bancs de test différents pour montrer l’intérêt de notre approche.

\subsection{Le banc de test SSB1}
\subsubsection{Sélection de la configuration de l'IJB}
Dans cette section nous allons nous baser sur les configurations d’index obtenues dans les deux dernières sections où l’approche \textsc{DynaClose} a retourné l’attribut \emph{P\_Brand} (cf. figure \ref{f2}). L’approche \textsc{Close} a donné l’attribut \emph{D\_year} comme le montre la figure \ref{f1} et notre approche \textsc{TM-IJB} a suggéré  \emph{P\_Brand} et \emph{D\_year} d’après la figure \ref{f3}.
\subsubsection{Validation sous Oracle}

Maintenant, nous passons à la validation sous le SGBD \textsc{Oracle} 12c. Au premier lieu nous allons exécuter toutes les requêtes sans aucun index et retenir leur temps d’exécution.

\begin{figure}[htbp]
\begin{center}
\includegraphics[scale=0.9]{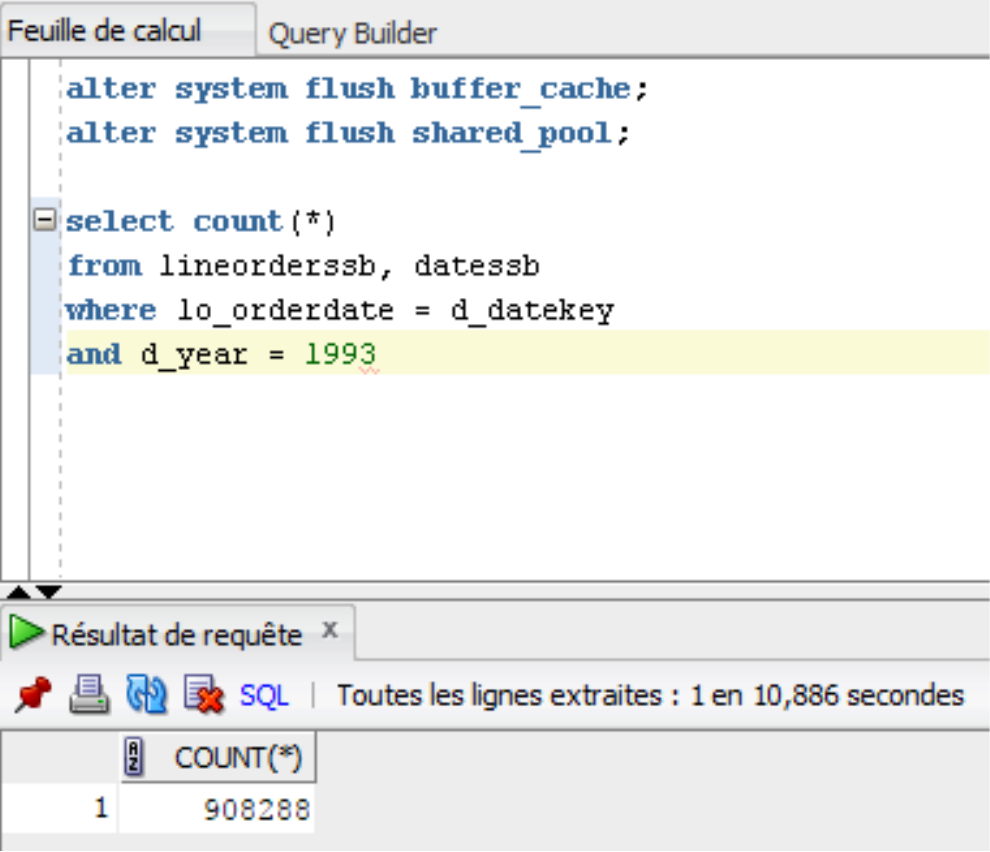}
\caption{Temps d'exécution d'une requête sans IJB \label{f4}}
\end{center}
\end{figure}

\begin{figure}[h!]
\begin{center}
\includegraphics[scale=0.8]{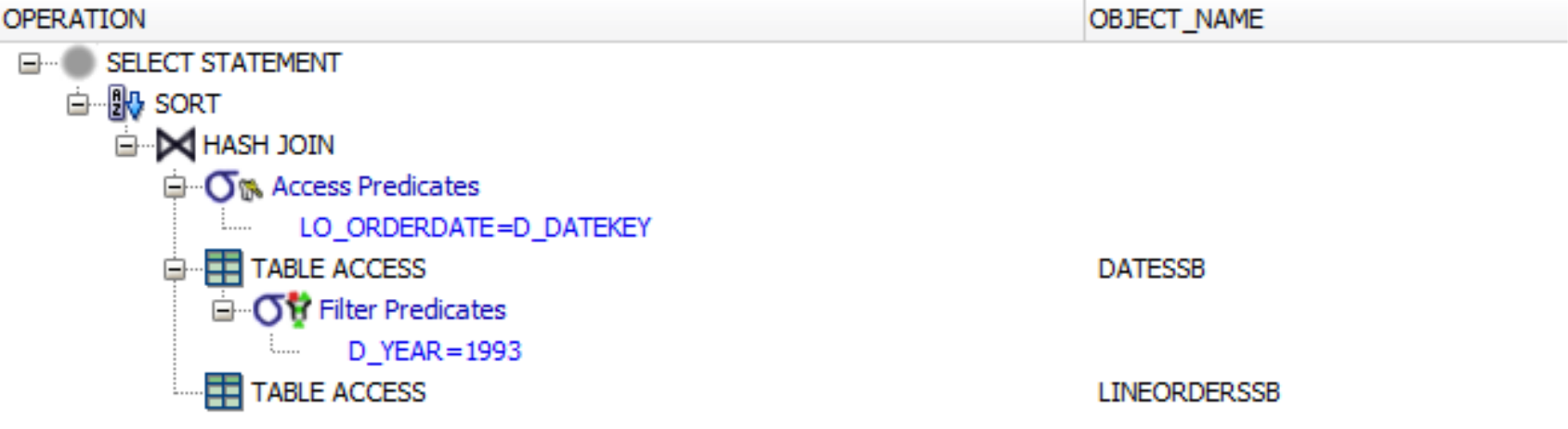}
\caption{Le plan d'exécution d'une requête sans IJB  \label{f5}}
\end{center}
\end{figure}

Sans oublier de vérifier le plan d’exécution de la requête en absence de l’$IJB$ (cf. figure \ref{f5})  pour pouvoir le comparer par la suite avec celui en présence de l’index.  

\indent Par la suite, nous créons physiquement les configurations générées par chaque algorithme et nous exécutons les requêtes en présence de l’indexe et retenir le temps d’exécution de chaque requête sans oublier de vérifier à chaque fois l’utilisation des $IJB$. Pour vérifier que les $IJB$ sont bien utilisés par \textsc{Oracle} pour exécuter une requête donnée, nous avons utilisé l’outil \textsc{Plan d’exécution} fournit avec \textsc{Oracle}. Cet outil permet de visualiser le plan d’exécution détaillé d’une requête.

\begin{figure}[htbp]
\begin{center}
\includegraphics[scale=0.9]{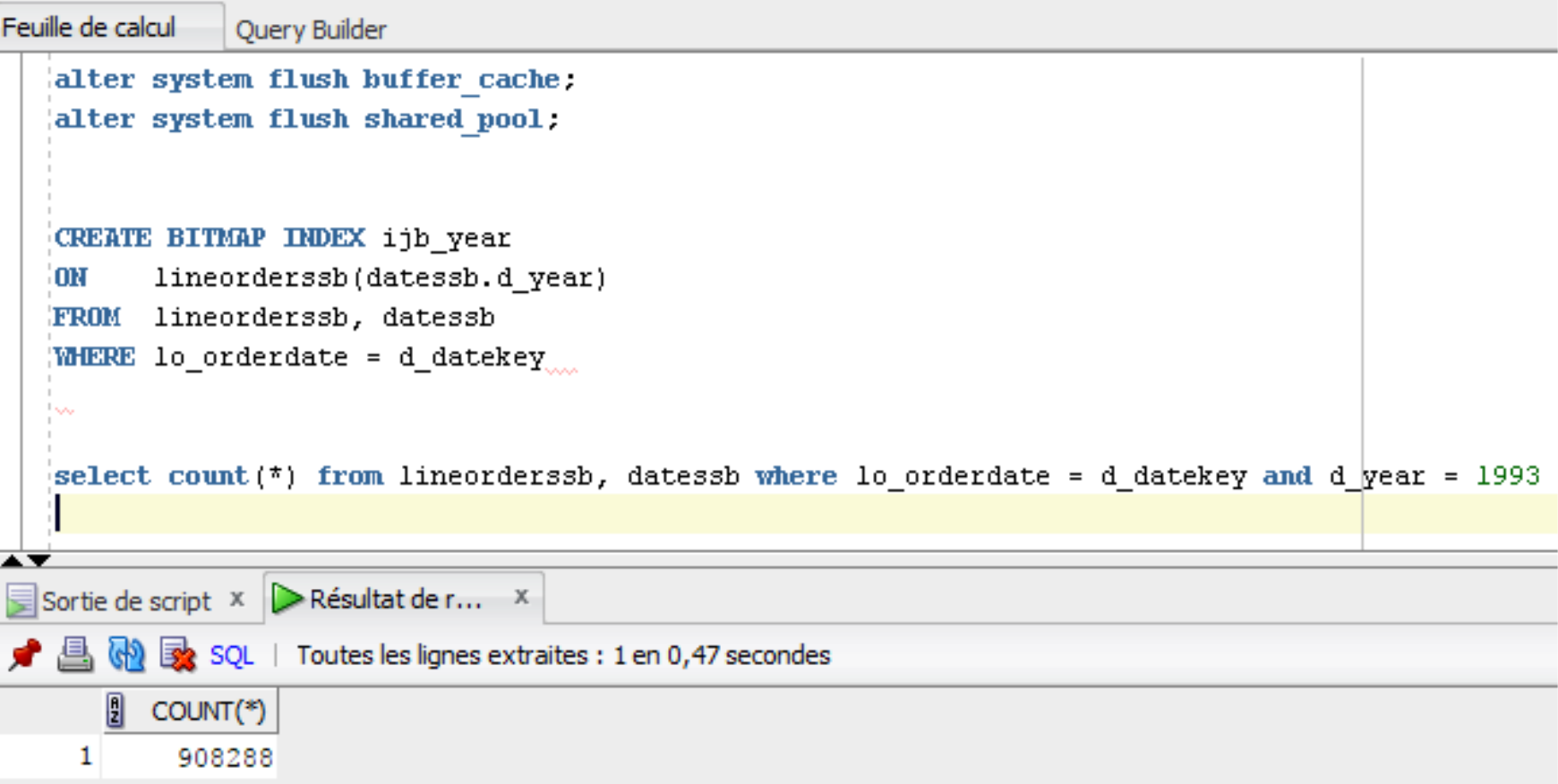}
\caption{Temps d'exécution d'une requête en présence d'IJB \label{f6}}
\end{center}
\end{figure}

\begin{figure}[htbp]
\begin{center}
\includegraphics[scale=0.8]{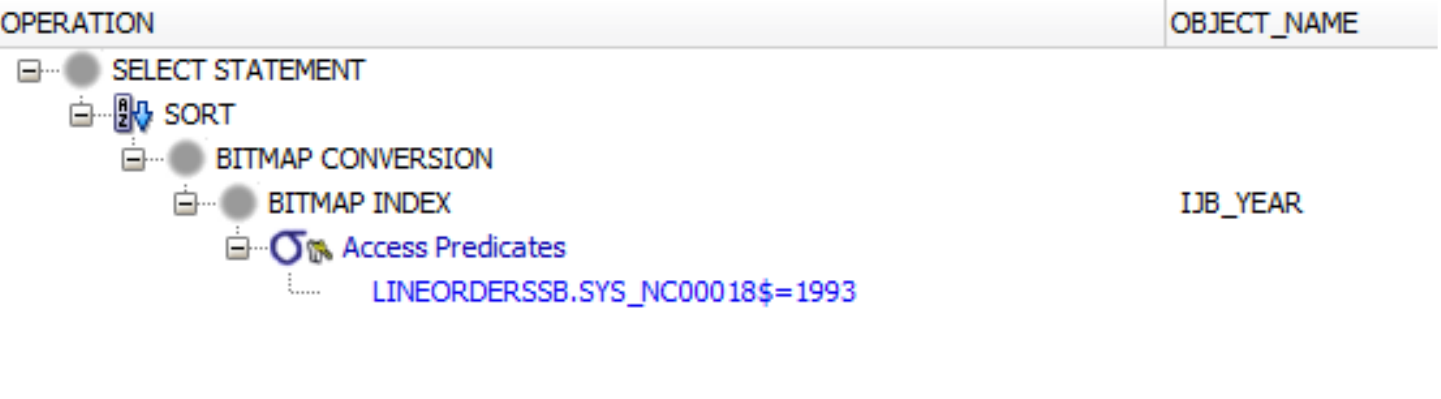}
\caption{Le plan d'exécution d'une requête sans IJB  \label{f7}}
\end{center}
\end{figure}

Comme les tâches sont répétitives pour les 30 requêtes, nous avons choisi de :
\begin{description}

  \item[$\bullet$] Montrer qu'un résultat prouvant l’amélioration d’une requête en utilisant un $IJB$ sur \emph{d\_year}. En effet, avant la création de l’index, la requête prenait \emph{10.886 secs} pour s’exécuter et après sa création, elle est devenue presque instantanée de l’ordre de quelques millisecondes comme le montre la figure \ref{f6}.
  \item[$\bullet$] S’assurer de l’utilisation de l’$IJB$ dans le plan d’exécution comme le montre la figure par rapport a la figure lors de l’exécution de la requête sans l’utilisation de l’index, (voir figure \ref{f7}).
\end{description}

\indent L’exécution des  requêtes sur toutes les configurations d’index nous a permet de confirmer la grande utilité des $IJB$ pour les requêtes de type Count(*). Le coût d’exécution de ces requêtes en utilisant les $IJB$ est négligeable devant le coût sans index. Ces requêtes sont celles les plus bénéficiaires des $IJB$, car aucun accès aux données n’est effectué (l’accès aux $IJB$ suffit pour répondre à ces requêtes). Les requêtes les moins bénéficiaires des $IJB$ sont celles référençant des attributs de dimension dans la clause SELECT ou celles ayant moins d’attributs indexés, car elles nécessitent des jointures supplémentaires entre les tables de dimension et la table de faits \cite{KB09}.

\subsubsection{Étude comparative}

\begin{figure}[htbp]
\begin{center}
\includegraphics[scale=0.7]{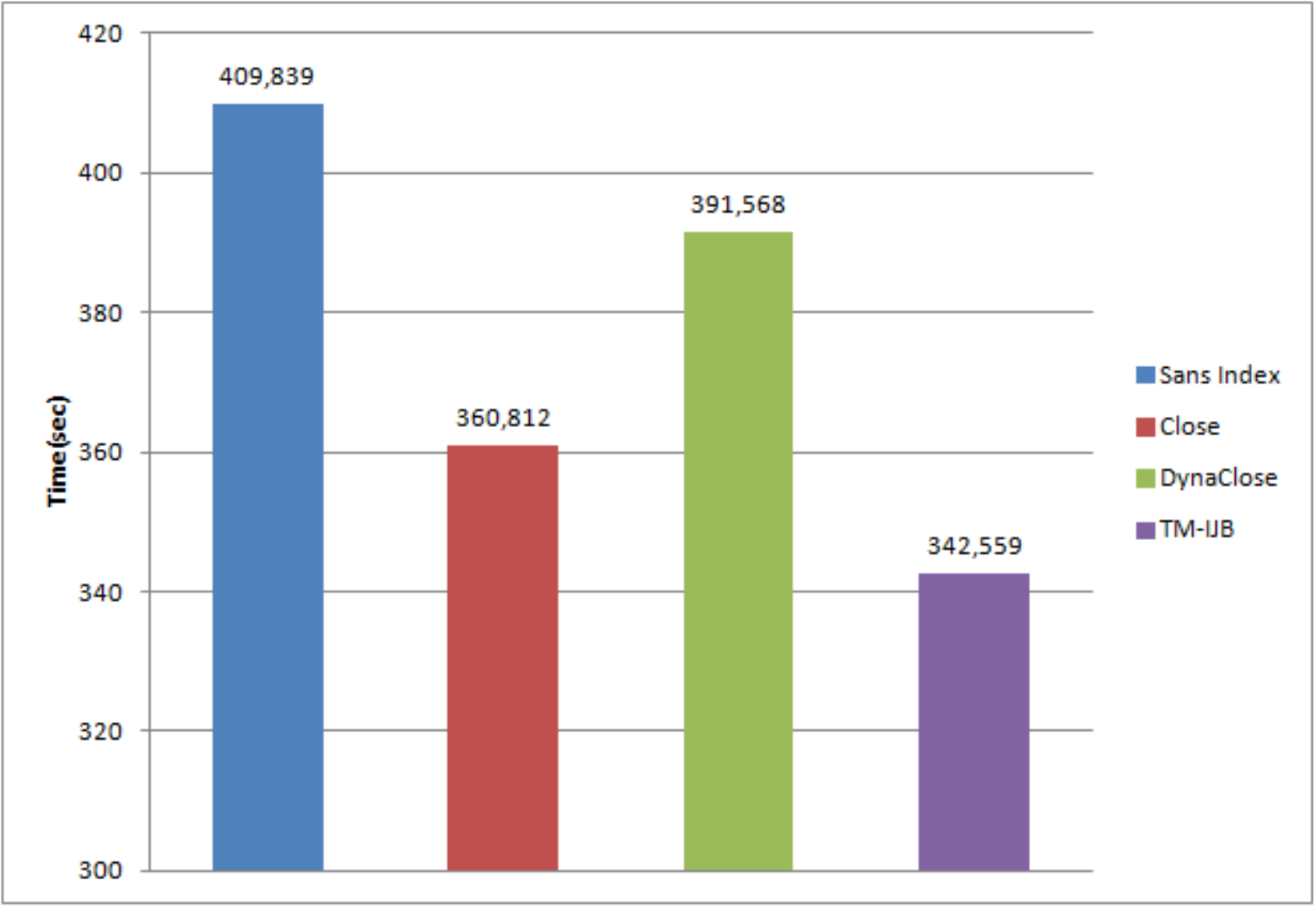}
\caption{Temps d'exécution des requêtes sur le banc de test SSB1 \label{f8}}
\end{center}
\end{figure}
\indent Après avoir exécuté toutes les requêtes, l’histogramme de la figure \ref{f8} montre la somme des temps d’exécution des 30 requêtes en absence des $IJB$ et en présence des différentes configurations de chaque approche. Cet histogramme montre que notre approche \textsc{TM-IJB} est meilleure en termes de temps d’exécution vu qu’elle possède le temps le plus petit par rapport aux autres approches. 
\begin{figure}[htbp]
\begin{center}
\includegraphics[scale=0.7]{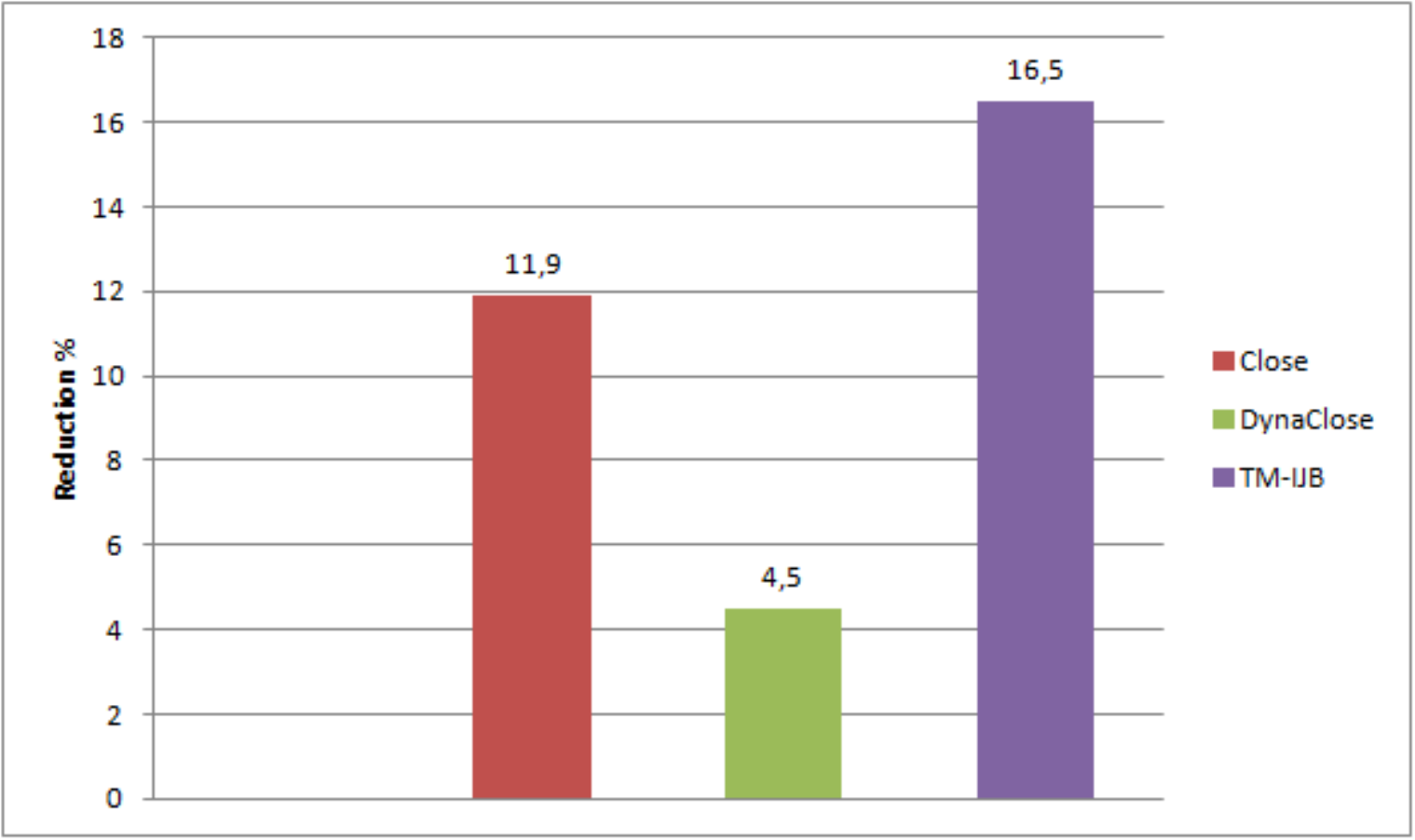}
\caption{Le taux de réduction des requêtes du banc de test SSB1\label{f11}}
\end{center}
\end{figure}

\begin{figure}[h!]
\begin{center}
\includegraphics[scale=0.7]{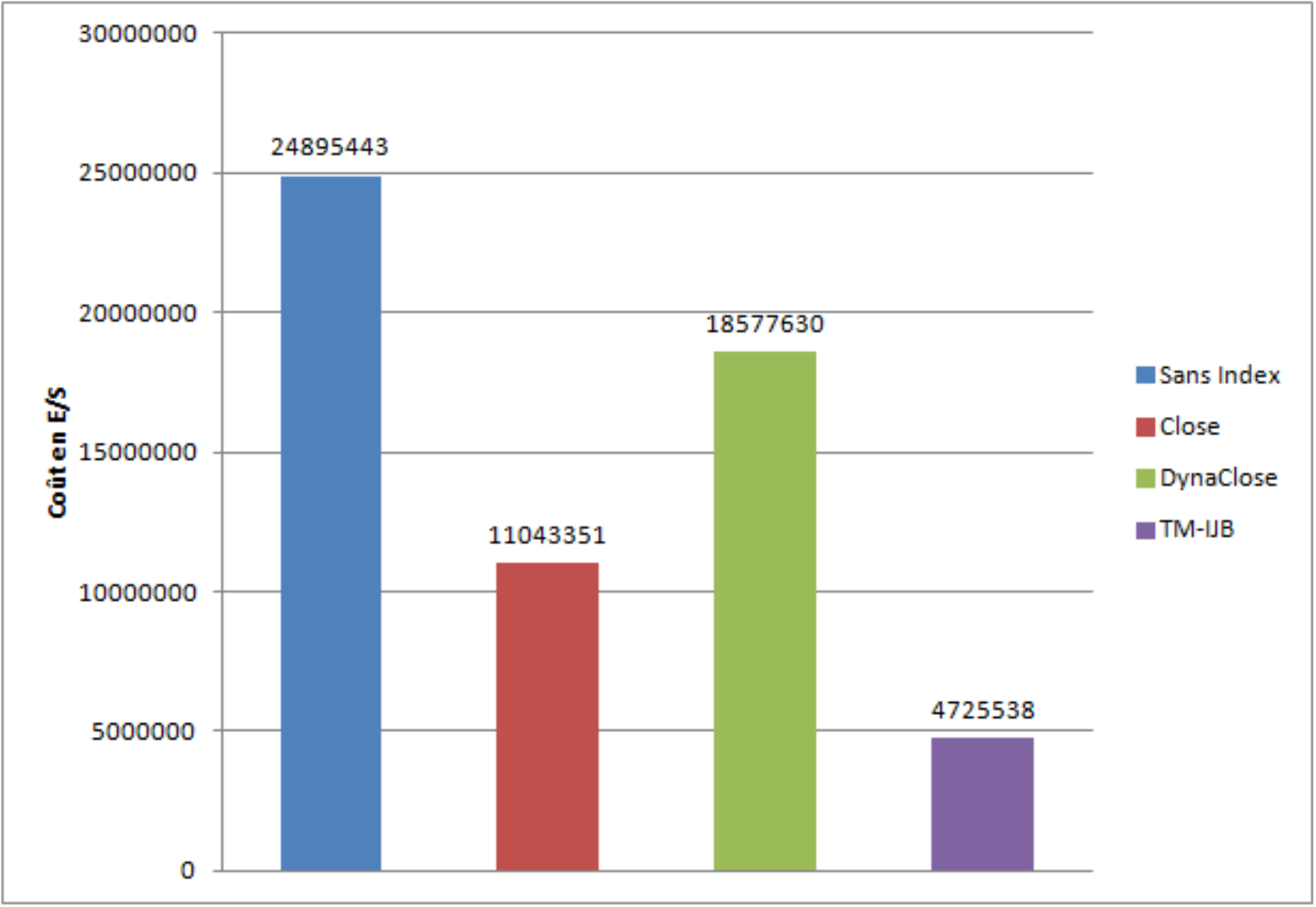}
\caption{Le coût d'exécution en E/S du banc de test SSB1\label{f12}}
\end{center}
\end{figure}
\indent L’histogramme de la figure \ref{f11} montre le taux de réduction total de chaque approche, notre approche possède le taux de plus élevé de 16.5\% quant à \textsc{Close} et \textsc{DynaClose} est de 11.9\% et 4.5\% respectivement.  \\

\indent La figure \ref{f12} montre comment les différentes approches d'indexation réduisent le coût d'exécution des requêtes. Le résultat principal est que \textsc{TM-IJB} surpasse toutes les approches.

\indent Comme nous pouvons le remarquer dans la figure \ref{f12}, grâce au modèle de coût développé, nous avons pu quantifier la qualité de la structure d’optimisation sélectionnée $IJB$ sur les attributs proposés par les différentes approches. Ainsi, nous pouvons constater qu’il est plus bénéfique de tenir compte de l’utilisation des traverses minimales pour extraire les $IJB$.
\subsection{Le banc de test TPC-H}
Nous allons répéter toutes les étapes précédentes mais cette fois avec le banc de test $TPC$-$H$. Au départ, nous avons effectué quelques tests pour définir la valeur appropriée de $minsup$, qui permet la génération d'un grand ensemble d'$IFF$ ensuite exécuter les algorithmes  \textsc{Close}, \textsc{DynaClose} et \textsc{TM-IJB} pour avoir les différentes configurations de chaque approche et à la fin passer à la validation sous \textsc{Oracle}.
\subsubsection{Sélection de la configuration de l'IJB}
\begin{enumerate}[label=\roman*]
\item \textbf{L'approche Close :} une fois l’algorithme \textsc{Close} exécuté avec un $minsup$ égal à 0.1, il retourne 2 items fréquents fermés, i.e., 2 et 41 avec un support relatif égal à 0.21035, qui formeront par la suite la configuration d’$IJB$. Les items 2 et 41 sont respectivement les attributs \emph{n\_name} et \emph{o\_orderdate} comme le montre la figure \ref{f13}. \\

\begin{figure}[h!]
\begin{center}
\includegraphics[scale=0.6]{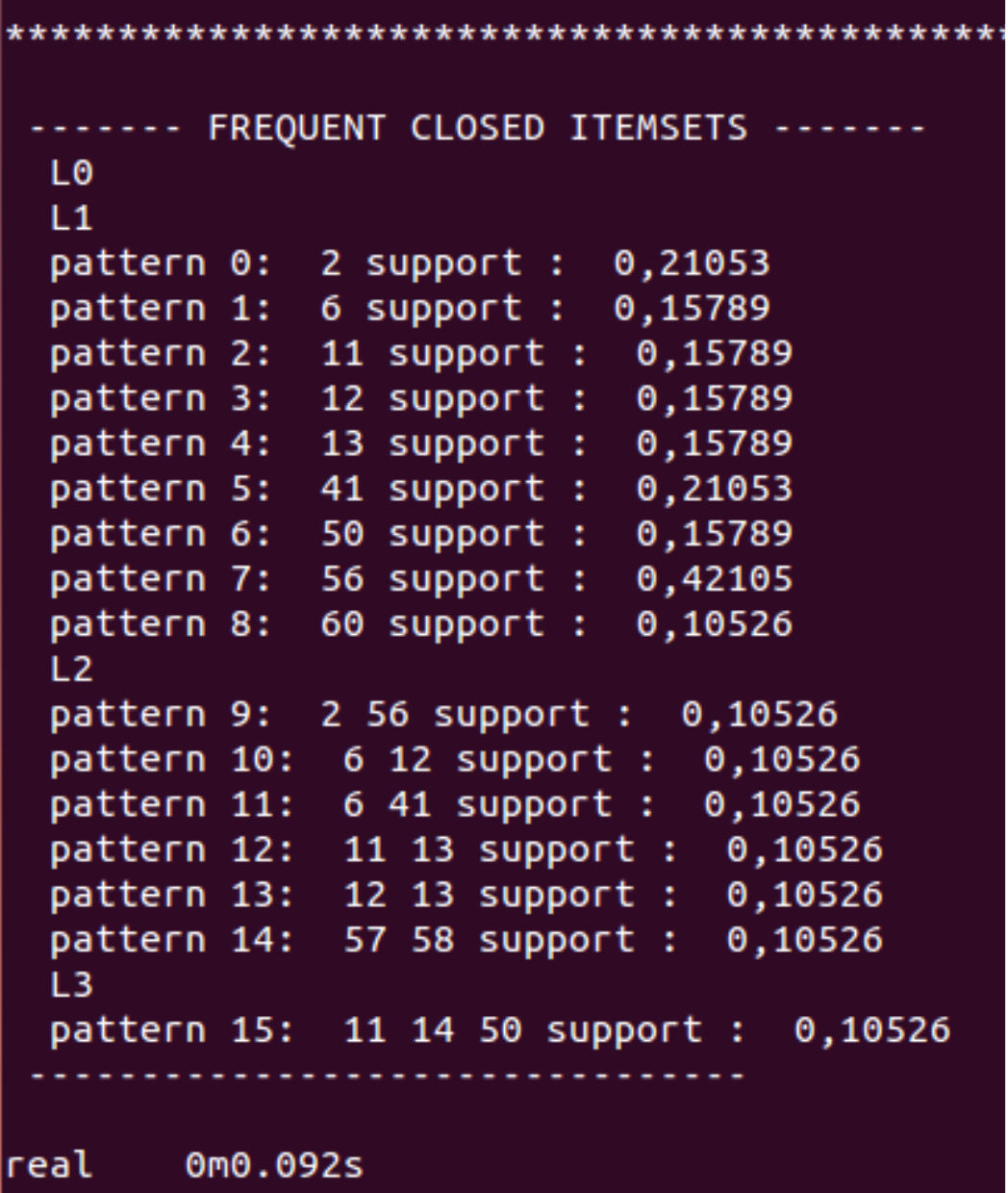}
\caption{L'exécution de l'aglorithme de Close sur TPC-H \label{f13}}
\end{center}
\end{figure}

\item \textbf{L'approche DynaClose :} une fois l’algorithme \textsc{DynaClose} exécuté avec un $minsup$ égal à 0.1, il retourne 2 items fréquents fermés avec leur fonction \textsc{Fitness}, i.e., 11 et 41 qui formeront par la suite la configuration d’$IJB$. Les items 11 et 41 sont respectivement les attributs \emph{p\_brand} et \emph{o\_orderdate} comme le montre la figure \ref{f14}.  \\
\begin{figure}[h!]
\begin{center}
\includegraphics[scale=0.4]{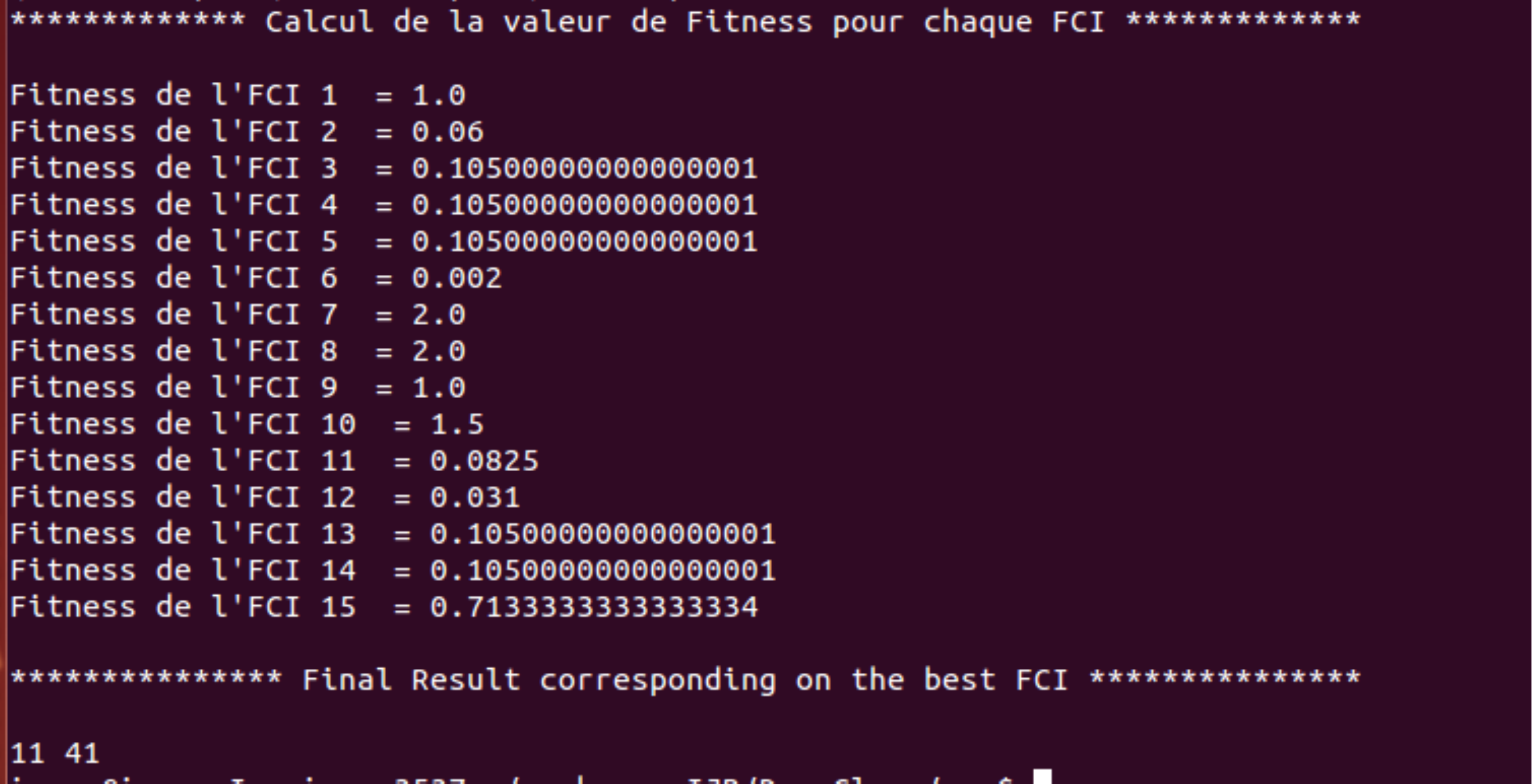}
\caption{L'exécution de l'aglorithme de DynaClose sur TPC-H \label{f14}}
\end{center}
\end{figure}

\item \textbf{L'approche TM-IJB }: l’algorithme commence par la détermination du nombre de transversalité qui est, selon l’hypergraphe de la figure \ref{f15}, égal à 6, ensuite l’algorithme continue par l’extraction de toutes les traverses minimales de taille 6 qui sont 54 $TMs$. Ensuite, il va calculer la fonction \textsc{Fitness} pour chacune de ces traverses et ne garde que celle qui a le meilleur résultat, vu que dans ce cas, la fonction \textsc{Fitness} retourne plus qu’une traverse minimale. Ainsi, l’algorithme calcule pour chacune de ces traverses minimales la somme de cardinalité de chaque attribut et ne retient que la $TM$ ayant la plus petite somme. 

\begin{figure}[htbp]
\begin{center}
\includegraphics[scale=0.7]{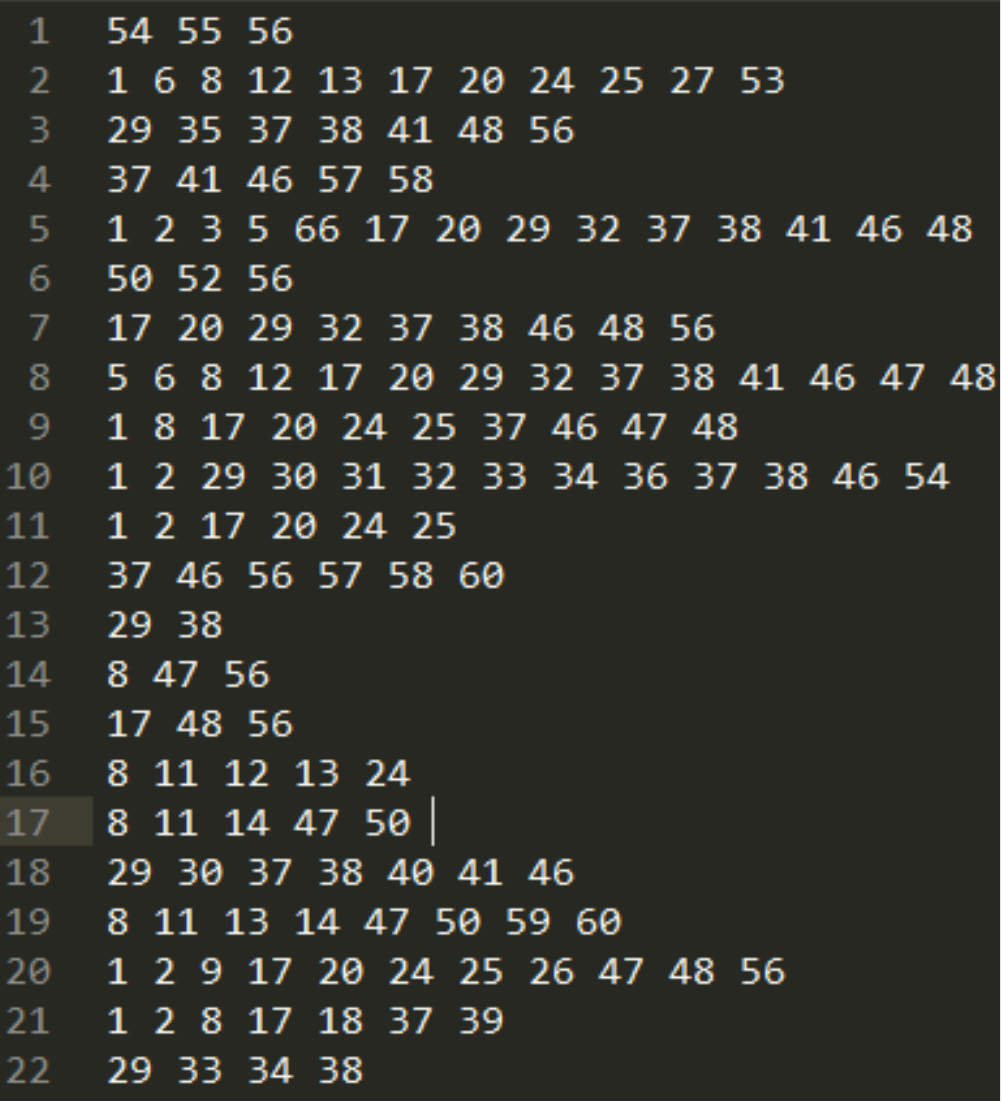}
\caption{L'hypergraphe de TPC-H \label{f15}}
\end{center}
\end{figure}

\indent À la fin, l’algorithme retourne la traverse minimale (2,13,34,41,50,56), et seuls les sommets 2,13,34 et 41 sont des attributs indexables qui sont respectivement \emph{n\_name, p\_size, c\_acctbal et o\_orderdate }comme le montre la figure \ref{f16}.  \\

\begin{figure}[htbp]
\begin{center}
\includegraphics[scale=0.7]{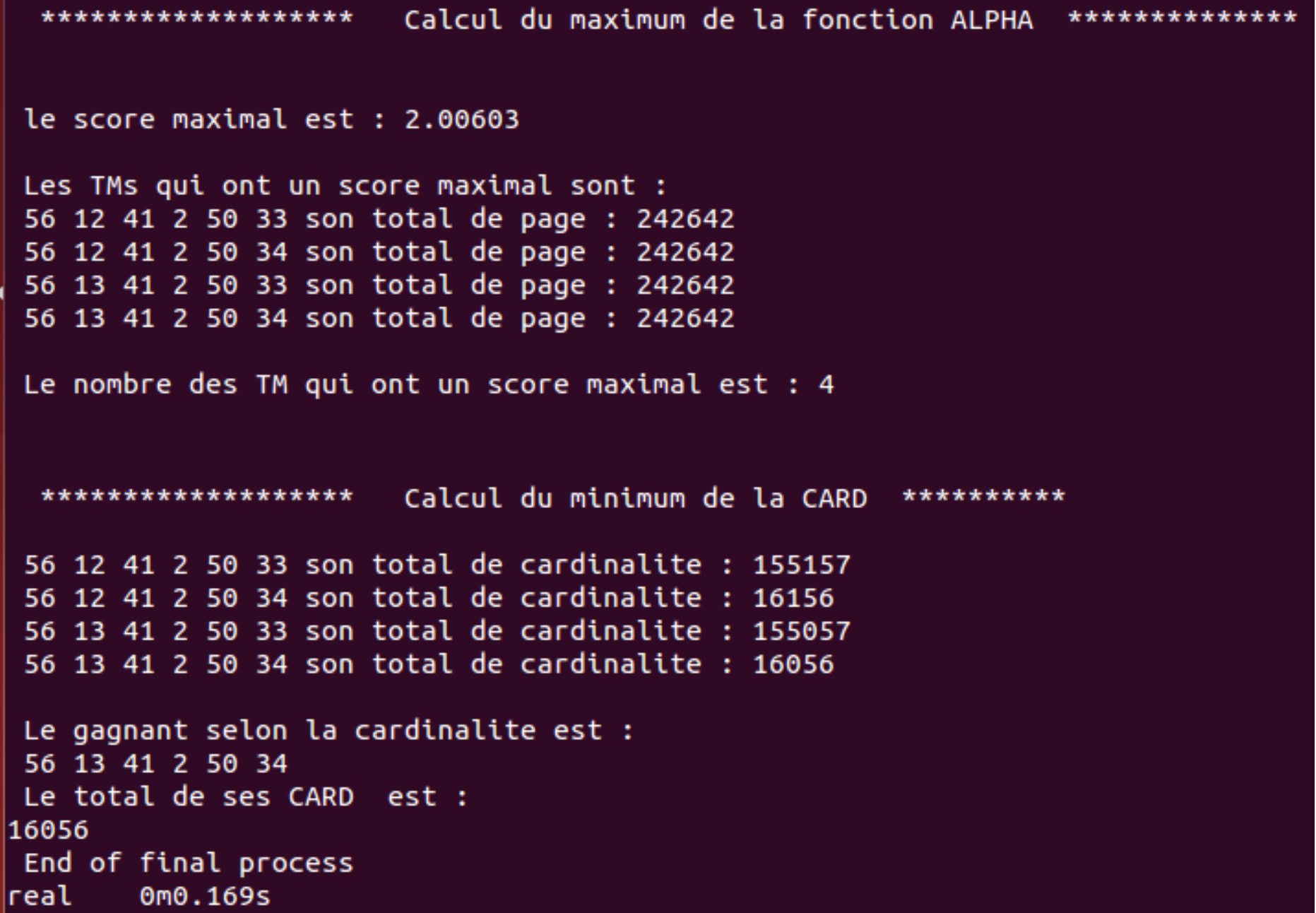}
\caption{Résultat de l'aglorithme de \textsc{TM-IJB} sur le banc de test TPC-H \label{f16}}
\end{center}
\end{figure}
\end{enumerate}

\subsubsection{ Validation sous Oracle}
\begin{figure}[htbp]
\begin{center}
\includegraphics[scale=0.7]{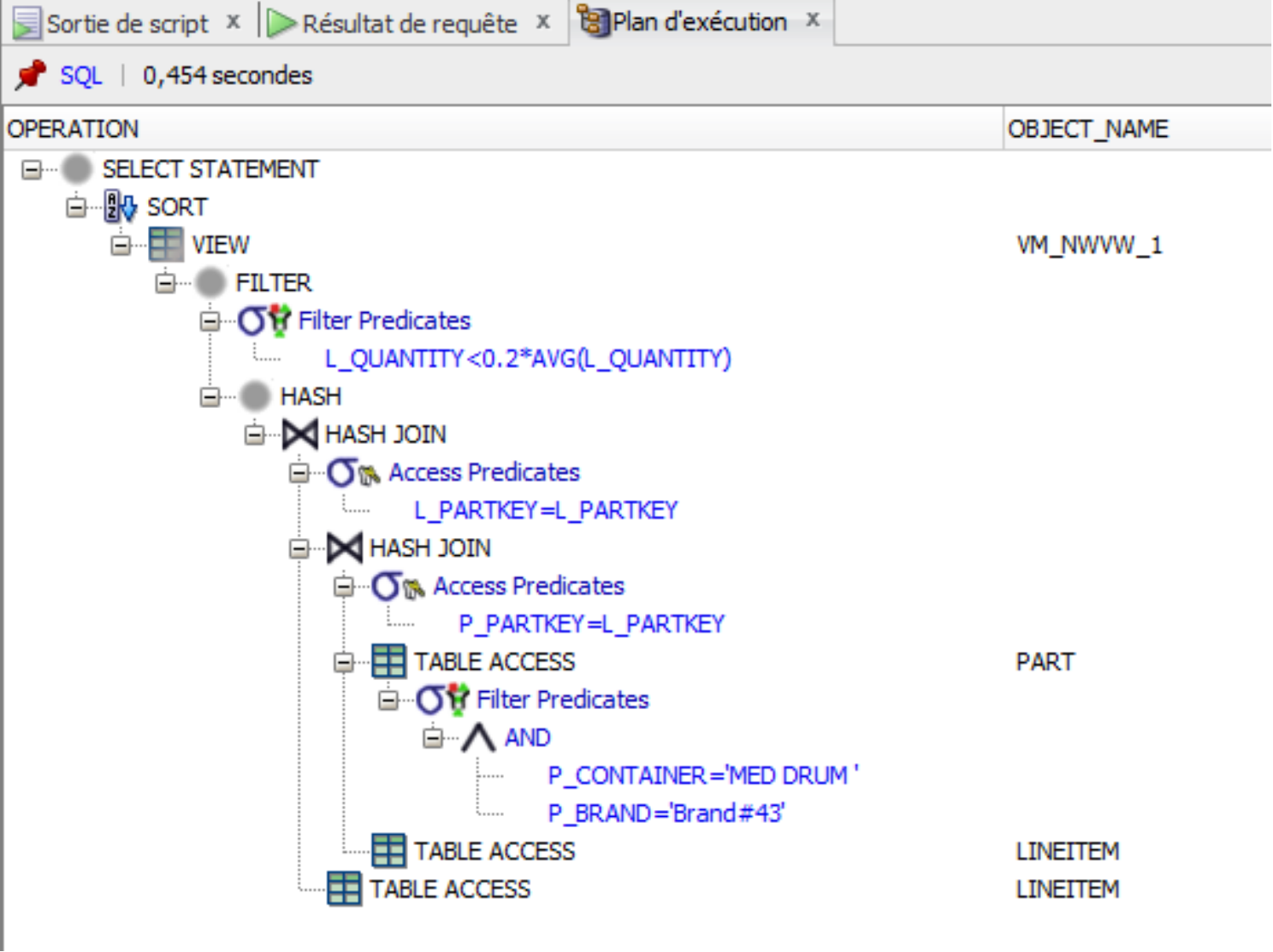}
\caption{Plan d'exécution d'une requête sans IJB \label{f17}}
\end{center}
\end{figure}

Maintenant, nous passons à la validation sous le SGBD \textsc{Oracle} 12c. Au premier lieu nous allons exécuter toutes les requêtes sans aucun index et retenir leur temps d’exécution. 
Sans oublier de vérifier le plan d’exécution de la requête en absence de l’$IJB$ (cf. figure \ref{f17})  pour pouvoir le comparer par la suite avec celui en présence de l’index.  

\begin{figure}[htbp]
\begin{center}
\includegraphics[scale=0.9]{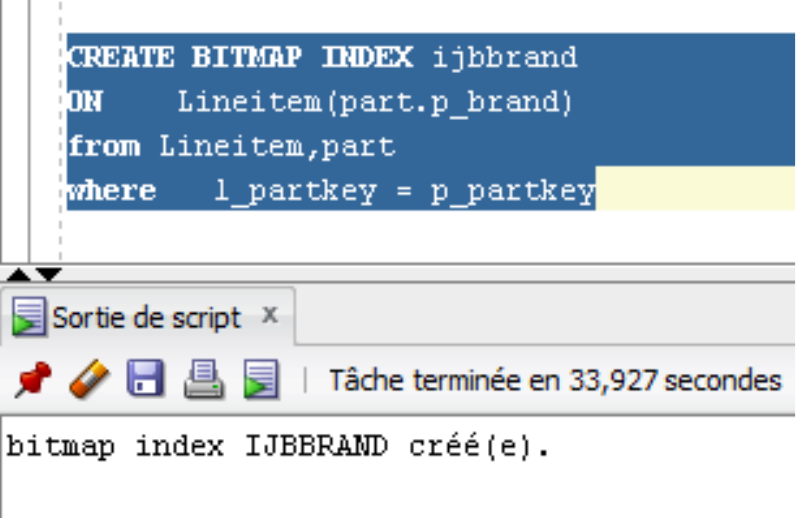}
\caption{La création d'un IJB \label{f18}}
\end{center}
\end{figure}

Par la suite, nous créons physiquement les configurations générées par chaque approche. La figure \ref{f18} montre un exemple de création d’un $IJB$, celui sur l’attribut \emph{p\_brand}.

\begin{figure}[htbp]
\begin{center}
\includegraphics[scale=0.7]{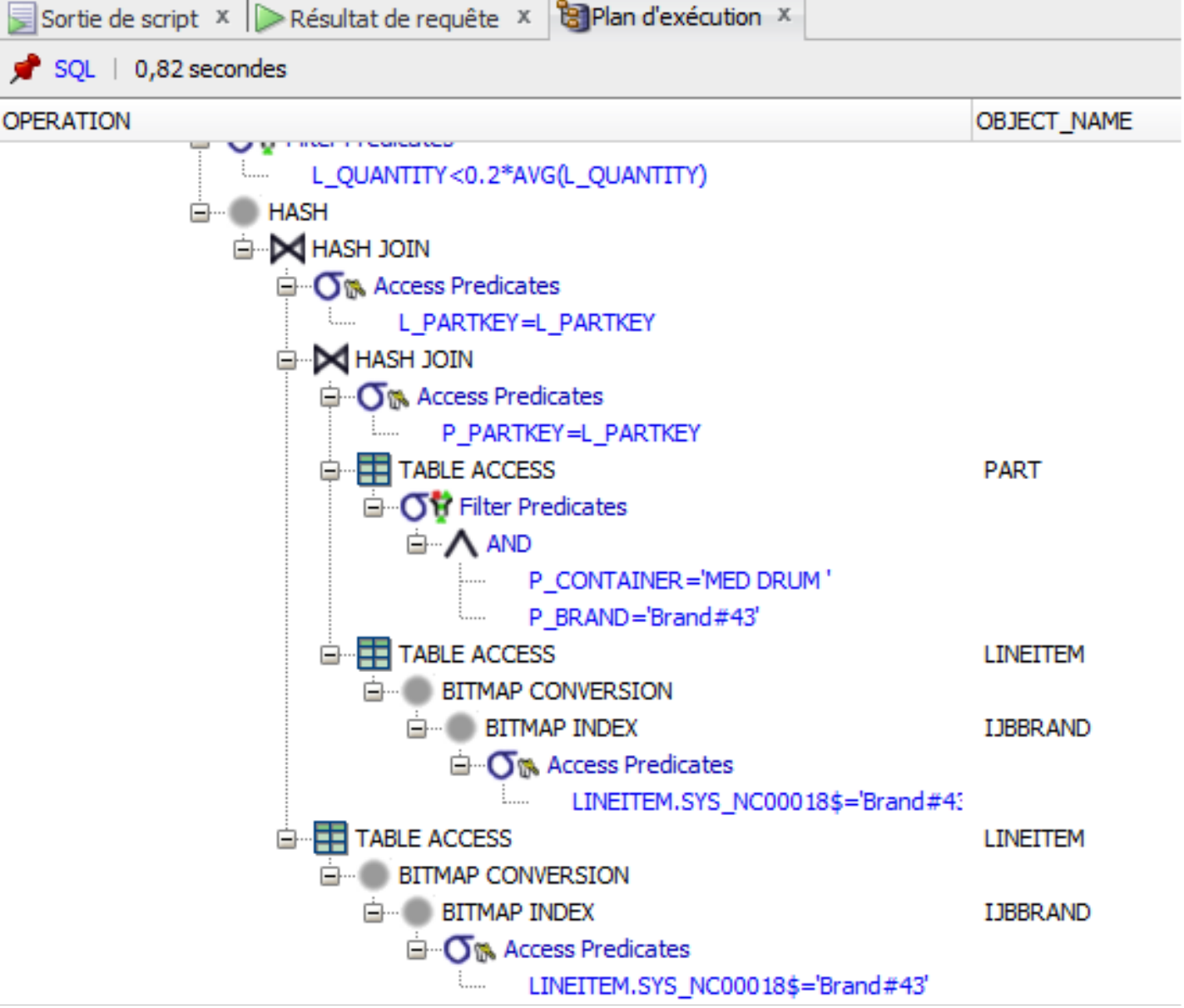}
\caption{Plan d'exécution d'une requête avec IJB sur le banc de test TPC-H  \label{f19}}
\end{center}
\end{figure}

Nous exécutons après les requêtes en présence de l’index et retenir le temps d’exécution de chaque requête sans oublier de vérifier à chaque fois l’utilisation des IJB. \\
 Nous prenons cette requête comme un exemple : \\
\begin{exemple}
\begin{verbatim}
Select 	sum(l_extendedprice) / 7.0 as avg_yearly
From 	lineitem, 	part,
        (select l_partkey as agg_partkey, 
        0.2 * avg(l_quantity) as avg_quantity 
        from lineitem 
        group by l_partkey) part_agg
Where 	p_partkey = l_partkey
    and agg_partkey = l_partkey
    and p_brand = 'Brand#43'
    and p_container = 'MED DRUM '
    and l_quantity < avg_quantity
\end{verbatim}
\end{exemple}
Cette requête s’est exécutée en \emph{18.766 secs} en présence de l’$IJB$ comme le montre le plan d’exécution de la figure \ref{f19} par contre elle a pris \emph{31.646 secs} pour s’exécuter en absence de l’$IJB$. Nous pouvons conclure ici que la présence de l’$IJB$ nous a permis d’avoir un gain de 41.8\% du temps d’exécution. 

\subsubsection{Étude comparative}
\begin{figure}[htbp]
\begin{center}
\includegraphics[scale=0.7]{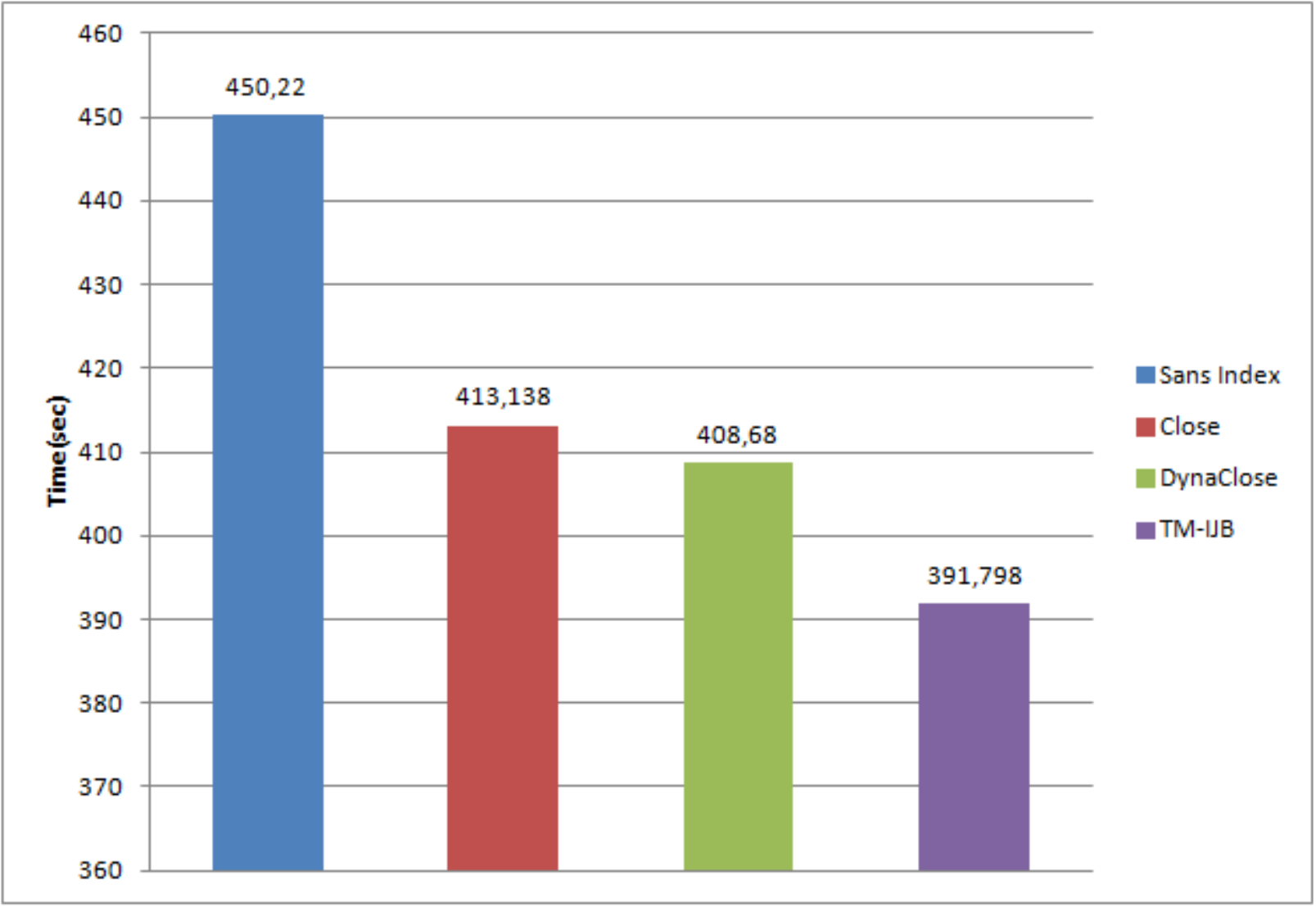}
\caption{Temps d'exécution des 22 requêtes du banc de test TPC-H\label{ff0}}
\end{center}
\end{figure}
Après avoir exécuté toutes les requêtes, l’histogramme de la figure \ref{ff0} montre la somme des temps d’exécution des 22 requêtes en absence des $IJB$ et en présence des différentes configurations de chaque approche. Cet histogramme montre que notre approche \textsc{TM-IJB} est meilleure en termes de temps d’exécution vu qu’elle a le temps le plus petit par rapport aux autres approches. 
\subsubsection{Étude comparative}
\begin{figure}[htbp]
\begin{center}
\includegraphics[scale=0.7]{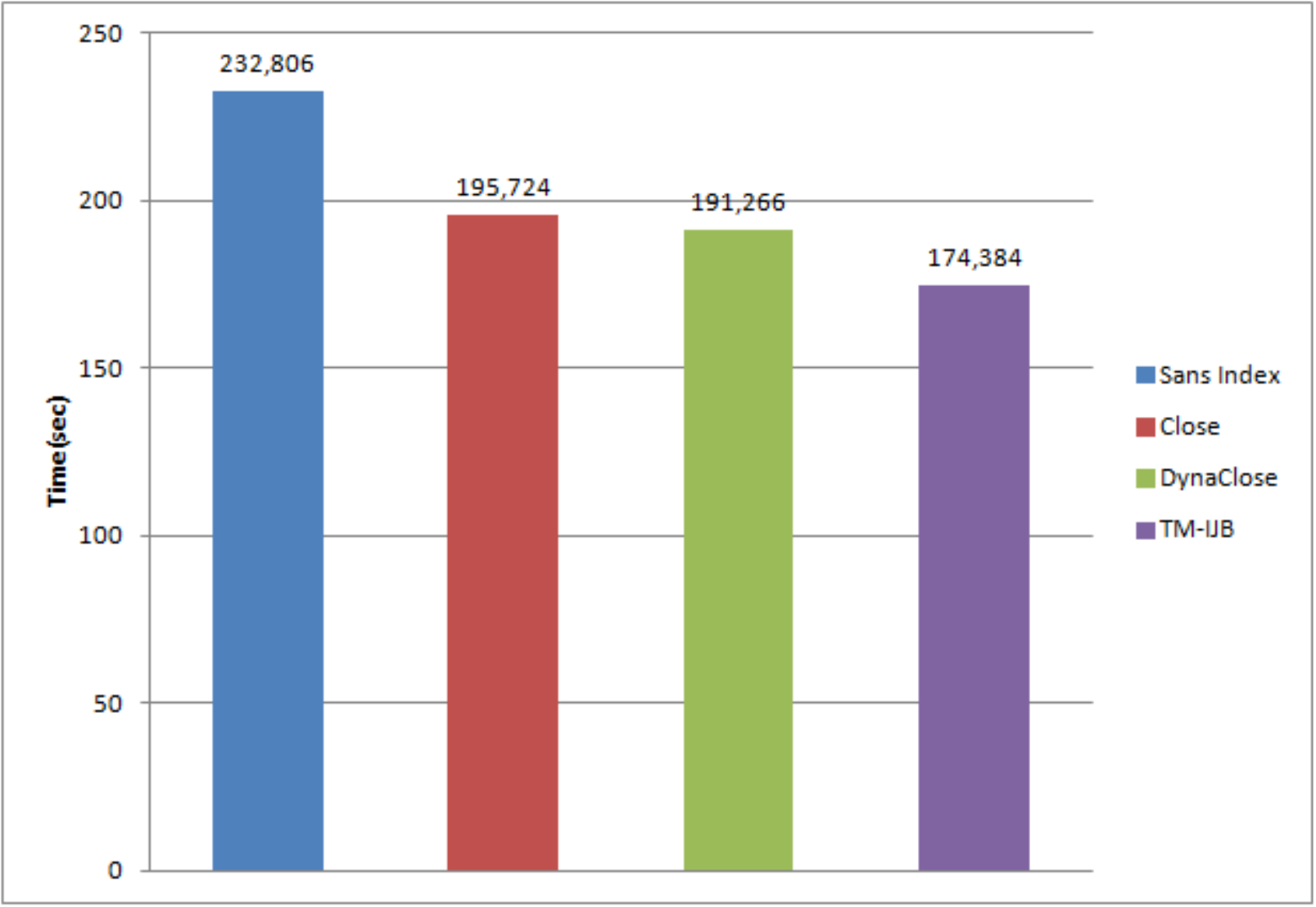}
\caption{Temps d'exécution des 8 requêtes du banc de test TPC-H\label{ff1}}
\end{center}
\end{figure}
\indent Vu que seules 8 des 22 requêtes fournies par le banc de test sont susceptibles d’avoir un $IJB$, dans le sens que ces requêtes interrogent seulement la tables de faits ou bien seulement les tables de dimension. Ainsi, nous avons tracé 2 histogrammes, un contenant les résultats des 22 requêtes illustré dans la figure \ref{ff0}. Et la figure \ref{ff1} illustre les 8 requêtes sur lesquelles nous pouvons exécuter l’$IJB$.

\begin{figure}[htbp]
\begin{center}
\includegraphics[scale=0.7]{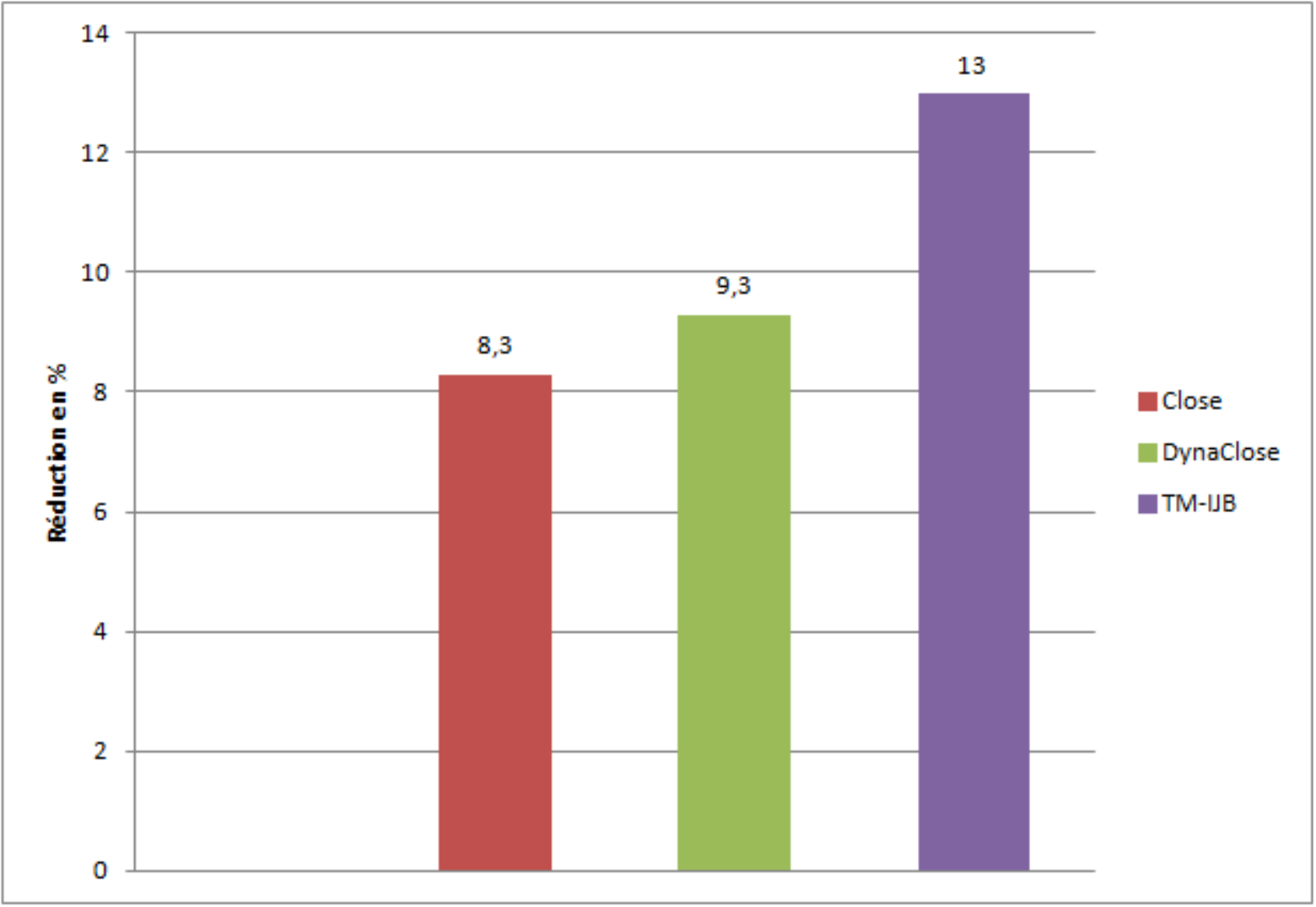}
\caption{Taux de réduction des 22 requêtes du banc de test TPC-H\label{ff2}}
\end{center}
\end{figure}

\begin{figure}[htbp]
\begin{center}
\includegraphics[scale=0.7]{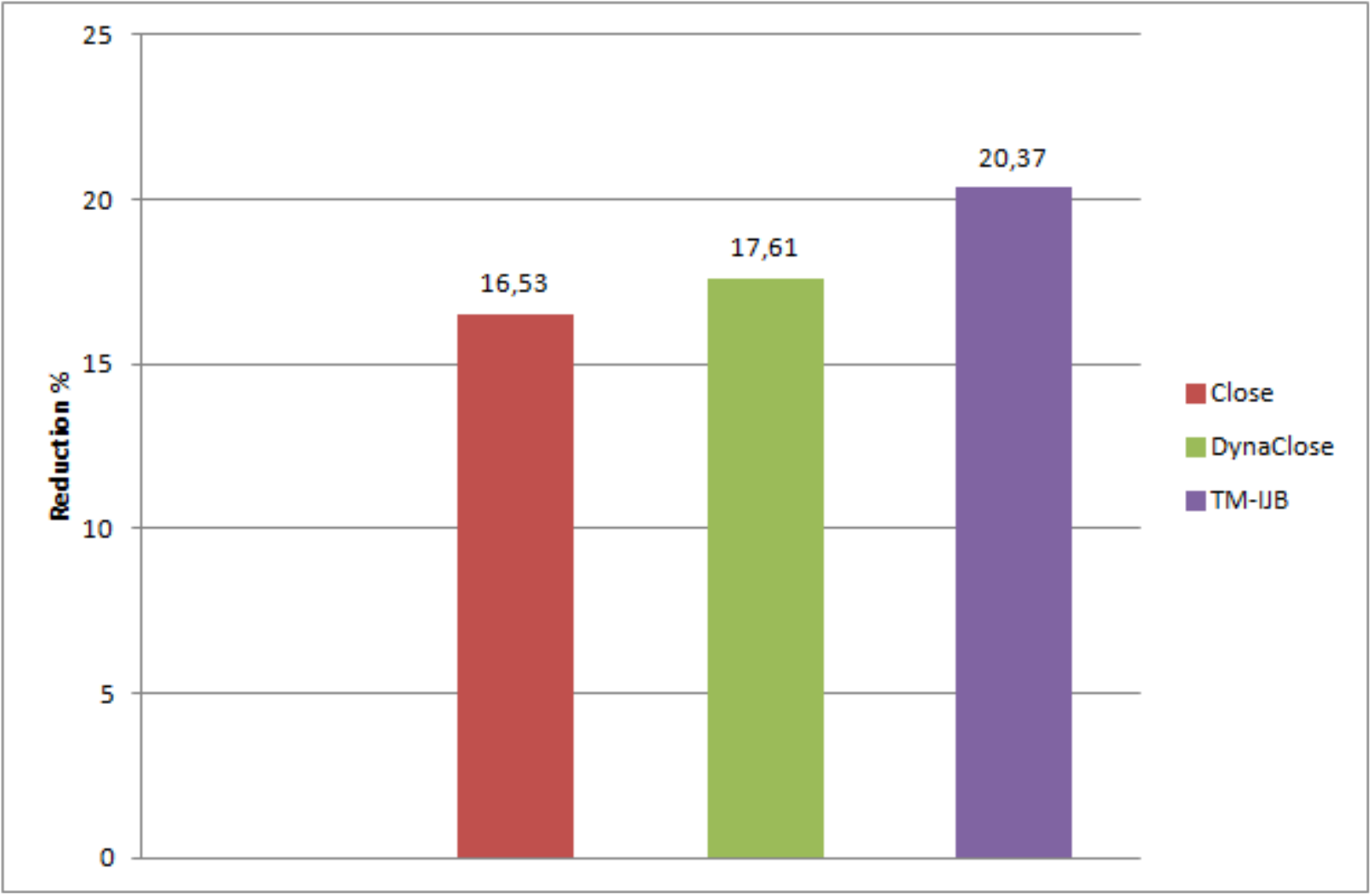}
\caption{Taux de réduction des 8 requêtes du banc de test TPC-H\label{ff3}}
\end{center}
\end{figure}

\indent L’histogramme de la figure  \ref{ff2}  montre le taux de réduction total de chaque approche. Notre approche a le taux de plus élevé de 13\% quant à \textsc{DynaClose} et \textsc{Close} est de 9.3\% et 8.3\% respectivement.  Ces taux s’élèvent considérablement, à presque le double, si nous calculons le gain seulement sur les 8 requêtes indexables comme l’illustre l’histogramme de la figure \ref{ff3}.

\begin{figure}[htbp]
\begin{center}
\includegraphics[scale=0.7]{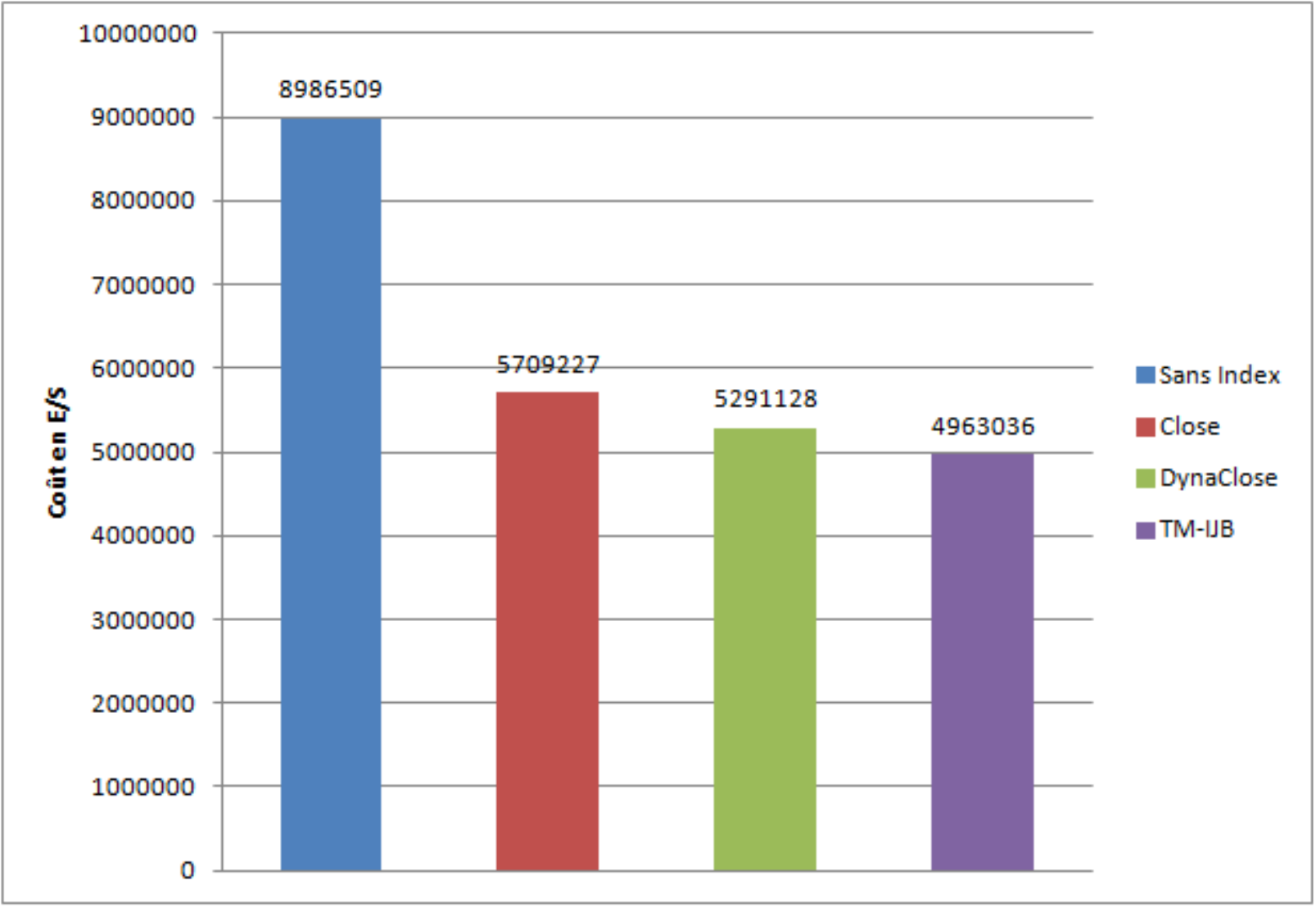}
\caption{Histogramme du coût d'exécution en E/S\label{ff4}}
\end{center}
\end{figure}

La figure \ref{ff4} montre comment les différentes approches d'indexation réduisent le coût d'exécution des 22 requêtes. Le résultat principal est que \textsc{TM-IJB} surpasse toutes les approches.

Comme nous pouvons le remarquer dans la figure \ref{ff4}, grâce au modèle de coût développé, nous avons pu quantifier la qualité de la structure d’optimisation sélectionnée $IJB$ sur les attributs proposés par les différentes approches. Ainsi, nous pouvons constater qu’il est plus bénéfique de tenir compte de l’utilisation des traverses minimales pour extraire les $IJB$.
\section*{Conclusion}

Dans ce chapitre, nous sommes passés de l’étude théorique à la partie expérimentale pour mettre en exergue le travail réalisé. Nous avons présenté les résultats des approches existantes et nos résultats en termes de temps d’exécution, taux de réduction et le coût en entrées/sorties pour valider notre approche, afin d’évaluer l’efficacité de cette dernière. Ce qui nous a permis de prouver l’importance de notre contribution, à savoir la pertinence d’utiliser les traverses minimales au service de sélection des $IJB$.

\clearpage \addcontentsline{toc}{chapter}{Conclusion générale et perspectives}
\chapter*{Conclusion générale et perspectives}
\markboth{Conclusion générale et perspectives}{Conclusion générale et perspectives}
Une des caractéristiques des entrepôts de données est leur volume de données important qui peut atteindre quelques milliers de giga-octets (téraoctet). Les entrepôts de données sont dédiés aux applications d’analyse et de prise de décision. Cette analyse est souvent réalisée par l’intermédiaire de requêtes complexes caractérisées par des opérations de sélections, de jointures et d’agrégations.  Pour optimiser ces requêtes, l’administrateur est amené à effectuer une tâche très importante : la conception physique durant laquelle, l’administrateur choisit un ensemble de techniques d’optimisation à sélectionner de manière isolée ou combinée. \\
\indent Dans ce travail, nous nous sommes ouvert au domaine le domaine des entrepôts de données son architecture, sa modélisation et ses différentes techniques d’optimisation. Nous avons décrit les indexs de jointure binaires, avec les problèmes liés à leurs sélections ainsi que les différents travaux proposés pour résoudre ces problèmes.\\

\indent Dans ce travail, nous avons essayé d'utiliser la technique d’index de jointure binaire qui est considérée comme une technique très pertinente pour l’optimisation des requêtes en étoile (OLAP) utilisées dans le cas des entrepôts de données. Nous avons proposé une approche de sélection d’une configuration finale d’index de jointure binaires. Notre approche repose sur l’utilisation d’une technique appelée les traverses minimales. La configuration finale obtenue est composée de plusieurs index qui permettent d'optimiser l’ensemble de requêtes. \\

\indent Nous avons effectué plusieurs expérimentations sur notre approche et sur des approches déjà proposés telles que \textsc{Close} et \textsc{DynaClose}, afin de prouver la qualité de notre approche par rapport à ces dernières.
À la fin, nous avons déduit que notre approche proposée a prouvé son efficacité quand à l’optimisation des performances des entrepôts de données. L’idée de proposer une nouvelle approche pour sélectionner une combinaison d’index de jointure binaire est très originale et prometteuse. Les résultats présentés dans ce mémoire montrent l’efficacité de notre approche tout en contribuant à un axe de recherche très prometteur et d’actualité.\\

\indent De nombreuses perspectives de recherche peuvent être envisagées. Nous présentons succinctement celles qui nous paraissent être les plus intéressantes.\\

\indent Les approches de sélection d’index de jointure binaires existantes sont contraintes seulement par l’espace de stockage, sera intéressant de considérer le coût de mise à jour des index sélectionnés comme étant une contrainte supplémentaire \cite{lpah22,lpah21}.\\

\indent Les approches de sélection que nous avons proposées se basent sur une charge de travail fixe construites à travers les requêtes considérées fréquentes. Cette charge peut évoluer et ainsi rendre la conception physique obsolète. Il serait intéressant de développer une stratégie permettant de détecter les changements intervenus au niveau de la charge de requêtes et de proposer un ensemble de recommandations à l’administrateur. L’identification des changements peut être effectuée en analysant les fichiers logs à travers l’implémentation d’un ensemble de déclencheurs (triggers). Les recommandations générées peuvent varier d’un simple avertissement sur un changement dans une requête à une proposition d’une nouvelle configuration d’index et une nouvelle conception physique de l’entrepôt \cite{lpah25,lpah24,lpah23}. \\

\indent Et enfin, nous avons identifié des similarités entre les indexs de jointure binaires et la  fragmentation horizontale dérivée, qui nous ont permis de les combiner. Il serait judicieux de refaire le même travail en considérant d’autres techniques comme la sélection des vues matérialisées\cite{lpah20,lpah21}. 

\chapter*{Annexe A}
\markboth{Annexe A}{Annexe A}
Nous présentons la charge de travail utilisée pour l’expérimentation de notre stratégie de sélection des $IJB$ sur le banc de tests $TPC$-$H$. 22 requêtes de plusieurs types ont été considérées : requêtes de type count(*) avec et sans agrégation, requêtes utilisant les fonctions d’agrégation comme \textit{Sum}, \textit{Min}, \textit{Max}, requêtes ayant des attributs de dimension dans la clause SELECT, etc.

\begin{scriptsize}

\begin{verbatim}

Q1 - SELECT L_RETURNFLAG, L_LINESTATUS, SUM(L_QUANTITY) AS SUM_QTY,
 SUM(L_EXTENDEDPRICE) AS SUM_BASE_PRICE, SUM(L_EXTENDEDPRICE*(1-L_DISCOUNT)) AS SUM_DISC_PRICE,
 SUM(L_EXTENDEDPRICE*(1-L_DISCOUNT)*(1+L_TAX)) AS SUM_CHARGE, AVG(L_QUANTITY) AS AVG_QTY,
 AVG(L_EXTENDEDPRICE) AS AVG_PRICE, AVG(L_DISCOUNT) AS AVG_DISC, COUNT(*) AS COUNT_ORDER
FROM LINEITEM
WHERE L_SHIPDATE <= dateadd(dd, -90, cast('1998-12-01' as date))
GROUP BY L_RETURNFLAG, L_LINESTATUS
ORDER BY L_RETURNFLAG,L_LINESTATUS


Q2 - SELECT TOP 100 S_ACCTBAL, S_NAME, N_NAME, P_PARTKEY, P_MFGR, S_ADDRESS, S_PHONE, S_COMMENT
FROM PART, SUPPLIER, PARTSUPP, NATION, REGION
WHERE P_PARTKEY = PS_PARTKEY AND S_SUPPKEY = PS_SUPPKEY AND P_SIZE = 15 AND
P_TYPE LIKE '%%BRASS' AND S_NATIONKEY = N_NATIONKEY AND N_REGIONKEY = R_REGIONKEY AND
R_NAME = 'EUROPE' AND
PS_SUPPLYCOST = (SELECT MIN(PS_SUPPLYCOST) FROM PARTSUPP, SUPPLIER, NATION, REGION
 WHERE P_PARTKEY = PS_PARTKEY AND S_SUPPKEY = PS_SUPPKEY
 AND S_NATIONKEY = N_NATIONKEY AND N_REGIONKEY = R_REGIONKEY AND R_NAME = 'EUROPE')
ORDER BY S_ACCTBAL DESC, N_NAME, S_NAME, P_PARTKEY



Q3 - SELECT TOP 10 L_ORDERKEY, SUM(L_EXTENDEDPRICE*(1-L_DISCOUNT)) AS REVENUE, O_ORDERDATE, O_SHIPPRIORITY
FROM CUSTOMER, ORDERS, LINEITEM
WHERE C_MKTSEGMENT = 'BUILDING' AND C_CUSTKEY = O_CUSTKEY AND L_ORDERKEY = O_ORDERKEY AND
O_ORDERDATE < '1995-03-15' AND L_SHIPDATE > '1995-03-15'
GROUP BY L_ORDERKEY, O_ORDERDATE, O_SHIPPRIORITY
ORDER BY REVENUE DESC, O_ORDERDATE


Q4 - SELECT O_ORDERPRIORITY, COUNT(*) AS ORDER_COUNT FROM ORDERS
WHERE O_ORDERDATE >= '1993-07-01' AND O_ORDERDATE < dateadd(mm,3, cast('1993-07-01' as date))
AND EXISTS (SELECT * FROM LINEITEM WHERE L_ORDERKEY = O_ORDERKEY AND L_COMMITDATE < L_RECEIPTDATE)
GROUP BY O_ORDERPRIORITY
ORDER BY O_ORDERPRIORITY


Q5 - SELECT N_NAME, SUM(L_EXTENDEDPRICE*(1-L_DISCOUNT)) AS REVENUE
FROM CUSTOMER, ORDERS, LINEITEM, SUPPLIER, NATION, REGION
WHERE C_CUSTKEY = O_CUSTKEY AND L_ORDERKEY = O_ORDERKEY AND L_SUPPKEY = S_SUPPKEY
AND C_NATIONKEY = S_NATIONKEY AND S_NATIONKEY = N_NATIONKEY AND N_REGIONKEY = R_REGIONKEY
AND R_NAME = 'ASIA' AND O_ORDERDATE >= '1994-01-01' 
AND O_ORDERDATE < DATEADD(YY, 1, cast('1994-01-01' as date))
GROUP BY N_NAME
ORDER BY REVENUE DESC


Q6 - SELECT SUM(L_EXTENDEDPRICE*L_DISCOUNT) AS REVENUE
FROM LINEITEM
WHERE L_SHIPDATE >= '1994-01-01' AND L_SHIPDATE < dateadd(yy, 1, cast('1994-01-01' as date))
AND L_DISCOUNT BETWEEN .06 - 0.01 AND .06 + 0.01 AND L_QUANTITY < 24


Q7 - SELECT SUPP_NATION, CUST_NATION, L_YEAR, SUM(VOLUME) AS REVENUE
FROM ( SELECT N1.N_NAME AS SUPP_NATION, N2.N_NAME AS CUST_NATION, datepart(yy, L_SHIPDATE) AS L_YEAR,
 L_EXTENDEDPRICE*(1-L_DISCOUNT) AS VOLUME
 FROM SUPPLIER, LINEITEM, ORDERS, CUSTOMER, NATION N1, NATION N2
 WHERE S_SUPPKEY = L_SUPPKEY AND O_ORDERKEY = L_ORDERKEY AND C_CUSTKEY = O_CUSTKEY
 AND S_NATIONKEY = N1.N_NATIONKEY AND C_NATIONKEY = N2.N_NATIONKEY AND
 ((N1.N_NAME = 'FRANCE' AND N2.N_NAME = 'GERMANY') OR
 (N1.N_NAME = 'GERMANY' AND N2.N_NAME = 'FRANCE')) AND
 L_SHIPDATE BETWEEN '1995-01-01' AND '1996-12-31' ) AS SHIPPING
GROUP BY SUPP_NATION, CUST_NATION, L_YEAR
ORDER BY SUPP_NATION, CUST_NATION, L_YEAR


Q8 - SELECT O_YEAR, SUM(CASE WHEN NATION = 'BRAZIL' THEN VOLUME ELSE 0 END)/SUM(VOLUME) AS MKT_SHARE
FROM (SELECT datepart(yy,O_ORDERDATE) AS O_YEAR, L_EXTENDEDPRICE*(1-L_DISCOUNT) AS VOLUME, N2.N_NAME AS NATION
 FROM PART, SUPPLIER, LINEITEM, ORDERS, CUSTOMER, NATION N1, NATION N2, REGION
 WHERE P_PARTKEY = L_PARTKEY AND S_SUPPKEY = L_SUPPKEY AND L_ORDERKEY = O_ORDERKEY
 AND O_CUSTKEY = C_CUSTKEY AND C_NATIONKEY = N1.N_NATIONKEY AND
 N1.N_REGIONKEY = R_REGIONKEY AND R_NAME = 'AMERICA' AND S_NATIONKEY = N2.N_NATIONKEY
 AND O_ORDERDATE BETWEEN '1995-01-01' AND '1996-12-31' AND P_TYPE= 'ECONOMY ANODIZED STEEL') AS ALL_NATIONS
GROUP BY O_YEAR
ORDER BY O_YEAR


Q9 - SELECT NATION, O_YEAR, SUM(AMOUNT) AS SUM_PROFIT
FROM (SELECT N_NAME AS NATION, datepart(yy, O_ORDERDATE) AS O_YEAR,
 L_EXTENDEDPRICE*(1-L_DISCOUNT)-PS_SUPPLYCOST*L_QUANTITY AS AMOUNT
 FROM PART, SUPPLIER, LINEITEM, PARTSUPP, ORDERS, NATION
 WHERE S_SUPPKEY = L_SUPPKEY AND PS_SUPPKEY= L_SUPPKEY AND PS_PARTKEY = L_PARTKEY AND
 P_PARTKEY= L_PARTKEY AND O_ORDERKEY = L_ORDERKEY AND S_NATIONKEY = N_NATIONKEY AND
 P_NAME LIKE '%%green%%') AS PROFIT
GROUP BY NATION, O_YEAR
ORDER BY NATION, O_YEAR DESC


Q10 - SELECT TOP 20 C_CUSTKEY, C_NAME, SUM(L_EXTENDEDPRICE*(1-L_DISCOUNT)) AS REVENUE, C_ACCTBAL,
N_NAME, C_ADDRESS, C_PHONE, C_COMMENT
FROM CUSTOMER, ORDERS, LINEITEM, NATION
WHERE C_CUSTKEY = O_CUSTKEY AND L_ORDERKEY = O_ORDERKEY AND O_ORDERDATE>= '1993-10-01' AND
O_ORDERDATE < dateadd(mm, 3, cast('1993-10-01' as date)) AND
L_RETURNFLAG = 'R' AND C_NATIONKEY = N_NATIONKEY
GROUP BY C_CUSTKEY, C_NAME, C_ACCTBAL, C_PHONE, N_NAME, C_ADDRESS, C_COMMENT
ORDER BY REVENUE DESC


Q11 - SELECT PS_PARTKEY, SUM(PS_SUPPLYCOST*PS_AVAILQTY) AS VALUE
FROM PARTSUPP, SUPPLIER, NATION
WHERE PS_SUPPKEY = S_SUPPKEY AND S_NATIONKEY = N_NATIONKEY AND N_NAME = 'GERMANY'
GROUP BY PS_PARTKEY
HAVING SUM(PS_SUPPLYCOST*PS_AVAILQTY) > (SELECT SUM(PS_SUPPLYCOST*PS_AVAILQTY) * 0.0001000000
 FROM PARTSUPP, SUPPLIER, NATION
 WHERE PS_SUPPKEY = S_SUPPKEY AND S_NATIONKEY = N_NATIONKEY AND N_NAME = 'GERMANY')
ORDER BY VALUE DESC


Q12 - SELECT L_SHIPMODE,
SUM(CASE WHEN O_ORDERPRIORITY = '1-URGENT' OR O_ORDERPRIORITY = '2-HIGH' THEN 1 ELSE 0 END) AS HIGH_LINE_COUNT,
SUM(CASE WHEN O_ORDERPRIORITY <> '1-URGENT' AND O_ORDERPRIORITY <> '2-HIGH' THEN 1 ELSE 0 END ) AS LOW_LINE_COUNT
FROM ORDERS, LINEITEM
WHERE O_ORDERKEY = L_ORDERKEY AND L_SHIPMODE IN ('MAIL','SHIP')
AND L_COMMITDATE < L_RECEIPTDATE AND L_SHIPDATE < L_COMMITDATE AND L_RECEIPTDATE >= '1994-01-01'
AND L_RECEIPTDATE < dateadd(mm, 1, cast('1995-09-01' as date))
GROUP BY L_SHIPMODE
ORDER BY L_SHIPMODE


Q13 - SELECT C_COUNT, COUNT(*) AS CUSTDIST
FROM (SELECT C_CUSTKEY, COUNT(O_ORDERKEY)
 FROM CUSTOMER left outer join ORDERS on C_CUSTKEY = O_CUSTKEY
 AND O_COMMENT not like '%%special%%requests%%'
 GROUP BY C_CUSTKEY) AS C_ORDERS (C_CUSTKEY, C_COUNT)
GROUP BY C_COUNT
ORDER BY CUSTDIST DESC, C_COUNT DESC


Q14 - SELECT 100.00* SUM(CASE WHEN P_TYPE LIKE 'PROMO%%' THEN L_EXTENDEDPRICE*(1-L_DISCOUNT)
ELSE 0 END) / SUM(L_EXTENDEDPRICE*(1-L_DISCOUNT)) AS PROMO_REVENUE
FROM LINEITEM, PART
WHERE L_PARTKEY = P_PARTKEY AND L_SHIPDATE >= '1995-09-01' AND L_SHIPDATE < dateadd(mm, 1, '1995-09-01')


Q15 - CREATE VIEW REVENUE0 (SUPPLIER_NO, TOTAL_REVENUE) AS
SELECT L_SUPPKEY, SUM(L_EXTENDEDPRICE*(1-L_DISCOUNT)) FROM LINEITEM
WHERE L_SHIPDATE >= '1996-01-01' AND L_SHIPDATE < dateadd(mm, 3, cast('1996-01-01' as date))
GROUP BY L_SUPPKEY

SELECT S_SUPPKEY, S_NAME, S_ADDRESS, S_PHONE, TOTAL_REVENUE
FROM SUPPLIER, REVENUE0
WHERE S_SUPPKEY = SUPPLIER_NO AND TOTAL_REVENUE = (SELECT MAX(TOTAL_REVENUE) FROM REVENUE0)
ORDER BY S_SUPPKEY
DROP VIEW REVENUE0


Q16 - SELECT P_BRAND, P_TYPE, P_SIZE, COUNT(DISTINCT PS_SUPPKEY) AS SUPPLIER_CNT
FROM PARTSUPP, PART
WHERE P_PARTKEY = PS_PARTKEY AND P_BRAND <> 'Brand#45' AND P_TYPE NOT LIKE 'MEDIUM POLISHED%%'
AND P_SIZE IN (49, 14, 23, 45, 19, 3, 36, 9) AND PS_SUPPKEY NOT IN (SELECT S_SUPPKEY FROM SUPPLIER
 WHERE S_COMMENT LIKE '%%Customer%%Complaints%%')
GROUP BY P_BRAND, P_TYPE, P_SIZE
ORDER BY SUPPLIER_CNT DESC, P_BRAND, P_TYPE, P_SIZE


Q17 - SELECT SUM(L_EXTENDEDPRICE)/7.0 AS AVG_YEARLY FROM LINEITEM, PART
WHERE P_PARTKEY = L_PARTKEY AND P_BRAND = 'Brand#23' AND P_CONTAINER = 'MED BOX'
AND L_QUANTITY < (SELECT 0.2*AVG(L_QUANTITY) FROM LINEITEM WHERE L_PARTKEY = P_PARTKEY)


Q18 - SELECT TOP 100 C_NAME, C_CUSTKEY, O_ORDERKEY, O_ORDERDATE, O_TOTALPRICE, SUM(L_QUANTITY)
FROM CUSTOMER, ORDERS, LINEITEM
WHERE O_ORDERKEY IN (SELECT L_ORDERKEY FROM LINEITEM GROUP BY L_ORDERKEY HAVING
 SUM(L_QUANTITY) > 300) AND C_CUSTKEY = O_CUSTKEY AND O_ORDERKEY = L_ORDERKEY
GROUP BY C_NAME, C_CUSTKEY, O_ORDERKEY, O_ORDERDATE, O_TOTALPRICE
ORDER BY O_TOTALPRICE DESC, O_ORDERDATE


Q19 - SELECT SUM(L_EXTENDEDPRICE* (1 - L_DISCOUNT)) AS REVENUE
FROM LINEITEM, PART
WHERE (P_PARTKEY = L_PARTKEY AND P_BRAND = 'Brand#12' 
AND P_CONTAINER IN ('SM CASE', 'SM BOX', 'SM PACK', 'SM PKG') 
AND L_QUANTITY >= 1 AND L_QUANTITY <= 1 + 10 AND P_SIZE BETWEEN 1 AND 5
AND L_SHIPMODE IN ('AIR', 'AIR REG') AND L_SHIPINSTRUCT = 'DELIVER IN PERSON')
OR (P_PARTKEY = L_PARTKEY AND P_BRAND ='Brand#23' 
AND P_CONTAINER IN ('MED BAG', 'MED BOX', 'MED PKG', 'MED PACK') 
AND L_QUANTITY >=10 AND L_QUANTITY <=10 + 10 AND P_SIZE BETWEEN 1 AND 10 
AND L_SHIPMODE IN ('AIR', 'AIR REG') AND L_SHIPINSTRUCT = 'DELIVER IN PERSON') 
OR (P_PARTKEY = L_PARTKEY AND P_BRAND = 'Brand#34' 
AND P_CONTAINER IN ( 'LG CASE', 'LG BOX', 'LG PACK', 'LG PKG') 
AND L_QUANTITY >=20 AND L_QUANTITY <= 20 + 10 AND P_SIZE BETWEEN 1 AND 15
AND L_SHIPMODE IN ('AIR', 'AIR REG') AND L_SHIPINSTRUCT = 'DELIVER IN PERSON')


Q20 - SELECT S_NAME, S_ADDRESS FROM SUPPLIER, NATION
WHERE S_SUPPKEY IN (SELECT PS_SUPPKEY FROM PARTSUPP
 WHERE PS_PARTKEY in (SELECT P_PARTKEY FROM PART WHERE P_NAME like 'forest%%') AND
 PS_AVAILQTY > (SELECT 0.5*sum(L_QUANTITY) FROM LINEITEM WHERE L_PARTKEY = PS_PARTKEY AND
  L_SUPPKEY = PS_SUPPKEY AND L_SHIPDATE >= '1994-01-01' AND
  L_SHIPDATE < dateadd(yy,1,'1994-01-01'))) AND S_NATIONKEY = N_NATIONKEY AND N_NAME = 'CANADA'
ORDER BY S_NAME


Q21 - SELECT TOP 100 S_NAME, COUNT(*) AS NUMWAIT
FROM SUPPLIER, LINEITEM L1, ORDERS, NATION WHERE S_SUPPKEY = L1.L_SUPPKEY AND
O_ORDERKEY = L1.L_ORDERKEY AND O_ORDERSTATUS = 'F' AND L1.L_RECEIPTDATE> L1.L_COMMITDATE
AND EXISTS (SELECT * FROM LINEITEM L2 WHERE L2.L_ORDERKEY = L1.L_ORDERKEY
 AND L2.L_SUPPKEY <> L1.L_SUPPKEY) AND
NOT EXISTS (SELECT * FROM LINEITEM L3 WHERE L3.L_ORDERKEY = L1.L_ORDERKEY AND
 L3.L_SUPPKEY <> L1.L_SUPPKEY AND L3.L_RECEIPTDATE > L3.L_COMMITDATE) AND
S_NATIONKEY = N_NATIONKEY AND N_NAME = 'SAUDI ARABIA'
GROUP BY S_NAME
ORDER BY NUMWAIT DESC, S_NAME


Q22 - SELECT CNTRYCODE, COUNT(*) AS NUMCUST, SUM(C_ACCTBAL) AS TOTACCTBAL
FROM (SELECT SUBSTRING(C_PHONE,1,2) AS CNTRYCODE, C_ACCTBAL
 FROM CUSTOMER WHERE SUBSTRING(C_PHONE,1,2) IN ('13', '31', '23', '29', '30', '18', '17') AND
 C_ACCTBAL > (SELECT AVG(C_ACCTBAL) FROM CUSTOMER WHERE C_ACCTBAL > 0.00 AND
  SUBSTRING(C_PHONE,1,2) IN ('13', '31', '23', '29', '30', '18', '17')) AND
 NOT EXISTS ( SELECT * FROM ORDERS WHERE O_CUSTKEY = C_CUSTKEY)) AS CUSTSALE
GROUP BY CNTRYCODE
ORDER BY CNTRYCODE

\end{verbatim}
\end{scriptsize} 
\chapter*{Annexe B}
\markboth{Annexe B}{Annexe B}
Nous présentons la charge de travail utilisée pour l’expérimentation de notre stratégie de sélection des $IJB$ sur le banc de tests $SSB1$. 30 requêtes de plusieurs types ont été considérées : requêtes de type count(*) avec et sans agrégation, requêtes utilisant les fonctions d’agrégation comme \textit{Sum}, \textit{Min}, \textit{Max}, requêtes ayant des attributs de dimension dans la clause SELECT, etc.

\begin{scriptsize}

\begin{verbatim}
Q1 - select sum(lo_extendedprice*lo_discount) as revenue 
from lineorder, dates 
where lo_orderdate = d_datekey and d_year = 1993 
and lo_discount >= 1 and lo_discount <= 3 and lo_quantity < 25

Q2 - select count(*) f
rom lineorder, dates 
where lo_orderdate = d_datekey and d_year = 1993 
and lo_discount >= 1 and lo_discount <= 3 and lo_quantity < 25

Q3 - select sum(lo_extendedprice*lo_discount) as revenue 
from lineorder, dates 
where lo_orderdate = d_datekey and d_year = 1993

Q4 - select count(*) 
from lineorder, dates 
where lo_orderdate = d_datekey and d_year = 1993


Q5 - select sum(lo_revenue), d_year 
from lineorder, dates, part, supplier 
where lo_orderdate = d_datekey and lo_partkey = p_partkey 
and lo_suppkey = s_suppkey and p_brand = 'MFGR#2221' and s_region = 'ASIA' 
group by d_year order by d_year


Q6 - select sum(lo_revenue) 
from lineorder, part 
where lo_partkey = p_partkey and p_brand = 'MFGR#2221'  


Q7 - select avg(lo_revenue), d_year 
from lineorder, dates, part, supplier 
where lo_orderdate = d_datekey and lo_partkey = p_partkey 
and lo_suppkey = s_suppkey and p_brand = 'MFGR#2221' and s_region = 'ASIA' 
group by d_year order by d_year


Q8 - select sum(lo_revenue), d_year 
from lineorder, dates, part, supplier 
where lo_orderdate = d_datekey and lo_partkey = p_partkey 
and lo_suppkey = s_suppkey and p_brand = 'MFGR#2221' and s_region = 'EUROPE' 
group by d_year, p_brand order by d_year, p_brand


Q9 - select sum(lo_revenue) 
from lineorder, part, supplier 
where lo_partkey = p_partkey and lo_suppkey = s_suppkey 
and p_brand = 'MFGR#2221' and s_region = 'EUROPE' 

Q10 - select count(*), d_year 
from lineorder, dates, part, supplier 
where lo_orderdate = d_datekey and lo_partkey = p_partkey 
and lo_suppkey = s_suppkey and p_brand = 'MFGR#2221' and s_region = 'EUROPE' 
group by d_year order by d_year

Q11 - select c_nation, s_nation, d_year, sum(lo_revenue) as revenue 
from lineorder,customer,  supplier, dates 
where lo_custkey = c_custkey and lo_suppkey = s_suppkey and lo_orderdate = d_datekey 
and c_region = 'ASIA' and s_region = 'ASIA' and d_year >= 1992 
and d_year <= 1997 group by c_nation, s_nation, d_year order by d_year asc, revenue desc

Q12 - select s_nation, sum(lo_revenue) as revenue 
from lineorder, supplier 
where lo_suppkey = s_suppkey and s_region = 'ASIA' group by s_nation order by revenue desc


Q13 - select s_nation, count(*) as revenue 
from lineorder, supplier 
where lo_suppkey = s_suppkey and s_region = 'ASIA' group by s_nation order by revenue desc

Q14 - select s_nation, d_year, sum(lo_revenue) as revenue 
from lineorder, supplier, dates 
where lo_suppkey = s_suppkey and lo_orderdate = d_datekey and s_region = 'ASIA' 
and d_year >= 1992 and d_year <= 1997 
group by s_nation, d_year order by d_year asc, revenue desc


Q15 - select c_nation, s_nation, d_year, avg(lo_revenue) as avg_revenue 
from lineorder,customer,  supplier, dates 
where lo_custkey = c_custkey and lo_suppkey = s_suppkey and lo_orderdate = d_datekey 
and c_region = 'ASIA' and s_region = 'ASIA' and d_year >= 1992 and d_year <= 1997 
group by c_nation, s_nation, d_year order by d_year asc, avg_revenue desc

Q16 - select c_nation, s_nation, d_year, count(*) 
from lineorder,customer,  supplier, dates 
where lo_custkey = c_custkey and lo_suppkey = s_suppkey and lo_orderdate = d_datekey 
and c_region = 'ASIA' and s_region = 'ASIA' and d_year >= 1992 and d_year <= 1997 
group by c_nation, s_nation, d_year order by d_year asc


Q17 - select c_city, s_city, d_year, sum(lo_revenue) as revenue 
from lineorder,customer,  supplier, dates 
where lo_custkey = c_custkey and lo_suppkey = s_suppkey and lo_orderdate = d_datekey 
and c_nation = 'UNITED STATES' and s_nation = 'UNITED STATES' and d_year >= 1992 and d_year <= 1997 
group by c_city, s_city, d_year order by d_year asc, revenue desc

Q18 - select s_city, sum(lo_revenue) as revenue 
from lineorder, supplier 
where lo_suppkey = s_suppkey and s_nation = 'UNITED STATES' 
group by s_city order by revenue desc

Q19 - select s_city, avg(lo_revenue) as avg_revenue 
from lineorder, supplier where 
lo_suppkey = s_suppkey and s_nation = 'UNITED STATES' 
group by s_city order by avg_revenue desc

Q20 - select c_city, s_city, count(*) 
from lineorder,customer,  supplier 
where lo_custkey = c_custkey and lo_suppkey = s_suppkey 
and c_nation = 'UNITED STATES' and s_nation = 'UNITED STATES' 
group by c_city, s_city 

Q21 - select c_city, s_city, d_year, avg(lo_revenue) as avg_revenue 
from  lineorder,customer, supplier, dates 
where lo_custkey = c_custkey and lo_suppkey = s_suppkey and lo_orderdate = d_datekey 
and c_nation = 'UNITED STATES' and s_nation = 'UNITED STATES' and d_year >= 1992 and d_year <= 1997 
group by c_city, s_city, d_year order by d_year asc, avg_revenue desc 

Q22 - select c_city, s_city, d_year, count(*) 
from lineorder,customer,  supplier, dates 
where lo_custkey = c_custkey and lo_suppkey = s_suppkey and lo_orderdate = d_datekey 
and c_nation = 'UNITED STATES' and s_nation = 'UNITED STATES' and d_year >= 1992 and d_year <= 1997 
group by c_city, s_city, d_year order by d_year asc

Q23 - select d_year, s_nation, p_category, sum(lo_revenue - lo_supplycost) as profit 
from lineorder,dates, customer, supplier, part  
where lo_custkey = c_custkey and lo_suppkey = s_suppkey and lo_partkey = p_partkey 
and lo_orderdate = d_datekey and c_region = 'AMERICA' and s_region = 'AMERICA' 
and (d_year = 1997 or d_year = 1998) and (p_mfgr = 'MFGR#1' or p_mfgr = 'MFGR#2') 
group by d_year, s_nation, p_category order by d_year, s_nation, p_category

Q24 - select d_year, s_nation, p_category, sum(lo_revenue - lo_supplycost) as profit 
from lineorder, dates, supplier, part  
where lo_suppkey = s_suppkey and lo_partkey = p_partkey and lo_orderdate = d_datekey and s_region = 'AMERICA' 
and (d_year = 1997 or d_year = 1998) and (p_mfgr = 'MFGR#1' or p_mfgr = 'MFGR#2') 
group by d_year, s_nation, p_category order by d_year, s_nation, p_category

Q25 - select d_year, s_nation, p_category, avg(lo_revenue - lo_supplycost) as avg_profit 
from lineorder, dates, customer, supplier, part   
where lo_custkey = c_custkey and lo_suppkey = s_suppkey and lo_partkey = p_partkey and lo_orderdate = d_datekey
and c_region = 'AMERICA' and s_region = 'AMERICA' and (d_year = 1997 or d_year = 1998) 
and (p_mfgr = 'MFGR#1' or p_mfgr = 'MFGR#2') 
group by d_year, s_nation, p_category order by d_year, s_nation, p_category

Q26 - select d_year, s_nation, p_category, count(*) 
from lineorder, dates, customer, supplier, part  
where lo_custkey = c_custkey and lo_suppkey = s_suppkey and lo_partkey = p_partkey and 
lo_orderdate = d_datekey and c_region = 'AMERICA' and s_region = 'AMERICA' and (d_year = 1997 or d_year = 1998) 
and (p_mfgr = 'MFGR#1' or p_mfgr = 'MFGR#2') 
group by d_year, s_nation, p_category order by d_year, s_nation, p_category

Q27 - select d_year, s_nation, count(*) 
from  lineorder,dates, supplier  
where lo_suppkey = s_suppkey and lo_orderdate = d_datekey and s_region = 'AMERICA' 
and (d_year = 1997 or d_year = 1998) 
group by d_year, s_nation order by d_year, s_nation

Q28 - select s_nation, count(*) from lineorder,supplier   
where lo_suppkey = s_suppkey and s_region = 'AMERICA' 
group by s_nation order by s_nation

Q29 - select d_year, s_nation, sum(lo_revenue) as revenue 
from lineorder,dates, supplier   
where lo_suppkey = s_suppkey and lo_orderdate = d_datekey and s_region = 'AMERICA' 
and (d_year = 1997 or d_year = 1998) group by d_year, s_nation order by d_year, s_nation

Q30 - select sum(lo_revenue), p_brand 
from lineorder, part, supplier 
where lo_partkey = p_partkey and lo_suppkey = s_suppkey and p_brand = 'MFGR#2221' 
and s_region = 'ASIA' group by p_brand order by p_brand

\end{verbatim}

\end{scriptsize}

\includepdf[pages=-]{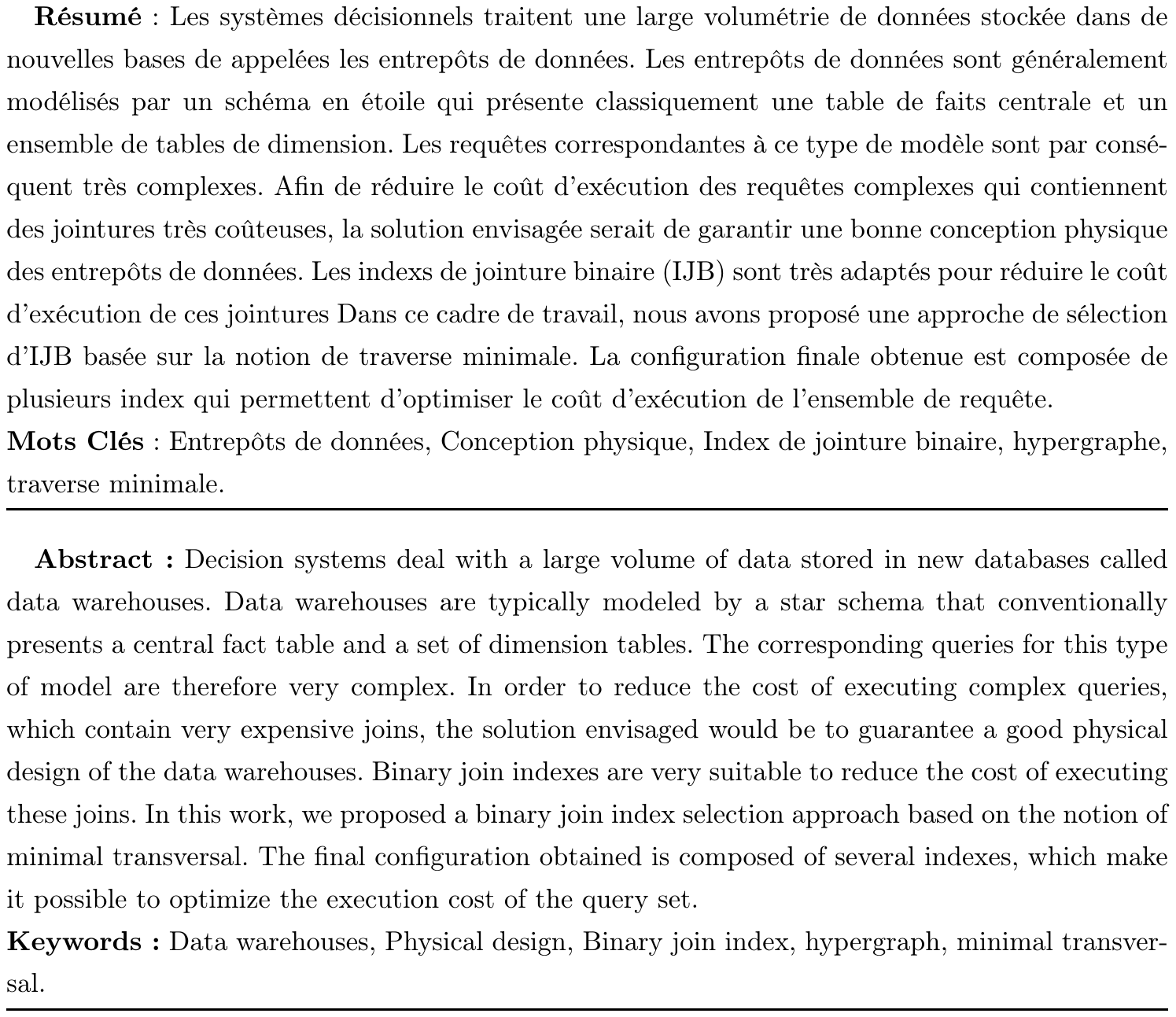}
\end{document}